\documentclass[english]{article}
\usepackage{amssymb}
\usepackage{fontenc}
\usepackage[latin1]{inputenc}
\usepackage{amsmath}
\usepackage{fontenc}
\usepackage{babel}

\input{tcilatex}
\begin{document}

\title{{\Large One-loop energy-momentum tensor in QED with electric-like
background}}
\author{S.P. Gavrilov\thanks{%
Department of General and Experimental Physics, Herzen State Pedagogical
University of Russia, Moyka emb. 48, 191186 St. Petersburg, Russia; e-mail:
gavrilovsergeyp@yahoo.com} and D.M. Gitman \thanks{%
Institute of Physics, University of São Paulo, Brazil; e-mail:
gitman@dfn.if.usp.br}}
\maketitle

\begin{abstract}
We have obtained nonperturbative one-loop expressions for the mean
energy-momentum tensor and current density of Dirac's field on a constant
electric-like background. One of the goals of this calculation is to give a
consistent description of back-reaction in such a theory. Two cases of
initial states are considered: the vacuum state and the thermal equilibrium
state. First, we perform calculations for the vacuum initial state. In the
obtained expressions, we separate the contributions due to particle creation
and vacuum polarization. The latter contributions are related to the
Heisenberg--Euler Lagrangian. Then, we study the case of the thermal initial
state. Here, we separate the contributions due to particle creation, vacuum
polarization, and the contributions due to the work of the external field on
the particles at the initial state. All these contributions are studied in
detail, in different regimes of weak and strong fields and low and high
temperatures. The obtained results allow us to establish restrictions on the
electric field and its duration under which QED with a strong constant
electric field is consistent. Under such restrictions, one can neglect the
back-reaction of particles created by the electric field. Some of the
obtained results generalize the calculations of Heisenberg--Euler for energy
density to the case of arbitrary strong electric fields.

\end{abstract}

\section{Introduction}

It is well-known that quantum field theory (QFT) in external backgrounds is
a good model for the study of quantum processes in cases when a part of a
quantized field is strong enough to be treated as a classical one. Here, it
is assumed that quantum processes under consideration do not affect
significantly the classical field (back-reaction is supposed to be small)
and the field is treated as an external one. Of course, it is
well-understood that, in principle, back-reaction must be calculated, at any
rate, to evaluate the limits of applicability of the obtained results.
However, the latter problem often remains open due to its high complexity.
From physical considerations, it is clear that back-reaction may be very
strong, so it must be taken into account for external backgrounds that can
create particles from vacuum (for theories with an unstable vacuum under the
action of an external field) and produce actual work on particles. The
effect of particle-creation from vacuum by an external field (vacuum
instability in external fields) ranks among the most intriguing nonlinear
phenomena in quantum theory. Its consideration is theoretically important,
since it demands that one should go beyond the scope of perturbation theory,
and its experimental observation would be able to verify the validity of a
theory in the domain of superstrong fields. Concerning the history of the
subject, the reader may refer to the books \cite{N79,GMM94,FGS91,GMR85} and
to the articles \cite{DunS05} for recent computational developments. Typical
problems examined in the models with an unstable vacuum are related to the
calculation of the density of particles created from vacuum. In some cases,
this allows one to make phenomenological conclusions about back-reaction;
see, e.g., \cite{KluES93}. However, a complete description of back-reaction
is related to the calculation of mean values for the current density and
energy-momentum tensor (EMT) of the matter field. This problem, is, in fact,
one of the goals of the present article. In this connection, we recall that
Heisenberg and Euler calculated the mean value of energy density of Dirac's
field in a constant electromagnetic background at zero temperature under the
condition of weakness of the electric field\footnote{$M$ is the electron
mass and $e$ is the absolute value of the electron charge.} $\mathbf{E},$ $%
\left| \mathbf{E}\right| \ll E_{c}=M^{2}c^{3}/e\hbar \simeq 1,3\cdot
10^{16}\;V/cm$, see \cite{HeiE36}. Having this energy density at hand, they
constructed a one-loop effective Lagrangian $\mathcal{L}.$ It turns out that
this Lagrangian had been re-introduced by Schwinger \cite{S51} (for
subsequent developments, see the review \cite{Dunn04}) to define the
vacuum-to-vacuum transition amplitude $c_{v}$ as follows:%
\begin{equation}
c_{v}=\exp \left( i\int dx\mathcal{L}\right) ,  \label{vacampl}
\end{equation}%
without any restrictions on the external field. Schwinger demonstrated that
the probability $P^{v}$ of a vacuum to remain a vacuum in a constant
electric field is related to the imaginary part of $\mathcal{L}$ as follows: 
\begin{equation}
P^{v}=|c_{v}|^{2}=\exp \{-VT2\func{Im}\mathcal{L}\}\,.  \label{vacver}
\end{equation}%
Here, $T$ is the field duration, and $V$ is the volume of observation.
Obviously, the effect can actually be observed as soon as the strength of
the external field approaches the critical value $E_{c}$. Given this, there
remains an open question of what is the form of the energy density, as well
as the form of the entire EMT, in arbitrary constant fields. In the present
article, we study the mean current density and EMT of a spinor field subject
to the action of an electric-like external background, while paying a
special attention to the initial state of thermal equilibrium. A more exact
and detailed setting of the problem is given in the next section.

Aside from the principal importance of describing back-reaction for an
understanding of the limits of applicability of QED in external fields, the
calculations of the article are closely related to a wide class of problems
(in the framework of quantum-field models in external fields) that recently
attract considerable attention. Below, we present a brief review of these
problems and of their importance, which may be regarded as an additional
motivation for the present study.

Even though an actual possibility of creating superstrong fields under
laboratory conditions does not exist at present, the $e^{+}e^{-}$-pair
production from vacuum by a slowly varying external electric field is
relevant to phenomenology with the advent of a new laser technology capable
of accessing the true strong-field domain. It is widely discussed \cite%
{Dre02} at the SLAC and TESLA X-ray laser facilities. Such strong fields may
be relevant in astrophysics, where characteristic values of electromagnetic
and gravitational fields near black holes are enormous. The Coulomb barrier
at the quark surface of a hot strange star may be a powerful source of $%
e^{+}e^{-}$-pairs, which are created in extremely strong constant electric
fields (dozens of times higher that the critical field $E_{c}$) of the
barrier and the flow away from the star \cite{Usov97}. Such emission could
be the main observational signature of quark stars. The electric field near
a cosmic string can also be extremely strong \cite{NCV99}.

Apart from pure QED problems, there exist some closely related QFT problems
in which vacuum instability in various electric-like external backgrounds
plays an important role, for example, phase transitions in non-Abelian
theories, the problem of boundary conditions, or the influence of topology
on vacuum, the problem of a consistent vacuum construction in QCD and GUT,
multiple particle-creation in the context of heavy-ion collisions, etc.
Recently, it has also been recognized that the presence of a background
electric field must be taken into account in string theory constructions;
see, e.g., \cite{string} and references therein.

In the early 1970s, finite-temperature effects were recognized as very
important in QFT, and then Dittrich \cite{Dit79} computed the one-loop
effective potential of finite-temperature QED in the presence of a constant
magnetic field, starting from the Schwinger relation (\ref{vacampl}). This
was followed by an intense study of finite-temperature and
finite-chemical-potential one-loop effects for fermions in a constant
uniform magnetic field; see for a review \cite{CanD96,ElmPS93}. This study
was motivated by its possible applications to various areas of physics,
including astrophysics, condensed matter, particle physics, etc., where
strong electromagnetic fields may exist. The presence of an electric-like
field implies a qualitative change (in comparison with a magnetic-like
external field alone) in the character of processes involved, due to vacuum
instability{\Large \ }and the work of the field on the particles at the
initial state.

In some of the above-mentioned examples with an electric-like external
field, the initial state is in thermal equilibrium. In the chromoelectric
flux-tube model \cite{CasNN79}, the back-reaction of created pairs induces a
gluon mean field and plasma oscillations (see \cite{KluES93} and references
therein). Various problems of cosmological QCD phase-transitions and dark
matter formation are discussed on the basis of the chromoelectric flux tube
model (see, for example, \cite{Bha99} and references therein). It appears
that for a calculation of particle creation in this model one needs to apply
the general formalism of QFT for pair-production at finite temperatures and
at zero temperature, both from vacuum and from many-particle states (see the
corresponding physical reasons in \cite{GanKP95,KluME98,Sch98}). A
consideration of various time scales in heavy-ion collisions shows that the
stabilization time (a time interval during which the characteristic
asymptotic form for differential mean numbers of particles created by a
constant chromoelectric field from vacuum is achieved) is far less than the
period of plasma and mean-field oscillations. Then, the approximation of a
strong quasiconstant chromoelectric field can be used in the treatment of
such collisions during a period when the partons produced can be regarded as
weakly coupled, due to the property of asymptotic freedom in QCD. It may
also be reasonable to neglect dynamical back-reaction effects \cite{GavGT06}%
. In the case of a strange star \cite{Usov97}, it has been argued that there
is a macroscopic time interval during which local thermal equilibrium is
achieved, but the strong electric field\ has not yet been depleted, due to $%
e^{+}e^{-}$-pair production. By this reason,\ there has been a considerable
interest in establishing a relevant formalism for finite-temperature QED
with an unstable vacuum. An adequate techniques for such a case has been
proposed in \cite{GavGF87}. The special case of thermally-influenced pair
production in a constant electric field has been studied at the one-loop
level in \cite{GanKP95,GavGT06,BuhGF80,HalL95,Gie99}. The results of \cite%
{GavGT06,BuhGF80,HalL95}, obtained by various methods in the framework of
the generalized Furry representation, are in mutual agreement. The authors
of \cite{GanKP95,Gie99} have arrived at different results, being, at the
same time, in contradiction with themselves. We believe that the
difficulties \cite{GanKP95,Gie99} are related to some attempts to generalize
relation (\ref{vacampl}) to nonzero temperatures; see Discussion and Summary.

The paper is organized as follows. In Section 2, we describe the setting of
the problem and recall some necessary details of the non-perturbative
techniques that we use in our calculations. In Section 3, we calculate the
current density and EMT for the vacuum initial state. In the obtained
expressions, we separate the contributions due to particle creation and
vacuum polarization. The latter contributions are related to the
Heisenberg--Euler Lagrangian. In Section 4, we calculate the current density
and EMT for the thermal initial state. In the obtained expressions, we
separate the contributions due to particle-creation, vacuum polarization,
and the contributions that appear due to the work of the external field on
the particles at the initial state. All these contributions are studied in
detail in different regimes, limits of weak and strong fields and low and
high temperatures. We have established restrictions on the electric field
and its duration under which QED with a strong constant electric field is
consistent. Within these restrictions, one can neglect the back-reaction of
particles created by the electric field. In the last section, we summarize
and discuss the main results. Some necessary but cumbersome calculations are
placed in the Appendix.

In what follows, we use the relativistic units $\hslash =c=1$\ in which the
fine structure constant is $\alpha =e^{2}/c\hslash =e^{2}$.

\section{The setting of the problem}

We consider a quantized spinor (Dirac) field $\psi (x),$ $x=\left( x^{0}=t,%
\mathbf{x}\right) $, in an external electromagnetic background specified by
a quasiconstant electric field and a constant magnetic field. The field
strength $F_{\mu \nu }$ has nonzero invariants. In this case, we can always
choose a reference frame in which the electric and magnetic fields are
parallel and directed along the $x^{3}$-axis: 
\begin{equation}
F_{\mu \nu }=F_{\mu \nu }^{E}+F_{\mu \nu }^{B}\,,\;\;F_{\mu \nu
}^{E}=E(\delta _{\mu }^{0}\delta _{\nu }^{3}-\delta _{\mu }^{3}\delta _{\nu
}^{0})\,,\;\;F_{\mu \nu }^{B}=B(\delta _{\mu }^{2}\delta _{\nu }^{1}-\delta
_{\mu }^{1}\delta _{\nu }^{2})\,.  \label{ext-field}
\end{equation}

Our aim is to study the mean values of electric current and EMT in states
that have evolved from vacuum and thermal equilibrium. It is well-known that
a uniform constant electric field acting during an infinite time creates an
infinite number density of pairs from vacuum and produces infinite work on
particles. There arise other divergences in the course of QED calculations
related to this phenomenon. This is why one needs a regularization to deal
with these divergences. One type of such a regularization is the following,
namely, we consider, from the beginning, a uniform electric field, which
efficiently acts only during a sufficiently large but finite time $T$. Such
a regularization has been used in our works \cite{FGS91,BagGitSh75,GavG96a}.
In these works, we studied particle-creation effects in the so-called $T$%
-constant field, being a uniform electric field, constant (and nonzero)
within a time interval $T$, and zero outside this interval\footnote{%
The physical interpretation in the case of a field that violates the
stability of vacuum (for instance, a constant electric field) is extremely
involved when the field is given by a space-dependent potential which does
not disappear asymptotically. This happens because the state space of
strong-field QFT is quite different from the state space of standard
scattering theory, in which such a gauge for an external field is
appropriate. In the general case of a space-dependent potential, one cannot
use the standard definition of vacuum in QFT because such a state is
well-defined only on a space-like hypersurface. It is then generally unclear
when a strong electric field is given by a space-dependent potential, what
particles and antiparticles are, what the vacuum state is, and what is the
connection of the definitions being applied with the standard QFT approach.
In this case, instead of the standard evolution in time, one has to use the
rather obscure ``evolution in space''.}.{\Large \ }A\ similar regularization
is used in the present article: we choose the\ $T$-constant field to switch
on at the instant $t_{1}=-T/2$ and to switch off at the instant $t_{2}=T/2$.
The results of our calculations will be presented in a form in which
contributions of different nature are separated.{\Large \ }These are
contributions due to vacuum polarization, particle-creation from vacuum, and
the contributions due to the work of the external field on the particles at
the initial state.{\Large \ }Some of these contributions have constant
parts, as well as parts depending on the time interval $t-t_{1}$ that passes
since the instant the electric field turns on. Among the contributions
depending on $t-t_{1}$, there exist contributions that can increase and
decrease as the interval $t-t_{1}$ increases.{\Large \ }Having the general
expressions at hand, we analyze their dependence on the magnitude $E$ of
electric field and\ the time interval $t-t_{1}$, as well as their behavior
at low and high temperatures. In the case of a strong electric field, when
among the parts depending on $t-t_{1}$ the leading ones are the
contributions due to particle-creation, we are interested in the divergent
parts related to a large $t-t_{1}$. In the latter parts, we retain only the
contributions leading in{\Large \ }$t-t_{1}$. In particular, we analyze the
dynamics of mean-energy growth due to particle creation from vacuum and due
to the work of electric field on real particles. Making a comparison of this
dynamics with the energy density of the external electric field, we can
establish the limits of a correct applicability of the concepts of a{\Large %
\ }constant external field. In our calculations, we use the technics and
general results outlined below in this section with the corresponding
citations.

We assume the initial state to be a state of free $in$-particles in thermal
equilibrium at a given temperature $\theta .$ In such a model, we are going\
to calculate the mean values of electric current and the EMT of the Dirac
field. In the general case, the background under consideration is intense,
time-dependent and violates the vacuum stability. Such a background must be
considered nonperturbatively. In doing this, we follow the formulation
proposed in \cite{Gitma77}. Our calculations can be treated as a one-loop
approximation within the method of complete QED, developed in \cite%
{FGS91,FraGi81} and \cite{GavGF87,GavGT06}. Namely, this method provides a
technique of calculating the mean values for systems with unstable vacuum,
which is based on the generalized Furry picture for systems in strong
external backgrounds (see \cite{Gitma77,FGS91}); this technique may be
interpreted as a generalization of the Schwinger--Keldysh technique (\cite%
{real-time}; see also the review \cite{LanW87}) on systems with unstable
vacuum.

It is assumed that in the Heisenberg picture (for notation, see \cite%
{FGS91,GavGT06}) there exists a set of creation and annihilation operators $%
a_{n}^{\dagger }(in)$, $a_{n}(in)$ of $in$-particles (electrons), and
analogous operators $b_{n}^{\dagger }(in)$, $b_{n}(in)$ of $in$%
-antiparticles (positrons), with the corresponding $in$-vacuum $|0,in\rangle
,$ and a set of creation and annihilation operators $a_{n}^{\dagger }(out)$, 
$a_{n}(out),$ of $out$-electrons and similar operators $b_{n}^{\dagger
}(out) $, $b_{n}(out)$ of $out$-positrons, with the corresponding $out$%
-vacuum $|0,out\rangle $. By $n$, we denote a complete set of possible
quantum numbers. The $in$- and $out$-operators obey the canonical
anticommutation relations%
\begin{equation}
\lbrack a_{n}(in),a_{m}^{\dagger }(in)]_{+}=[a_{n}(out),a_{m}^{\dagger
}(out)]_{+}=[b_{n}(in),b_{m}^{\dagger }(in)]_{+}=[b_{n}(out),b_{m}^{\dagger
}(out)]_{+}=\delta _{nm}\,.  \label{a.1}
\end{equation}

The $in$-particles ($\zeta =+$ for electrons and $\zeta =-$ for positrons)
are associated with a complete set of solutions $\left\{ _{\zeta }\psi
_{n}(x)\right\} $ (we call it an $in$-set in what follows) of the Dirac
equation with the external electromagnetic field under consideration with
the asymptotics $\left\{ _{\zeta }\psi _{n}(t_{1},\mathbf{x})\right\} $ at
the initial time-instant $t_{1}.$ The latter functions are eigenvectors of
the one-particle Dirac Hamiltonian $\mathcal{H}(t)$ (with external field
under consideration; see its exact definition in the next section) at $%
t=t_{1}$, 
\begin{equation}
\mathcal{H}(t_{1})_{\zeta }\psi _{n}(t_{1},\mathbf{x})=\zeta \varepsilon
_{n}^{(\zeta )}{}_{\zeta }\psi _{n}(t_{1},\mathbf{x})\,,\ \mathcal{H}%
(t)=\gamma ^{0}(\left[ M+\boldsymbol{\gamma }\left( i\boldsymbol{\nabla }-q%
\mathbf{A}\left( t,\mathbf{x}\right) \right) \right] ),  \label{asy1}
\end{equation}%
where $\gamma ^{\mu }=\left( \gamma ^{0},\boldsymbol{\gamma }\right) $ are
Dirac's gamma-matrices; $q$ is the particle charge (for an electron $q=-e$); 
$\mathbf{A}$ is the vector potential of the external electromagnetic field ($%
A_{0}$ is zero in the case under consideration); $\varepsilon _{n}^{(\zeta
)} $\ are the energies of $in$-particles in a state specified by a complete
set of quantum numbers $n$, and $\varepsilon _{n}^{(\pm )}>0$. The $out$%
-particles are associated with a complete $out$-set of solutions $\left\{
^{\zeta }\psi _{n}\left( x\right) \right\} $ of the Dirac equation with the
asymptotics $^{\zeta }\psi _{n}(t_{2},\mathbf{x})$ at $t_{2}$ being
eigenvectors of the one-particle Dirac Hamiltonian at $t_{2}$, namely, 
\begin{equation}
\mathcal{H}(t_{2})\,^{\zeta }\psi _{n}(t_{2},\mathbf{x})=\zeta \tilde{%
\varepsilon}_{n}^{(\zeta )}\,^{\zeta }\psi _{n}(t_{2},\mathbf{x})\,,
\label{asy2}
\end{equation}%
where $\tilde{\varepsilon}_{n}^{(\pm )}$ are the energies of $out$-particles
in a state specified by a complete set of quantum numbers $n$, and $\tilde{%
\varepsilon}_{n}^{(\pm )}>0$.

The $out$-set can be decomposed in the $in$-set as follows: 
\begin{equation}
{}^{\zeta }\psi (x)={}_{+}\psi (x)G\left( {}_{+}|{}^{\zeta }\right)
+{}_{-}\psi (x)G\left( {}_{-}|{}^{\zeta }\right) \,,  \label{emt18}
\end{equation}%
where the decomposition coefficients $G\left( {}_{\zeta }|{}^{\zeta ^{\prime
}}\right) $ are expressed via the inner products of these sets. These
coefficients obey unitary conditions that follow from normalization
conditions for the solutions.

The Hamiltonian $H\left( t\right) $ of the quantized Dirac field\ is
time-dependent due to the external field. It is diagonalized (and has a
canonical form) in terms of the first set at the initial time instant, and
is diagonalized (and has a canonical form) in terms of the second set at the
final time instant. For example, 
\begin{equation}
H\left( t_{1}\right) =\sum_{n}\left[ \varepsilon _{n}^{(+)}a_{n}^{\dagger
}\left( in\right) a_{n}\left( in\right) +\varepsilon
_{n}^{(-)}b_{n}^{\dagger }\left( in\right) b_{n}\left( in\right) \right] \,.
\label{emt2}
\end{equation}
Correspondingly, the initial vacuum is defined by $a_{n}(in)|0,in\rangle
=b_{n}(in)|0,in\rangle =0$ for every $n$.

All the information about the processes of particle-creation, annihilation,
and scattering is contained in the elementary probability amplitudes%
\begin{eqnarray}
&&w\left( +|+\right) _{mn}=c_{v}^{-1}<0,out\left| a_{m}(out)a_{n}^{\dagger
}(in)\right| 0,in>=G^{-1}\left( {}_{+}|{}^{+}\right) _{mn}\,,  \notag \\
&&w\left( -|-\right) _{nm}=c_{v}^{-1}<0,out\left| b_{m}(out)b_{n}^{\dagger
}(in)\right| 0,in>=\left[ G^{-1}\left( {}_{-}|{}^{-}\right) \right]
_{nm}^{\dagger },  \notag \\
&&w\left( 0|-+\right) _{nm}=c_{v}^{-1}<0,out\left| b_{n}^{\dagger
}(in)a_{m}^{\dagger }(in)\right| 0,in>=-\left[ G\left( {}_{+}|{}^{-}\right)
G^{-1}\left( {}_{-}|{}^{-}\right) \right] _{nm}^{\dagger }\,,  \notag \\
&&w\left( +-|0\right) _{mn}=c_{v}^{-1}<0,out\left|
a_{m}(out)b_{n}(out)\right| 0,in>=\left[ G^{-1}\left( {}_{+}|{}^{+}\right)
G\left( {}_{+}|{}^{-}\right) \right] _{mn}\,,  \notag \\
&&c_{v}=\langle 0,out|0,in\rangle \,,  \label{a.2b}
\end{eqnarray}%
where $c_{v}$ is the vacuum-to-vacuum transition amplitude.

The sets of $in$ and $out$-operators are related to each other by a linear
canonical transformation (sometimes called the Bogolyubov transformation).
It has been demonstrated that in the general case such a relation has the
form (see \cite{Gitma77}) 
\begin{equation}
V\left( a^{\dagger }(out),a(out),b^{\dagger }(out),b(out)\right) V^{\dagger
}=\left( a^{\dagger }(in),a(in),b^{\dagger }(in),b(in)\right)
\,,\;|0,in\rangle =V|0,out\rangle \,,  \label{a.3}
\end{equation}%
where the unitary operator $V$ has the form\footnote{%
Here and elsewhere, we use condenced notations, for example, 
\begin{equation*}
bw\left( 0|-+\right) a=\sum_{n,m}b_{n}w\left( 0|-+\right) _{nm}a_{m}\,.
\end{equation*}%
} 
\begin{eqnarray}
V &=&\exp \left\{ -a^{\dagger }(out)w\left( +-|0\right) b^{\dagger
}(out)\right\} \exp \left\{ -b(out)\ln w\left( -|-\right) b^{\dagger
}(out)\right\}  \notag \\
&&\times \exp \left\{ a^{\dagger }(out)\ln w\left( +|+\right) a(out)\right\}
\exp \left\{ -b(out)w\left( 0|-+\right) a(out)\right\} ,  \label{a.3a}
\end{eqnarray}%
and 
\begin{equation}
c_{v}=\langle 0,out|V|0,out\rangle =\exp \left\{ -\mathrm{Tr}\ln w\left(
-|-\right) \right\} \,.  \label{a.3c}
\end{equation}

Note that the exact formula (\ref{a.3c}) holds true for an arbitrary
external field. In \cite{Nik70}, has been demonstrated that in the case of a
constant electric-like field the Schwinger formula (\ref{vacampl}) leads to
the same result.

As our final purpose, we are going to analyze the following mean values:%
\begin{equation}
\langle j_{\mu }\left( t\right) \rangle =\mathrm{Tr\,}\left[ \rho
_{in}j_{\mu }\right] \,,\;\;\langle T_{\mu \nu }\left( t\right) \rangle =%
\mathrm{Tr\,}\left[ \rho _{in}T_{\mu \nu }\right] \,,  \label{emt22}
\end{equation}%
where $\rho _{in}$ is a density operator of the initial state in the
Heisenberg picture; the operators of current density $j_{\mu }$, and
energy-momentum tensor $T_{\mu \nu }$ have the form%
\begin{eqnarray*}
&&j_{\mu }=\frac{q}{2}\left[ \bar{\psi}(x),\gamma _{\mu }\;\psi (x)\right]
\,,\;\;T_{\mu \nu }=\frac{1}{2}\left( T_{\mu \nu }^{can}+T_{\nu \mu
}^{can}\right) \,, \\
&&T_{\mu \nu }^{can}=\frac{1}{4}\left\{ \left[ \bar{\psi}(x),\gamma _{\mu
}P_{\nu }\;\psi (x)\right] +\left[ P_{\nu }^{\ast }\bar{\psi}(x),\gamma
_{\mu }\;\psi (x)\right] \right\} \,.
\end{eqnarray*}%
Here, $P_{\mu }=i\partial _{\mu }-qA_{\mu }(x)$ and $\psi (x)$ are Dirac
field operators in the Heisenberg representation that obey the Dirac
equation with the external field $A_{\mu }(x).$

In the next section, we calculate and analyze the mean values (\ref{emt22})
for the initial vacuum state. We denote them as follows:%
\begin{equation}
\langle j_{\mu }\left( t\right) \rangle ^{0}=\langle 0,in|j_{\mu
}|0,in\rangle \,,\;\;\langle T_{\mu \nu }\left( t\right) \rangle
^{0}=\langle 0,in|T_{\mu \nu }|0,in\rangle \,.  \label{vac}
\end{equation}%
Then, we analyze the more complicated case (\ref{emt22}), in which the
initial state of the system under consideration is prepared as an
equilibrium state of noninteracting $in$-particles at temperature $\theta $
with the chemical potentials $\mu ^{(\zeta )}$, and is characterized by the
density operator $\rho _{in}$ in the Heisenberg picture,%
\begin{equation}
\rho _{in}=Z^{-1}\exp \left\{ \beta \left( \sum_{\zeta =\pm }\mu ^{(\zeta
)}N^{(\zeta )}-H\left( t_{1}\right) \right) \right\} ,\;\;\mathrm{Tr\,}\rho
_{in}=1\,,  \label{emt1}
\end{equation}%
where $Z$ is a normalization constant, $\beta =\theta ^{-1}\,,$ and $%
N^{(\zeta )}$ are the operators of $in$-paricles numbers, 
\begin{equation*}
N^{(+)}=\sum_{n}a_{n}^{\dagger }\left( in\right) a_{n}\left( in\right)
\,,\;N^{(-)}=\sum_{n}b_{n}^{\dagger }\left( in\right) b_{n}\left( in\right)
\,.
\end{equation*}

In our calculations, we also need the matrix elements%
\begin{eqnarray}
&\langle j_{\mu }\left( t\right) \rangle ^{c}=&\langle 0,out|j_{\mu
}|0,in\rangle c_{v}^{-1}\,,\;\;\langle T_{\mu \nu }\left( t\right) \rangle
^{c}=\langle 0,out|T_{\mu \nu }|0,in\rangle c_{v}^{-1}\,,  \notag \\
&\langle j_{\mu }\left( t\right) \rangle _{out}^{0}=&\langle 0,out|j_{\mu
}|0,out\rangle ,\;\;\langle T_{\mu \nu }\left( t\right) \rangle
_{out}^{0}=\langle 0,out|T_{\mu \nu }|0,out\rangle .  \label{emt23}
\end{eqnarray}

The matrix elements (\ref{emt22}), (\ref{vac}), and (\ref{emt23}) can be
expressed via relevant singular functions of Dirac's fields, as follows: 
\begin{eqnarray}
&&i\mathrm{Tr\,}\left\{ \rho _{in}T\psi \left( x\right) \bar{\psi}\left(
x^{\prime }\right) \right\} =S_{in}^{c}\left( x,x^{\prime }\right)
+S^{\theta }\left( x,x^{\prime }\right) \,,  \notag \\
&&i\mathrm{Tr\,}\left\{ \rho _{in}\bar{\psi}\left( x^{\prime }\right) \psi
\left( x\right) \right\} =S_{in}^{+}\left( x,x^{\prime }\right) -S^{\theta
}\left( x,x^{\prime }\right) \,,  \notag \\
&&i\mathrm{Tr\,}\left\{ \rho _{in}\psi \left( x\right) \bar{\psi}\left(
x^{\prime }\right) \right\} =S_{in}^{-}\left( x,x^{\prime }\right)
+S^{\theta }\left( x,x^{\prime }\right) \,,  \notag \\
&&S_{in}^{c}\left( x,x^{\prime }\right) =\theta \left( x_{0}-x_{0}^{\prime
}\right) S_{in}^{-}\left( x,x^{\prime }\right) -\theta \left( x_{0}^{\prime
}-x_{0}\right) S_{in}^{+}\left( x,x^{\prime }\right) \,,  \label{emt11}
\end{eqnarray}%
where $S_{in}^{(\ldots )}$ are singular functions at zero temperature, 
\begin{eqnarray}
S_{in}^{c}(x,x^{\prime }) &=&i\langle 0,in|T\psi (x)\bar{\psi}(x^{\prime
})|0,in\rangle \,,  \notag \\
S_{in}^{-}(x,x^{\prime }) &=&i\langle 0,in|\psi (x)\bar{\psi}(x^{\prime
})|0,in\rangle \,,  \notag \\
S_{in}^{+}(x,x^{\prime }) &=&i\langle 0,in|\bar{\psi}(x^{\prime })\psi
(x)|0,in\rangle \,.  \label{w25}
\end{eqnarray}

All the singular functions can be expressed in terms of the $in$-set of
solutions as follows:%
\begin{eqnarray}
&&S_{in}^{\mp }\left( x,x^{\prime }\right) =i\sum_{n}\,_{\pm }\psi
_{n}\left( x\right) _{\pm }\bar{\psi}_{n}\left( x^{\prime }\right) \,, 
\notag \\
&&S^{\theta }\left( x,x^{\prime }\right) =\tilde{S}_{\theta }^{-}\left(
x,x^{\prime }\right) -\tilde{S}_{\theta }^{+}\left( x,x^{\prime }\right) \,,
\notag \\
&&\tilde{S}_{\theta }^{\mp }\left( x,x^{\prime }\right) =-i\sum_{n}\ _{\pm
}\psi _{n}\left( x\right) _{\pm }\bar{\psi}_{n}\left( x^{\prime }\right)
N_{n}^{(\pm )}\left( in\right) \,,  \label{emt13}
\end{eqnarray}%
where%
\begin{equation}
N_{n}^{\left( \zeta \right) }\left( in\right) =\left[ \exp \left\{ \beta
\left( \varepsilon _{n}^{(\zeta )}-\mu ^{(\zeta )}\right) \right\} +1\right]
^{-1}\,.  \label{in}
\end{equation}

It is important to emphasize that, in the general case, $S_{in}^{c}$ is
different from the causal Green function,%
\begin{equation}
S^{c}(x,x^{\prime })=i\langle 0,out|T\psi (x)\bar{\psi}(x^{\prime
})|0,in\rangle c_{v}^{-1}\,.  \label{emt17}
\end{equation}%
From (\ref{emt17}), it follows that%
\begin{eqnarray}
&&S^{c}\left( x,x^{\prime }\right) =\theta \left( x_{0}-x_{0}^{\prime
}\right) S^{-}\left( x,x^{\prime }\right) -\theta \left( x_{0}^{\prime
}-x_{0}\right) S^{+}\left( x,x^{\prime }\right) \,,  \notag \\
&&S^{-}\left( x,x^{\prime }\right) =i\sum_{n,m}{}^{+}\psi _{n}\left(
x\right) G\left( \left. _{+}\right| ^{+}\right) _{nm}^{-1}\,_{+}\bar{\psi}%
_{m}\left( x^{\prime }\right) \,,  \notag \\
&&S^{+}\left( x,x^{\prime }\right) =i\sum_{n,m}\,_{-}\psi _{n}\left(
x\right) \left[ G\left( \left. _{-}\right| ^{-}\right) ^{-1}\right]
_{nm}^{\dagger }\,^{-}\bar{\psi}_{m}\left( x^{\prime }\right) \,,
\label{emt19}
\end{eqnarray}%
see \cite{Gitma77}. Then the difference $S^{p}(x,x^{\prime })=\
S_{in}^{c}(x,x^{\prime })-S^{c}(x,x^{\prime })$ has the form 
\begin{equation}
\ S^{p}(x,x^{\prime })=i\sum_{nm}\,_{-}{\psi }_{n}(x)\,\left[
G(_{+}|^{-})G(_{-}|^{-})^{-1}\right] _{nm}^{\dagger }{_{+}\bar{\psi}}%
_{m}(x^{\prime })\,.  \label{emt21}
\end{equation}%
This function vanishes in the case of a stable vacuum, since it contains the
coefficients $G(_{+}|^{-})$ related to the mean number of created particles.

Another kind of Green's function that will be used later on is given by%
\begin{equation}
S_{out}^{c}(x,x^{\prime })=i\langle 0,out|T\psi (x)\bar{\psi}(x^{\prime
})|0,out\rangle \,.  \label{emt25-1}
\end{equation}%
It is related to $S^{c}(x,x^{\prime })$ as follows: 
\begin{eqnarray}
&&S_{out}^{c}(x,x^{\prime })=S^{c}(x,x^{\prime })+S^{\bar{p}}(x,x^{\prime
})\,,  \notag \\
&&S^{\bar{p}}(x,x^{\prime })=-i\sum_{nm}\,^{+}{\psi }_{n}(x)\,\left[
G(_{+}|^{+})^{-1}G(_{+}|^{-})\right] _{nm}{^{-}\bar{\psi}}_{m}(x^{\prime
})\,.  \label{emt25}
\end{eqnarray}

We can see that there exist relations between the mean values (\ref{emt22}),
(\ref{vac}) and matrix elements (\ref{emt23}), namely, 
\begin{eqnarray}
&\langle j_{\mu }\left( t\right) \rangle =&\langle j_{\mu }\left( t\right)
\rangle ^{0}+\langle j_{\mu }\left( t\right) \rangle ^{\theta },\;\;\langle
T_{\mu \nu }\left( t\right) \rangle =\langle T_{\mu \nu }\left( t\right)
\rangle ^{0}+\langle T_{\mu \nu }\left( t\right) \rangle ^{\theta },  \notag
\\
&\langle j_{\mu }\left( t\right) \rangle ^{0}=&\langle j_{\mu }\left(
t\right) \rangle ^{c}+\langle j_{\mu }\left( t\right) \rangle
^{p},\;\;\langle T_{\mu \nu }\left( t\right) \rangle ^{0}=\langle T_{\mu \nu
}\left( t\right) \rangle ^{c}+\langle T_{\mu \nu }\left( t\right) \rangle
^{p},  \notag \\
&\langle j_{\mu }\left( t\right) \rangle _{out}^{0}=&\langle j_{\mu }\left(
t\right) \rangle ^{c}+\langle j_{\mu }\left( t\right) \rangle ^{\bar{p}%
}\,,\;\;\langle T_{\mu \nu }\left( t\right) \rangle _{out}^{0}=\langle
T_{\mu \nu }\left( t\right) \rangle ^{c}+\langle T_{\mu \nu }\left( t\right)
\rangle ^{\bar{p}}\,.  \label{emt24.1}
\end{eqnarray}%
These relations involve the quantities 
\begin{eqnarray}
&&\,\langle j_{\mu }\left( t\right) \rangle ^{c,p,\bar{p},\theta }=iq\left. 
\mathrm{tr}\left[ \gamma _{\mu }S^{c,p,\bar{p},\theta }(x,x^{\prime })\right]
\right| _{x=x^{\prime }}\,,  \notag \\
&&\,\langle T_{\mu \nu }\left( t\right) \rangle ^{c,p,\bar{p},\theta
}=i\left. \mathrm{tr}\left[ A_{\mu \nu }S^{c,p,\bar{p},\theta }(x,x^{\prime
})\right] \right| _{x=x^{\prime }}\,,  \notag \\
&&A_{\mu \nu }=1/4\left[ \gamma _{\mu }\left( P_{\nu }+P\mathcal{^{\prime }}%
_{\nu }^{\ast }\right) +\gamma _{\nu }\left( P_{\mu }+P\mathcal{^{\prime }}%
_{\mu }^{\ast }\right) \right] \,,  \notag \\
&&P_{\mu }^{\prime \ast }=-i\frac{\partial }{\partial x^{\prime }{}^{\mu }}%
-q\,A_{\mu }(x^{\prime })\,,  \label{emt24.2}
\end{eqnarray}%
where $\mathrm{tr}\left[ \cdots \right] $ is the trace in the space of $%
4\times 4$ matrices, and the equality $x=x^{\prime }$ is understood as
follows:%
\begin{equation*}
\left. \mathrm{tr}\left[ \cdots (x,x^{\prime })\right] \right| _{x=x^{\prime
}}=\frac{1}{2}\left. \left[ \lim_{t\rightarrow t^{\prime }-0}\mathrm{tr}%
\left[ \cdots (x,x^{\prime })\right] +\lim_{t\rightarrow t^{\prime }+0}%
\mathrm{tr}\left[ \cdots (x,x^{\prime })\right] \right] \right| _{\mathbf{x=x%
}^{\prime }}\,.
\end{equation*}

The quantities (\ref{emt24.2}) are expressed in terms of $S^{c}(x,x^{\prime
})$, which is the causal Green function (propagator) of the Dirac equation
with an external field, and in terms of $S^{p,\bar{p},\theta }$, which are
solutions of the same equation. The quantities $<j_{\mu }\left( t\right)
>^{0}$, $<T_{\mu \nu }\left( t\right) >^{0},$ $<j_{\mu }\left( t\right)
>_{out}^{0}$, and $<T_{\mu \nu }\left( t\right) >_{out}^{0}$ are
contributions to the corresponding mean values at zero temperature and
density. The quantities $<j_{\mu }\left( t\right) >^{\theta }$and $<T_{\mu
\nu }\left( t\right) >^{\theta }$ present the contributions due to the
existence of the initial thermal distribution.

The quantities $\langle j_{\mu }\left( t\right) \rangle ^{0}$, $\langle
T_{\mu \nu }\left( t\right) \rangle ^{0}$, $\langle j_{\mu }\left( t\right)
\rangle _{out}^{0}$, $\langle T_{\mu \nu }\left( t\right) \rangle _{out}^{0}$%
, $\langle j_{\mu }\left( t\right) \rangle ^{\theta }$, and $\langle T_{\mu
\nu }\left( t\right) \rangle ^{\theta }$ are real-valued by construction,
due to the properties of the singular functions $S_{in}^{c}$, $S_{out}^{c}$
and $S^{\theta }$. On the contrary, the quantities $\langle j_{\mu }\left(
t\right) \rangle ^{c,p,\bar{p}}$ and $\langle T_{\mu \nu }\left( t\right)
\rangle ^{c,p,\bar{p}}$ are not necessarily real-valued. For the purpose of
the following consideration, it is useful to rewrite $\langle j_{\mu }\left(
t\right) \rangle ^{0}$ and $\langle T_{\mu \nu }\left( t\right) \rangle ^{0}$
in the form%
\begin{eqnarray}
&\langle j_{\mu }\left( t\right) \rangle ^{0}=&\func{Re}\,\langle j_{\mu
}\left( t\right) \rangle ^{c}+\func{Re}\,\langle j_{\mu }\left( t\right)
\rangle ^{p}\,,  \notag \\
&\langle T_{\mu \nu }\left( t\right) \rangle ^{0}=&\func{Re}\,\langle T_{\mu
\nu }\left( t\right) \rangle ^{c}+\func{Re}\,\langle T_{\mu \nu }\left(
t\right) \rangle ^{p}\,,  \notag \\
&\langle j_{\mu }\left( t\right) \rangle _{out}^{0}=&\func{Re}\,\langle
j_{\mu }\left( t\right) \rangle ^{c}+\func{Re}\,\langle j_{\mu }\left(
t\right) \rangle ^{\bar{p}}\,,  \notag \\
&\langle T_{\mu \nu }\left( t\right) \rangle _{out}^{0}=&\func{Re}\,\langle
T_{\mu \nu }\left( t\right) \rangle ^{c}+\func{Re}\,\langle T_{\mu \nu
}\left( t\right) \rangle ^{\bar{p}}\,.  \label{emt24.3}
\end{eqnarray}

\section{The initial state as vacuum}

\subsection{T-constant field regularization}

In this section, we examine the quantities $<j_{\mu }\left( t\right) >^{0}$
and $<T_{\mu \nu }\left( t\right) >^{0}$ that represent contributions to the
corresponding mean values at zero temperature. In order to calculate such
quantities, the $T$-constant field regularization is necessary. In
particular, we are going to study the leading contributions to the divergent
parts of the quantities at $T\rightarrow \infty $. To make the consideration
complete, and to provide the reader with some formulas necessary for the
further consideration, we start by reproducing some of the results on the
mean numbers of created particle that were obtained in \cite{GavG96a}.

We chose the $T$-constant field potentials $A_{\mu }^{E}$ as follows, $%
A_{0}^{E}=A_{1}^{E}=A_{2}^{E}=0$: 
\begin{equation*}
A_{3}^{E}(t)=\left\{ 
\begin{array}{ll}
Et_{1}\,, & t\in (-\infty ,t_{1})\,, \\ 
Et\,, & t\in \lbrack t_{1},t_{2}], \\ 
Et_{2}\,, & t\in (t_{2},+\infty )\,,%
\end{array}%
\right.
\end{equation*}%
where $t_{2}=-t_{1}=T/2.\ $The corresponding electric field $E(t)$ is given
by 
\begin{equation}
E(t)=\left\{ 
\begin{array}{ll}
0\,, & t\in (-\infty ,t_{1})\,, \\ 
E\,, & t\in \lbrack t_{1},t_{2}], \\ 
0\,, & t\in (t_{2},+\infty )\,,%
\end{array}%
\right.  \label{el-field}
\end{equation}

First, let us suppose that the magnetic field is absent.

For the purpose of $T$-regularization, it is sufficient to choose, for $%
\left\{ _{\zeta }\psi _{n}(x)\right\} $ and $\left\{ ^{\zeta }\psi
_{n}(x)\right\} $, defined above, the following orthonormalized sets of
solutions of the Dirac equation with a constant electric field: 
\begin{eqnarray}
&&_{\pm }\psi _{\mathbf{p},r}(x)=(\gamma P+M){}_{\pm }\phi _{\mathbf{p},\pm
1,r}(x)\,,\;\;^{\pm }\psi _{\mathbf{p},r}(x)=(\gamma P+M){}^{\pm }\phi _{%
\mathbf{p},\mp 1,r}(x)\,,  \notag \\
&&_{\pm }\phi _{\mathbf{p},s,r}(x)=\,_{\pm }\phi _{\mathbf{p},s}(t)\exp \{i%
\mathbf{p}\mathbf{x}\}v_{s,r}\,,\;\;^{\pm }\phi _{\mathbf{p},s,r}(x)=\;^{\pm
}\phi _{\mathbf{p},s}(t)\exp \{i\mathbf{p}\mathbf{x}\}v_{s,r}\,,  \notag \\
&&_{+}^{-}\phi _{\mathbf{p},s}(t)=CD_{\nu -\frac{1+s}{2}}(\pm (1-i)\xi
)\,,\;\;_{-}^{+}\phi _{\mathbf{p},s}(t)=CD_{-\nu -\frac{1-s}{2}}(\pm
(1+i)\xi )\,,\;s=\pm 1\,,  \label{emt36}
\end{eqnarray}%
where $\mathbf{p}$ is the momentum and $r=\pm 1$ is the spin projection;%
{\Large \ }$D_{\nu }(z)$ is the Weber parabolic cylinder (WPC) function \cite%
{BatE74}, and 
\begin{eqnarray*}
&&\nu =\frac{i\lambda }{2},\;\;\lambda =\frac{M^{2}+\mathbf{p}_{\bot }^{2}}{%
\left\vert qE\right\vert },\;\mathbf{p}_{\bot }=(p^{1},p^{2},0),\;\;\xi =\xi
(t)=\frac{qEt-p_{3}}{\sqrt{\left\vert qE\right\vert }}\mathrm{sgn}\left(
qE\right) , \\
&&C=(2\pi )^{-3/2}(2\left\vert qE\right\vert )^{-1/2}\exp \left( -\pi
\lambda /8\right) \,,
\end{eqnarray*}%
$v_{s,r}$ are constant orthonormal spinors, $v_{s,r}^{\dagger
}v_{s,r^{\prime }}=\delta _{r,r^{\prime }}$, subject to the supplementary
condition $(1\pm \gamma ^{0}\gamma ^{3})v_{\mp 1,r}=0$. Using an asymptotic
expansion of the WPC-function \cite{BatE74}, 
\begin{equation}
D_{\nu }(z)=z^{\nu }\exp \left( -\frac{z^{2}}{4}\right) \left( \sum_{n=0}^{N}%
\frac{(-\frac{1}{2}\nu )_{n}(\frac{1}{2}-\frac{1}{2}\nu )_{n}}{n!(-\frac{1}{2%
}z^{2})^{n}}+O(|z|^{-2(N+1)})\right) ,\;|\arg z|<3\pi /4,  \label{e40ab}
\end{equation}%
we can obtain the asymptotics of particle energies as $T\rightarrow \infty $,%
\begin{equation}
\varepsilon _{\mathbf{p}}^{(\zeta )}=|\frac{qET}{2}+p_{3}|\,,\;\;\tilde{%
\varepsilon}_{\mathbf{p}}^{(\zeta )}=|\frac{qET}{2}-p_{3}|\,.  \label{energ}
\end{equation}

One can see that the matrices $\ G\left( {}_{\zeta }|{}^{\zeta ^{\prime
}}\right) $ (\ref{emt18}) are diagonal:%
\begin{equation}
G\left( _{\zeta }|^{\zeta ^{\prime }}\right) _{\mathbf{p},r,\mathbf{p}%
^{\prime },r^{\prime }}=\delta _{r,r^{\prime }}\delta (\mathbf{p}-\mathbf{p}%
^{\prime })g\left( {}_{\zeta }|{}^{\zeta ^{\prime }}\right) .  \label{e37a}
\end{equation}

The differential mean numbers of electrons (equal to the corresponding
differential mean number of pairs) with a given momentum $\mathbf{p}$ and
spin projections $r$ created from vacuum are%
\begin{equation}
\aleph _{\mathbf{p},r}=\langle 0,in|a_{\mathbf{p},r}^{\dagger }(out)a_{%
\mathbf{p},r}(out)|0,in\rangle =\left| g\left( _{-}|^{+}\right) \right| ^{2},
\label{e11}
\end{equation}%
were{\LARGE \ }the standard volume regularization is used, so that $\delta (%
\mathbf{p}-\mathbf{p}^{\prime })\rightarrow \delta _{\mathbf{p},\mathbf{p}%
^{\prime }}$. Note that $\aleph _{\mathbf{p},r}$ is even with respect to the
sign of $qE$.

If the time $T$ is sufficiently large, $T>>T_{0}=(1+\lambda )/\sqrt{\left|
qE\right| }$, the differential mean numbers $\aleph _{\mathbf{p},r}$ have
the form%
\begin{equation}
\aleph _{\mathbf{p},r}=\left\{ 
\begin{array}{l}
e^{-\pi \lambda }\left[ 1+O\left( \left[ \frac{1+\lambda }{K_{p}}\right]
^{3}\right) \right] \,,\;\;-\sqrt{\left| qE\right| }\frac{T}{2}\leq \bar{\xi}%
\leq -K_{p}\,, \\ 
O\left( 1\right) ,\;\;-K_{p}\,<\bar{\xi}\leq K_{p}\,, \\ 
O\left( \left[ \frac{1+\lambda }{\xi ^{2}}\right] ^{3}\right) ,\;\;\bar{\xi}%
>K_{p}\,,%
\end{array}%
\right.  \label{e40a}
\end{equation}%
where 
\begin{equation*}
\bar{\xi}=\frac{\left| p_{3}\right| -\left| qE\right| T/2\,}{\sqrt{\left|
qE\right| }}\,,
\end{equation*}%
and $K_{p}$ is a sufficiently large arbitrary constant, $K_{p}>>1+\lambda $;
see \cite{GavG96a}. In the limit $T\rightarrow \infty $, the differential
mean numbers have a simple form: 
\begin{equation}
\aleph _{\mathbf{p},r}=e^{-\pi \lambda }\,,  \label{e40ad}
\end{equation}%
which is identical with the one obtained for a constant electric field by
Nikishov \cite{Nik70}. One can see that the stabilization of the
differential mean numbers to the asymptotic form (\ref{e40ad}) for finite
longitudinal momenta is reached at $T>>T_{0}$. The characteristic time $%
T_{0} $ is called the stabilization time.

In order to study the effects of switching the electric field on and off, at 
$T>>T_{0}$, one can study another example of a quasiconstant electric field:%
\begin{equation}
E(t)=E\cosh ^{-2}\left( \frac{t}{\alpha }\right) \,.  \label{e2a}
\end{equation}%
This field switches on and off adiabatically as $t\rightarrow \pm \infty $
and is quasi-constant at finite times. The differential mean numbers of
particles created by such a field have been found in \cite{NarN70}. As shown
in \cite{GavG96a}, the differential mean numbers in the field (\ref{e2a})
have the asymptotic form (\ref{e40ad}) for sufficiently large $\alpha $, $%
\alpha >>\alpha _{0}=(1+\sqrt{\lambda })/\sqrt{\left| qE\right| }$ and for $%
\left| qE\right| \alpha \gg |p_{3}|$. Thus, $\alpha _{0}$ can be interpreted
as the stabilization time for such a field. At the same time, the latter
implies that the effects of switching on and off are not essential at large
times and finite longitudinal momenta for both fields. Extrapolating this
conclusion, one may suppose that in any electric field being quasiconstant $%
\approx E$ at least for a time period $T>>T_{0}$ and switching on and off
outside this period particle-creation effects have no dependence on the
details of switching on and off. Therefore, calculations in a $T$-constant
field are representative for a large class of quasiconstant electric fields.

\subsection{Contributions due to particle-creation}

Let us apply the above results to a calculation of the leading terms in $%
\func{Re}\langle j_{\mu }(t)\rangle ^{p,\bar{p}}$ and $\func{Re}\langle
T_{\mu \nu }(t)\rangle ^{p,\bar{p}}$ from (\ref{emt24.3}) for{\Large \ }%
sufficiently large times $t$,\ more exactly, for large values of $%
t-t_{1}=t+T/2$, and of $T$. Being written in a dimensionless form these two
conditions have the form%
\begin{eqnarray}
&&\sqrt{\left| qE\right| }\left( t+T/2\right) \gg 1+M^{2}/\left| qE\right|
\,,\;\mathrm{\;}t\leq T/2\,,  \notag \\
&&\sqrt{\left| qE\right| }T\gg 1+M^{2}/\left| qE\right| \,,\;\mathrm{\;}%
t>T/2\,.  \label{unistab}
\end{eqnarray}

We note that the dimensionless parameter%
\begin{equation*}
\sqrt{\left\vert qE\right\vert }T=\sqrt{\frac{c}{\hslash }\left\vert
qE\right\vert }T
\end{equation*}%
that has emerged here will play an important role in the further
considerations, and will enter all the consistency conditions that are
obtained in this article.

We can see that (\ref{unistab}) is a stabilization condition when the
leading terms do not depend on the details of switching on and off. As
follows from (\ref{emt24.2}), we need the singular functions $S^{p}$ (\ref%
{emt21}) and $S^{\bar{p}}$ (\ref{emt25}) at $x\approx x^{\prime }$ in such
an approximation, which provides the required contribution for $\func{Re}%
\langle j_{\mu }(t)\rangle ^{p,\bar{p}}$ and $\func{Re}\langle T_{\mu \nu
}(t)\rangle ^{p,\bar{p}}$ when (\ref{unistab}) is valid. Time-dependence
arises due to integration over $p_{3}$ under the condition $-\sqrt{%
\left\vert qE\right\vert }\frac{T}{2}\leq \bar{\xi}\leq -K$ with the
subsidiary condition that $\left\vert \xi (t)\right\vert $ must be
sufficiently large, $\left\vert \xi (t)\right\vert \geq K$, where $K$ is
subject to the condition $K\gg 1+m^{2}/\left\vert qE\right\vert $. The
distribution $\aleph _{\mathbf{p},r}$ plays the role of a cut-off factor in
the integral over $p_{\bot }$, the $T$-dependent contribution of $S^{p,\bar{p%
}}$ being thus convergent.

The range of integration over momenta in $S^{p}$ that determines the leading
contributions\ can be defined as 
\begin{equation*}
D:\left\{ 
\begin{array}{c}
\left\vert \mathbf{p}_{\bot }\right\vert \leq \sqrt{\left\vert qE\right\vert 
}\left[ \sqrt{\left\vert qE\right\vert }\left( t+T/2\right) -K\right] ^{1/2}
\\ 
-T/2+K/\sqrt{\left\vert qE\right\vert }\leq p_{3}/qE\leq t-K/\sqrt{%
\left\vert qE\right\vert }\,%
\end{array}%
\right. .
\end{equation*}

Using (\ref{emt18}) to express the solutions $_{\pm }\psi _{n}$ via $^{\pm
}\psi _{n}$, and taking into account the asymptotic expansion (\ref{e40ab}),
we can calculate the leading contributions to the function $%
S^{p}(x,x^{\prime })$ (\ref{emt21}), 
\begin{equation}
S^{p}(x,x^{\prime })=-i\int_{D}d\mathbf{p}\sum_{r=\pm 1}\aleph _{\mathbf{p}%
,r}\left[ ^{+}{\psi }_{\mathbf{p},r}(x)^{+}{\bar{\psi}}_{\mathbf{p}%
,r}(x^{\prime })-\,^{-}{\psi }_{\mathbf{p},r}(x)^{-}{\bar{\psi}}_{\mathbf{p}%
,r}(x^{\prime })\right] \,.  \label{emt40}
\end{equation}%
Taking summation over $r$, we represent (\ref{emt40}) as follows:%
\begin{eqnarray}
S^{p}(x,x^{\prime }) &=&(\gamma P+M)\Delta ^{p}(x,x^{\prime })\,,  \notag \\
\Delta ^{p}(x,x^{\prime }) &=&-i\int_{D}\aleph _{\mathbf{p},0}\exp \{i%
\mathbf{p}\left( \mathbf{x-x}^{\prime }\right) \}\left[ ^{+}\phi _{\mathbf{p}%
,0}(t)^{+}\phi _{\mathbf{p},0}^{\ast }(t^{\prime })+^{-}\phi _{\mathbf{p}%
,0}(t)^{-}\phi _{\mathbf{p},0}^{\ast }(t^{\prime })\right] d\mathbf{p\,}.
\label{emt40.1}
\end{eqnarray}%
Then, integrating over $\mathbf{p}_{\perp }$ in (\ref{emt40.1}), we obtain%
\begin{equation}
\Delta ^{p}(x,x^{\prime })=-i\int_{-T/2+K/\sqrt{\left\vert qE\right\vert }%
}^{t-K/\sqrt{\left\vert qE\right\vert }}h_{\parallel }\left( x_{\parallel
},x_{\parallel }^{\prime }\right) h_{\perp }\left( \mathbf{x}_{\perp },%
\mathbf{x}_{\perp }^{\prime }\right) d\left( \frac{p_{3}}{qE}\right) \,,
\label{emt44b}
\end{equation}%
where 
\begin{eqnarray}
&&h_{\parallel }\left( x_{\parallel },x_{\parallel }^{\prime }\right) =\frac{%
\left\vert qE\right\vert }{2\pi \left\vert qEt-p_{3}\right\vert }\exp \left[
ip_{3}\left( x_{3}-x_{3}^{\prime }\right) \right] \cos \left\{ \frac{1}{2}%
\left[ \xi (t^{\prime })^{2}-\xi (t)^{2}\right] \right\} ,  \notag \\
&&h\left( \mathbf{x}_{\perp },\mathbf{x}_{\perp }^{\prime }\right) =(2\pi
)^{-2}\left\vert qE\right\vert \exp \left( -\frac{\pi m^{2}}{\left\vert
qE\right\vert }-\frac{\left( \mathbf{x}_{\perp }-\mathbf{x}_{\perp }^{\prime
}\right) ^{2}\left\vert qE\right\vert }{4\pi }\right) \,,  \label{emt44c}
\end{eqnarray}%
and the notation $x_{\parallel }^{\mu }=\left( x^{0},0,0,x^{3}\right) $ and $%
\mathbf{x}_{\perp }=\left( x^{1},x^{2},0\right) $\ has been used.

In the same manner, we calculate the leading contributions to the function $%
S^{\bar{p}}(x,x^{\prime })$\ (\ref{emt25}), 
\begin{eqnarray}
S^{\bar{p}}(x,x^{\prime }) &=&(\gamma P+M)\Delta ^{\bar{p}}(x,x^{\prime })\,,
\notag \\
\Delta ^{\bar{p}}(x,x^{\prime }) &=&-i\int_{\bar{D}}\aleph _{\mathbf{p}%
,0}\exp \{i\mathbf{p}\left( \mathbf{x-x}^{\prime }\right) \}\left[ _{+}\phi
_{\mathbf{p},0}(t)_{+}\phi _{\mathbf{p},0}^{\ast }(t^{\prime })+_{-}\phi _{%
\mathbf{p},0}(t)_{-}\phi _{\mathbf{p},0}^{\ast }(t^{\prime })\right] d%
\mathbf{p\,}.  \label{emt43}
\end{eqnarray}%
The range of integration $\bar{D}$ reads%
\begin{equation*}
\bar{D}:\left\{ 
\begin{array}{c}
\left\vert \mathbf{p}_{\bot }\right\vert \leq \sqrt{\left\vert qE\right\vert 
}\left[ \sqrt{\left\vert qE\right\vert }\left( t+T/2\right) -K\right] ^{1/2}
\\ 
t+K/\sqrt{\left\vert qE\right\vert }\leq p_{3}/qE\leq T/2-K/\sqrt{\left\vert
qE\right\vert }\,%
\end{array}%
\right. .
\end{equation*}%
Integrating over $\mathbf{p}_{\perp }$ in (\ref{emt43}), we find 
\begin{equation}
\Delta ^{\bar{p}}(x,x^{\prime })=-i\int_{t+K/\sqrt{\left\vert qE\right\vert }%
}^{T/2-K/\sqrt{\left\vert qE\right\vert }}h_{\parallel }\left( x_{\parallel
},x_{\parallel }^{\prime }\right) h_{\perp }\left( \mathbf{x}_{\perp },%
\mathbf{x}_{\perp }^{\prime }\right) d\left( \frac{p_{3}}{qE}\right) \,,
\label{emt43.1}
\end{equation}

The following results will be outlined in the presence of a constant
magnetic field,$\;B\neq 0$. For such a field, we select the nonzero
potential as $A_{1}^{B}=Bx^{2}$. Now, the complete set of quantum numbers
that describes particles is $(p_{1},n_{B},p_{3},r)$, $n_{B}=0,1,\ldots $, $%
\lambda =\left( M^{2}+|qB|(2n_{B}+1-r)\right) \left| qE\right| ^{-1}$. The
space-time dependent part of the function $\phi _{\mathbf{p},s,r}(x)$, see (%
\ref{emt36}), is modified as follows:%
\begin{eqnarray}
&&\left( 2\pi \right) ^{-3/2}\exp \{i\mathbf{p}\mathbf{x}\}\rightarrow
\left( 2\pi \right) ^{-1/2}\exp \{-ip_{3}x^{3}\}\phi _{p_{1},n_{B},r}(%
\mathbf{x}_{\perp })\,,  \notag \\
&&\phi _{p_{1},n_{B},r}(\mathbf{x}_{\perp })=\left( \frac{\sqrt{|qB|}}{%
2^{n_{B}+1}\pi ^{3/2}n_{B}!}\right) ^{1/2}\exp \left\{ -ip_{1}x^{1}-\frac{%
|qB|}{2}\left( x^{2}-\frac{p_{1}}{qB}\right) ^{2}\right\}  \notag \\
&&\times \mathcal{H}_{n_{B}}\left[ \sqrt{|qB|}\left( x^{2}-\frac{p_{1}}{qB}%
\right) \right] \,.  \label{mag}
\end{eqnarray}%
Here, $\mathcal{H}_{n_{B}}(x)$ are the Hermite polynomials. Then,
integration over $\mathbf{p}_{\perp }$ in the above formulas must be
replaced by integration over $p_{1}$ from $-\infty $ to $+\infty $ with
summation over the integer quantum numbers $n_{B}$ in the interval 
\begin{equation*}
n_{B}\leq \left[ \sqrt{\left| qE\right| }\left( t+T/2\right) -K\right]
\left| E/2B\right| \,.
\end{equation*}%
After this, one can see that formula (\ref{emt44b}) needs only one
modification. Namely, the new (in the presence of magnetic field) value of $%
h_{\perp }\left( \mathbf{x}_{\perp },\mathbf{x}_{\perp }^{\prime }\right) $
reads 
\begin{equation}
h_{\perp }\left( \mathbf{x}_{\perp },\mathbf{x}_{\perp }^{\prime }\right) =%
\frac{qB\exp \left( \pi \Sigma ^{3}B/E\right) }{4\pi \sinh (\pi B/E)}\exp
\left\{ -\frac{\pi M^{2}}{\left| qE\right| }-\left( \mathbf{x}_{\perp }-%
\mathbf{x}_{\perp }^{\prime }\right) ^{2}\frac{qB}{4}\coth \left( \pi
B/E\right) \right\} .  \label{emt44d}
\end{equation}

Using (\ref{emt44b}) and (\ref{emt43.1}), one can represent $\langle j_{\mu
}\left( t\right) \rangle $ $^{p,\bar{p}}$ and $\langle T_{\mu \nu }\left(
t\right) \rangle $ $^{p,\bar{p}}$\ in (\ref{emt24.2}) as follows: 
\begin{eqnarray*}
&\langle j_{\mu }\left( t\right) \rangle ^{p,\bar{p}}=&iq\left. \mathrm{tr}%
\left[ \gamma _{\mu }\gamma P\Delta ^{p,\bar{p}}(x,x^{\prime })\right]
\right\vert _{x=x^{\prime }}\,, \\
&\langle T_{\mu \nu }\left( t\right) \rangle ^{p,\bar{p}}=&i\left. \mathrm{tr%
}\left[ A_{\mu \nu }\gamma P\Delta ^{p,\bar{p}}(x,x^{\prime })\right]
\right\vert _{x=x^{\prime }}\,.
\end{eqnarray*}%
Then, using (\ref{emt44b}), (\ref{emt44c}), (\ref{emt43.1}), and (\ref%
{emt44d}), we find%
\begin{eqnarray}
&\langle j_{\mu }(t)\rangle ^{p,\bar{p}}=&iq\left. P_{\mu }\mathrm{tr\,}%
\Delta ^{p,\bar{p}}(x,x^{\prime })\right\vert _{x=x^{\prime }}\,,  \notag \\
&\langle T_{\mu \mu }(t)\rangle ^{p,\bar{p}}=&i\left. P_{\mu }^{2}\mathrm{tr}%
\,\Delta ^{p,\bar{p}}(x,x^{\prime })\right\vert _{x=x^{\prime }}\,.
\label{emt44a}
\end{eqnarray}%
The off-diagonal matrix elements of $\langle T_{\mu \nu }\rangle ^{p}$ and $%
\langle T_{\mu \nu }\rangle ^{\bar{p}}$ are all equal to zero.

Taking derivatives, calculating traces, and integrating over $p_{3}$, we
obtain the leading contributions at large $T$. First of all, 
\begin{eqnarray}
&&\,\langle j_{\mu }\left( t\right) \rangle ^{p}=-2\delta _{\mu }^{3}q\,%
\mathrm{sgn}\left( qE\right) \left( 1/2+t/T\right) n^{cr},  \notag \\
&&\,\langle j_{\mu }\left( t\right) \rangle ^{\bar{p}}=-2\delta _{\mu
}^{3}q\,\mathrm{sgn}\left( qE\right) \left( 1/2-t/T\right) n^{cr},  \notag \\
&&\,\langle T_{00}\left( t\right) \rangle ^{p}=\langle T_{33}\left( t\right)
\rangle ^{p}=\left| qE\right| T\left( 1/2+t/T\right) ^{2}n^{cr},  \notag \\
&&\,\langle T_{00}\left( t\right) \rangle ^{\bar{p}}=\langle T_{33}\left(
t\right) \rangle ^{\bar{p}}=\left| qE\right| T\left( 1/2-t/T\right)
^{2}n^{cr},  \label{emt.44}
\end{eqnarray}%
where 
\begin{equation}
n^{cr}=\frac{q^{2}}{4\pi ^{2}}EBT\,\coth (\pi B/E)\left[ \exp \left( -\pi 
\frac{M^{2}}{\left| qE\right| }\right) +O\left( \frac{K}{\sqrt{\left|
qE\right| }T}\right) \right] \,.  \label{emt.45}
\end{equation}%
Secondly, 
\begin{eqnarray}
&&\,\langle T_{11}\left( t\right) \rangle ^{p}=\langle T_{22}\left( t\right)
\rangle ^{p}\,,\;\langle T_{11}\left( t\right) \rangle ^{\bar{p}}=\langle
T_{22}\left( t\right) \rangle ^{\bar{p}}\,,  \notag \\
&&\,\langle T_{11}\left( t\right) \rangle ^{p}=\widetilde{n}\left\{ 
\begin{array}{l}
\ln \left[ \sqrt{\left| qE\right| }\left( T/2+t\right) \right] +O\left( \ln
K\right) ,\,\mathrm{\;}\sqrt{\left| qE\right| }\left( T/2+t\right) >K\, \\ 
O\left( \ln K\right) \,,\;\mathrm{\;}\sqrt{\left| qE\right| }\left(
T/2+t\right) \leq K\,%
\end{array}%
\right. ,  \notag \\
&&\,\langle T_{11}\left( t\right) \rangle ^{\bar{p}}=\widetilde{n}\left\{ 
\begin{array}{l}
-\ln \left[ \sqrt{\left| qE\right| }\left( T/2-t\right) \right] +O\left( \ln
K\right) ,\;\sqrt{\left| qE\right| }\left( T/2-t\right) >K\, \\ 
-O\left( \ln K\right) \,,\;\mathrm{\;}\sqrt{\left| qE\right| }\left(
T/2-t\right) \leq K\,%
\end{array}%
\right. ,  \label{emt44}
\end{eqnarray}%
where 
\begin{equation}
\widetilde{n}=\frac{(qB)^{2}}{4\pi ^{2}\sinh ^{2}\left( \pi B/E\right) }\exp
\left( -\pi \frac{M^{2}}{\left| qE\right| }\right) \,.  \label{emt45}
\end{equation}

We can see that all the leading contributions are real-valued. Note that the
current density $\langle j_{\mu }\left( t\right) \rangle ^{p}$ and the
component of EMT $\langle T_{\mu \nu }\left( t\right) \rangle ^{p}$ are zero
for $t\leq t_{1}$ and change with time until $t_{2}$, $\langle j_{3}\left(
t\right) \rangle ^{p}$ being linear, $\langle T_{00}\left( t\right) \rangle
^{p}=\langle T_{33}\left( t\right) \rangle ^{p}$ quadratic, and $\langle
T_{11}\left( t\right) \rangle ^{p}=\langle T_{22}\left( t\right) \rangle
^{p}\,$logarithmic, respectively. As will be demonstrated, the rate $n^{cr}$
in (\ref{emt.44}) is the total number-density of pairs created by the $T$%
-constant electric field.

After switching on of the electric field (at $t>t_{2}$), all the mean values
(\ref{emt.44}), (\ref{emt44}) are constant and retain their values at $%
t_{2}. $

We should mention that formulas (\ref{emt.44}), (\ref{emt44}) have been
obtained for the first time.

At large times $t-t_{1}=t+T/2\gg K/\sqrt{\left| qE\right| }$, it is useful
to examine the quantities 
\begin{equation}
j_{\mu }^{cr}\left( t\right) =\langle j_{\mu }\left( t\right) \rangle
^{0}-\langle j_{\mu }\left( t\right) \rangle _{out}^{0}\,,\;\;T_{\mu \nu
}^{cr}\left( t\right) =\langle T_{\mu \nu }\left( t\right) \rangle
^{0}-\langle T_{\mu \nu }\left( t\right) \rangle _{out}^{0}\,,  \label{emt46}
\end{equation}%
which can be interpreted as those corresponding to $out$-particles at any
large $t$. At the final time instant $t_{2}$, they coincide with the
physical quantities $\langle j_{\mu }\left( t_{2}\right) \rangle ^{p}$ and $%
\langle T_{\mu \nu }\left( t_{2}\right) \rangle ^{p}$. Then it follows from (%
\ref{emt24.1}) and (\ref{emt.44}), (\ref{emt44}) that 
\begin{eqnarray}
&&j_{\mu }^{cr}\left( t\right) =\langle j_{\mu }\left( t\right) \rangle
^{p}-\langle j_{\mu }\left( t\right) \rangle ^{\bar{p}}=-2\delta _{\mu
}^{3}q\,\mathrm{sgn}\left( qE\right) \left( 2t/T\right) n^{cr}\,,  \notag \\
&&T_{\mu \nu }^{cr}\left( t\right) =\langle T_{\mu \nu }\left( t\right)
\rangle ^{p}-\langle T_{\mu \nu }\left( t\right) \rangle ^{\bar{p}}\,,
\label{emt47}
\end{eqnarray}%
and 
\begin{eqnarray}
&\langle T_{00}\left( t\right) \rangle ^{cr}=&\langle T_{33}\left( t\right)
\rangle ^{cr}=2\left| qE\right| tn^{cr},\;\;\langle T_{11}\left( t\right)
\rangle ^{cr}=\langle T_{22}\left( t\right) \rangle ^{cr}\,,  \label{emt48}
\\
&\langle T_{11}\left( t\right) \rangle ^{cr}=&\widetilde{n}\left\{ 
\begin{array}{l}
\ln \left[ \left| qE\right| \left( \left( T/2\right) ^{2}-t^{2}\right) %
\right] +O\left( \ln K\right) ,\;\sqrt{\left| qE\right| }\left( T/2-t\right)
>K\, \\ 
\ln \left[ \sqrt{\left| qE\right| }\left( T/2+t\right) \right] +O\left( \ln
K\right) ,\;\sqrt{\left| qE\right| }\left( T/2-t\right) \leq K%
\end{array}%
\right. .  \notag
\end{eqnarray}

At $t\geq t_{2}$, the production of pairs terminates, and the quantities (%
\ref{emt47}) and (\ref{emt48}) maintain their values (\ref{emt47}) and (\ref%
{emt48}) at $t=t_{2},$ and present the current density and EMT of created
particles. For example, from $j_{\mu }^{cr}$ in (\ref{emt47}) we can see
that $n^{cr}$ is the total number-density of pairs created during the entire
time of action of the electric field.

We emphasize that expressions (\ref{emt.44}), (\ref{emt44}) for $\langle
j_{\mu }(t)\rangle ^{p}$ and $\langle T_{\mu \nu }(t)\rangle ^{p}$ represent
contributions to $\langle j_{\mu }(t)\rangle ^{0}$ and $\langle T_{\mu \nu
}(t)\rangle ^{0}$, respectively, at any time instant $t$ if the
stabilization condition (\ref{unistab}) is valid. On the contrary,
expressions (\ref{emt47}), (\ref{emt48}) are valid only for time instants $%
\bar{t}$ that are sufficiently close to $t_{2}$, 
\begin{equation*}
\bar{t}:\;\left( T-2\bar{t}\right) /T\ll 1,
\end{equation*}%
when the interpretation in terms of final particles already makes sense.

\subsection{Vacuum polarization contributions}

We are now going to calculate the real-valued parts of the quantities $%
\langle j_{\mu }\left( t\right) \rangle ^{c}$ and $\langle T_{\mu \nu
}\left( t\right) \rangle ^{c}$ defined in (\ref{emt24.2}), which do not have
any $T$-divergences. In these expressions, we can use the causal Green
function $S^{c}(x,x^{\prime })$ (\ref{emt19}) in the constant field, since
the above quantities do not need any regularization.

For such a Green function, the so-called Fock--Schwinger proper time
representation holds true:%
\begin{equation}
S^{c}(x,x^{\prime })=(\gamma P+M)\Delta ^{c}(x,x^{\prime }),\;\;\Delta
^{c}(x,x^{\prime })=\int_{0}^{\infty }f(x,x^{\prime },s)ds\,;  \label{emt31}
\end{equation}%
see \cite{Nik70} and \cite{GGG98}, where the Fock--Schwinger kernel $%
f(x,x^{\prime },s)$ reads%
\begin{eqnarray}
\ f(x,x^{\prime },s) &=&\exp \left( -i\frac{q}{2}\sigma ^{\mu \nu }F_{\mu
\nu }s\right) f^{(0)}(x,x^{\prime },s),\;\sigma ^{\mu \nu }=\frac{i}{2}%
[\gamma ^{\mu },\gamma ^{\nu }],  \notag \\
\ f^{(0)}(x,x^{\prime },s) &=&\frac{q^{2}EB\exp \left( iq\Lambda \right) }{%
(4\pi )^{2}\sinh (qEs)\sin (qBs)}\exp \left[ -iM^{2}s-i\frac{\left(
x-x^{\prime }\right) qF\coth (qFs)\left( x-x^{\prime }\right) }{4}\right] ;
\label{emt33}
\end{eqnarray}%
see \cite{S51,F37}.\ Singularities of the kernel are situated at the origin
and on the imaginary axis, $\left| qE\right| s=-i\pi n$, $n=1,2,3,\ldots $.

It should be noted that in this expression\ the only term 
\begin{equation}
\ \Lambda =\Lambda _{\parallel }+\Lambda _{\perp },\;\;\Lambda _{\parallel
}=\left( x_{0}+x_{0}^{\prime }\right) \left( x_{3}-x_{3}^{\prime }\right)
E/2,\;\;\Lambda _{\perp }=-\int_{x^{\prime }}^{x}A_{\mu }^{B}dx^{\mu }\,,
\label{emt34}
\end{equation}%
is potential-dependent. Here, $A_{\mu }^{B}$ is the potential of magnetic
field $F_{\mu \nu }^{B},$ and the integral is taken along the line
connecting the points $x$ and $x^{\prime }.$

Using this representation, we calculate $\func{Re}\langle j_{\mu }\left(
t\right) \rangle ^{c}$ and $\func{Re}\langle T_{\mu \nu }\left( t\right)
\rangle ^{c}$. It is easy to see that $\langle j_{\mu }\left( t\right)
\rangle ^{c}=0$, as should be expected due to translational symmetry, and $%
\langle T_{\mu \nu }(t)\rangle ^{c}=0\,,\;\mu \neq \nu \,.$

Calculating the diagonal terms $\langle T_{\mu \mu }(t)\rangle ^{c}$, we
discover that it can be derived from the Heisenberg--Euler Lagrangian $%
\mathcal{L}$, 
\begin{equation*}
\mathcal{L=}\frac{1}{2}\mathrm{tr}\int_{0}^{\infty }s^{-1}\,f(x,x,s)\,ds\,.
\end{equation*}

Subtracting the zero external field contribution from $\mathcal{L}$ and
performing the standard renormalizations of $\mathcal{L-}\left. \mathcal{L}%
\right| _{F=0}\,$, that leave $qF_{\mu \nu }$ invariant, i.e., the standard
renormalizations of the coupling constant $q^{2}$ and potentials $A_{\mu }$,
we obtain a finite expression: 
\begin{equation}
\mathcal{L}_{ren}=\int_{0}^{\infty }\frac{ds\exp \left( -iM^{2}s\right) }{%
8\pi ^{2}s}\left[ q^{2}EB\coth \left( qEs\right) \cot \left( qBs\right) -%
\frac{1}{s^{2}}-\frac{q^{2}}{3}\left( E^{2}-B^{2}\right) \right] ;
\label{HEL1}
\end{equation}%
see \cite{HeiE36}. Making the same renormalization for $\langle T_{\mu \mu
}(t)\rangle ^{c}$, we can see that there holds the relation 
\begin{eqnarray}
&&\,\langle T_{00}(t)\rangle _{ren}^{c}=-\langle T_{33}(t)\rangle
_{ren}^{c}=E\frac{\partial \mathcal{L}\left( t\right) _{ren}}{\partial E}-%
\mathcal{L}\left( t\right) _{ren}\,,  \notag \\
&&\,\langle T_{11}(t)\rangle _{ren}^{c}=\langle T_{22}(t)\rangle _{ren}^{c}=%
\mathcal{L}\left( t\right) _{ren}-B\frac{\partial \mathcal{L}\left( t\right)
_{ren}}{\partial B}\,,  \label{HEL}
\end{eqnarray}%
where $\mathcal{L}\left( t\right) _{ren}=\left. \mathcal{L}_{ren}\right|
_{E\rightarrow E(t)}\,$, and $E(t)$ is defined by\footnote{%
In the absence of electric field, the Green function $S^{c}(x,x^{\prime })$, 
$S_{in}^{c}(x,x^{\prime })$, and $S_{out}^{c}(x,x^{\prime })$ coincide.
Therefore, in case $t<t_{1}$ we have $\langle T_{\mu \nu }(t)\rangle ^{0}=$.$%
\left. \langle T_{\mu \nu }(t)\rangle ^{0}\right| _{E=0}=\left. \langle
T_{\mu \nu }(t)\rangle ^{c}\right| _{E=0}$. For $t>t_{2}$, relation (\ref%
{emt46}) implies 
\begin{equation*}
\langle T_{\mu \nu }(t)\rangle ^{0}=\langle T_{\mu \nu }(t)\rangle
_{out}^{0}+T_{\mu \nu }^{cr}\left( t\right) \;,
\end{equation*}%
where $T_{\mu \nu }^{cr}\left( t\right) \;$is given by (\ref{emt47}),(\ref%
{emt48}) and $\langle T_{\mu \nu }(t)\rangle _{out}^{0}=$.$\left. \langle
T_{\mu \nu }(t)\rangle ^{0}\right| _{E=0}=\left. \langle T_{\mu \nu
}(t)\rangle ^{c}\right| _{E=0}$, i.e., this is a vacuum polarization
contribution.} (\ref{el-field}).

Finally, the vacuum mean values of the current density and EMT have the form 
\begin{equation}
\langle j_{\mu }(t)\rangle ^{0}=\func{Re}\langle j_{\mu }(t)\rangle
^{p},\;\langle T_{\mu \nu }(t)\rangle _{ren}^{0}=\mathrm{\func{Re}}\langle
T_{\mu \nu }(t)\rangle _{ren}^{c}+\func{Re}\langle T_{\mu \nu }(t)\rangle
^{p},  \label{emt52}
\end{equation}%
where $\func{Re}\langle j_{\mu }(t)\rangle ^{p}$ and $\func{Re}\langle
T_{\mu \nu }(t)\rangle ^{p}$ are presented by (\ref{emt.44}) and (\ref{emt44}%
). We can see that the $T$-dependent contributions $\func{Re}\langle j_{\mu
}(t)\rangle ^{p}$ and $\func{Re}\langle T_{\mu \nu }(t)\rangle ^{p}$ arise
due to vacuum instability; they are global physical quantities and have the
factor $\exp \left\{ -\pi M^{2}/\left| qE\right| \right\} $. This factor is
exponentially small for a weak electric field, $M^{2}/\left| qE\right| \gg 1$%
, and these quantities can actually be observed as soon as the external
field strength approaches the critical value $E_{c}$. On the other hand, the
term $\mathrm{\func{Re}}\langle T_{\mu \nu }(t)\rangle _{ren}^{c}$ is $t$%
-dependent and $T$-independent; it is therefore local and does exist in
arbitrary electric fields. When the $T$-constant electric field switches
off, the local contribution from the electric field to $\mathrm{\func{Re}}%
\langle T_{\mu \nu }(t)\rangle _{ren}^{c}$ vanishes, but the global one,
given by $\func{Re}\langle T_{\mu \nu }(t_{2})\rangle ^{p}$, does remain.
Thus, in the general case, both kinds of contributions are important. We
stress that in order to make $\langle T_{\mu \nu }(t)\rangle _{ren}^{0}$
finite the only term $\langle T_{\mu \nu }(t)\rangle ^{c}$ has to be
regularized and renormalized due to the standard ultraviolet divergences,
which is consistent with the fact that the ultraviolet divergences have a
local nature.

Using expression (\ref{emt52}), we can find a condition which imposes
restrictions on the strength of classical constant electric field. Note that
for a strong electric field ($B=0$), $M^{2}/\left| qE\right| \ll 1$, and for
a large $T$, there is a well-known asymptotic expression for the vacuum
energy density $\mathrm{\func{Re}}\langle T_{00}(t)\rangle _{ren}^{c}$, 
\begin{equation}
\mathrm{\func{Re}\,}\langle T_{00}(t)\rangle _{ren}^{c}=-\frac{q^{2}}{24\pi
^{2}}E^{2}\ln \frac{\left| qE\right| }{M^{2}}\,.  \label{mm}
\end{equation}%
It does not depend on $T$. The energy density of a classical electric field
is $E^{2}/8\pi $. Then (see \cite{GreRo73}) one makes the conclusion that
the notion of a strong constant{\Large \ }electric field is physically
meaningful when (\ref{mm}) is much less than $E^{2}/8\pi $, which implies%
\footnote{%
The same inequality has been obtained by Ritus as a condition to avoid the
breakdown of the loop expansion in a strong field; see, e.g., \cite{Rit87}.} 
\begin{equation}
\frac{q^{2}}{3\pi }\ln \frac{\left| qE\right| }{M^{2}}\ll 1\,.  \label{aa}
\end{equation}

However, if $T$ is large, one also has to take into account the $T$%
-dependent term $\func{Re}\langle T_{00}(t)\rangle ^{p}$ in (\ref{emt52}).
At $t=t_{2}$, and for an arbitrary magnitude of the electric field,{\Large \ 
}we have 
\begin{equation}
\func{Re}\,\langle T_{00}(t_{2})\rangle ^{p}=\frac{q^{2}E^{2}}{4\pi ^{3}}%
\left\vert qE\right\vert T^{2}\,\exp \left( -\pi \frac{M^{2}}{\left\vert
qE\right\vert }\right) .  \label{nn}
\end{equation}%
One can neglect the back-reaction of created pairs in electric field only in
case (\ref{nn}) is far less than $E^{2}/8\pi $, which implies 
\begin{equation}
\left\vert qE\right\vert T^{2}\ll \frac{\pi ^{2}}{2q^{2}}\exp \left( \pi 
\frac{M^{2}}{\left\vert qE\right\vert }\right) ,  \label{val}
\end{equation}%
or, in Gaussian units,%
\begin{equation*}
\frac{c}{\hslash }\left\vert qE\right\vert T^{2}\ll \frac{\pi ^{2}c\hslash }{%
2q^{2}}\exp \left( \pi \frac{c^{3}M^{2}}{\hslash \left\vert qE\right\vert }%
\right) .
\end{equation*}

The restriction (\ref{val}) is completely new, it is much more restrictive
than (\ref{aa}). It imposes restrictions both on the maximal strength of
electric field and its duration due to the back-reaction of created
particles.

\section{Initial state as thermal equilibrium}

We are now going to calculate the quantities $<j_{\mu }\left( t\right) >$
and $<T_{\mu \nu }\left( t\right) >$ (\ref{emt22}) with the density operator
(\ref{emt1}), by using relations (\ref{emt24.1}). In fact, the contributions 
$<j_{\mu }\left( t\right) >^{0}$ and $<T_{\mu \nu }\left( t\right) >^{0}$ to
these quantities have been calculated in the previous section. Here, we
study the remaining temperature-dependent contributions $\langle j_{\mu
}\left( t\right) \rangle ^{\theta }$ and $\langle T_{\mu \nu }\left(
t\right) \rangle ^{\theta }$ to these quantities.

\subsection{Contributions due to the particle creation}

Let us examine the function $S^{\theta }\left( x,x^{\prime }\right) $\ in (%
\ref{emt13}). The form of the functions $\tilde{S}_{\theta }^{\mp }$ in (\ref%
{emt13}) for $t<t_{1},$ and $t^{\prime }<t_{1}$ is{\Large \ }known, they are
functions for zero{\Large \ }electric field. Therefore, we have to study the
case $t>t_{1},\ t^{\prime }>t_{1}$. First of all, we separate the
contributions from particle creation in the same manner as it has been done
for the function $S_{in}^{c}(x,x^{\prime })$ in Sec. 2; for details, see
Subsec. 1 of the Appendix.

Using (\ref{emt24.1}), (\ref{emt52}), and (\ref{emt64}) (see the Appendix),
we present the mean values $\langle j_{\mu }\left( t\right) \rangle $ and $%
\langle T_{\mu \nu }(t)\rangle _{ren}$: 
\begin{eqnarray}
&&\,\langle j_{\mu }\left( t\right) \rangle =\mathrm{\func{Re}}\langle
j_{\mu }\left( t\right) \rangle _{\theta }^{c}+J_{\mu }^{p}\left( t\right)
\,,  \notag \\
&&J_{\mu }^{p}\left( t\right) =\func{Re}\langle j_{\mu }(t)\rangle ^{p}+%
\mathrm{\func{Re}}\langle j_{\mu }\left( t\right) \rangle _{\theta \,}^{p}\,,
\notag \\
&&\,\langle T_{\mu \nu }(t)\rangle _{ren}=\func{Re}\langle T_{\mu \nu
}(t)\rangle _{ren}^{c}+\mathrm{\func{Re}}\langle T_{\mu \nu }\left( t\right)
\rangle _{\theta }^{c}+\tau _{\mu \nu }^{p}\left( t\right) \,,  \notag \\
&&\tau _{\mu \nu }^{p}\left( t\right) =\func{Re}\langle T_{\mu \nu
}(t)\rangle ^{p}+\mathrm{\func{Re}}\langle T_{\mu \nu }\left( t\right)
\rangle _{\theta }^{p}\,.  \label{emt71}
\end{eqnarray}%
Now, $\langle T_{\mu \nu }(t)\rangle $ is replaced by the renormalized (with
respect to ultraviolet divergences) quantity $\langle T_{\mu \nu }(t)\rangle
_{ren}$ by inserting $\langle T_{\mu \nu }(t)\rangle _{ren}^{c}$ in the
r.h.s. instead of $\langle T_{\mu \nu }(t)\rangle ^{c}.$ The terms $J_{\mu
}^{p}\left( t\right) $ and $\tau _{\mu \nu }^{p}\left( t\right) $ with the
upper script $p$ represent the contributions from particle-creation, the
terms with the upperscript $c$ represent the remaining contributions. One
ought to recall that the vacuum polarization contribution $\langle T_{\mu
\nu }(t)\rangle _{ren}^{c}$ has been already calculated and has the form (%
\ref{HEL}). In the absence of electric field (for $t<t_{1}$ and $t>t_{2}$),
the quantity (\ref{emt71}) does not depend on time, in particular, $J_{\mu
}^{p}\left( t\right) =\tau _{\mu \nu }^{p}\left( t\right) =0$ for $t<t_{1}$.
As has been already mentioned, the interpretation in terms of $out$%
-particles makes sense only for $t>\bar{t}$, where $\bar{t}$ is sufficiently
close to $t_{2}$, $\left( T-2\bar{t}\right) /T\ll 1$. At the same time, for $%
t>\bar{t}$ the quantities $J_{\mu }^{p}\left( t\right) $ and $\tau _{\mu \nu
}^{p}\left( t\right) $ are the current density and EMT of $out$-particles
created by the electric field. The quantities $\mathrm{\func{Re}}\langle
j_{\mu }\left( t\right) \rangle _{\theta }^{c}$ and $\mathrm{\func{Re}}%
\langle T_{\mu \nu }\left( t\right) \rangle _{\theta }^{c}$ describe the
corresponding contributions related to the existence of real $in$-particles
at the initial state.

The quantities $\mathrm{\func{Re}}\langle j_{\mu }\left( t\right) \rangle
_{\theta }^{p}$ and $\mathrm{\func{Re}}\langle T_{\mu \nu }\left( t\right)
\rangle _{\theta }^{p}$ present the contributions from particle-creation at
finite temperature. They are $T$-divergent. The leading contributions to
these quantities at large $T$ can be found by analogy with the case of zero
temperature, which has been investigated above. We examine the leading
contributions to $S_{\theta ,\zeta }^{p}$, which are given by the expression%
\begin{equation}
S_{\theta ,\zeta }^{p}(x,x^{\prime })=i\int_{D}d\mathbf{p}\sum_{r=\pm
1}N_{p_{3}}^{(\zeta )}\left( in\right) \aleph _{\mathbf{p},r}\left( ^{+}{%
\psi }_{\mathbf{p},r}(x)^{+}{\bar{\psi}}_{\mathbf{p},r}(x^{\prime })-\,^{-}{%
\psi }_{\mathbf{p},r}(x)^{-}{\bar{\psi}}_{\mathbf{p},r}(x^{\prime })\right) .
\label{emt67}
\end{equation}%
The integral $\int_{D}d\mathbf{p}$ is defined in (\ref{emt40}); the
differential numbers $\aleph _{\mathbf{p},r}$ are defined in (\ref{e40ad}),
and 
\begin{equation}
N_{p_{3}}^{(\zeta )}\left( in\right) =\left\{ \exp \left[ \beta \left(
\varepsilon _{\mathbf{p}}^{(\zeta )}-\mu ^{(\zeta )}\right) \right]
+1\right\} ^{-1}\,,  \label{NNN}
\end{equation}%
where $\varepsilon _{\mathbf{p}}^{(\zeta )}$ in (\ref{NNN}) are
quasi-energies at $t_{1}$; see (\ref{energ}).

In order to obtain a generalization of (\ref{emt67}) to the presence of a
constant magnetic field,$\;B\neq 0$, we have to follow the way described in
Subsec. 3.2. Then, taking summation and integration, we obtain the leading
contribution to the function $S_{\theta ,\zeta }^{p}\left( x,x^{\prime
}\right) $ in the case under consideration:%
\begin{eqnarray}
&&S_{\theta ,\zeta }^{p}\left( x,x^{\prime }\right) =(\gamma P+M)\Delta
_{\theta ,\zeta }^{p}\left( x,x^{\prime }\right) \,,  \notag \\
&&\Delta _{\theta ,\zeta }^{p}(x,x^{\prime })=i\int_{-T/2+K/\sqrt{\left\vert
qE\right\vert }}^{t-K/\sqrt{\left\vert qE\right\vert }}N_{p_{3}}^{(\zeta
)}\left( in\right) h_{\parallel }\left( x_{\parallel },x_{\parallel
}^{\prime }\right) h_{\perp }\left( \mathbf{x}_{\perp },\mathbf{x}_{\perp
}^{\prime }\right) d\left( \frac{p_{3}}{qE}\right) \,.  \label{emt67.2}
\end{eqnarray}%
The functions $h_{\parallel }\left( x_{\parallel },x_{\parallel }^{\prime
}\right) $ and $h_{\perp }\left( \mathbf{x}_{\perp },\mathbf{x}_{\perp
}^{\prime }\right) $ are given by expressions (\ref{emt44c}) and (\ref%
{emt44d}), respectively.

Therefore, we obtain the leading contributions to $\langle j_{\mu }\left(
t\right) \rangle _{\theta }^{p}\;$and $\langle T_{\mu \nu }\left( t\right)
\rangle _{\theta }^{p}$ , namely, 
\begin{eqnarray}
&\langle j_{\mu }\left( t\right) \rangle _{\theta }^{p}=&iq\sum_{\zeta =\pm
}\left. P_{\mu }\mathrm{tr}\,\Delta _{\theta ,\zeta }^{p}(x,x^{\prime
})\right\vert _{x=x^{\prime }}\,,  \notag \\
&\langle T_{\mu \nu }\left( t\right) \rangle _{\theta }^{p}=&0\,,\;\mu \neq
\nu ;\;\langle T_{\mu \mu }\left( t\right) \rangle _{\theta
}^{p}=i\sum_{\zeta =\pm }\left. P_{\mu }^{2}\mathrm{tr}\,\Delta _{\theta
,\zeta }^{p}(x,x^{\prime })\right\vert _{x=x^{\prime }}\,.  \label{emt68}
\end{eqnarray}

In order to examine the temperature-dependent contributions to $J_{\mu
}^{p}\left( t\right) $ and $\tau _{\mu \nu }^{p}\left( t\right) $ in (\ref%
{emt71}), we have to investigate the quantities (\ref{emt68}). At low
temperatures, $\beta \left| \left( \varepsilon _{\mathbf{p}}^{(\zeta )}-\mu
^{(\zeta )}\right) \right| \gg 1$, and assuming $\varepsilon _{\mathbf{p}%
}^{(\zeta )}>\left| \mu ^{(\zeta )}\right| $, one can see that the
contributions from $\mathrm{\func{Re}}\langle j_{\mu }\left( t\right)
\rangle _{\theta }^{p}$ and $\mathrm{\func{Re}}\langle T_{\mu \nu }\left(
t\right) \rangle _{\theta }^{p}$ are very small in comparison with the
vacuum contributions $\func{Re}\langle j_{\mu }(t)\rangle ^{p}$ and $\func{Re%
}\langle T_{\mu \nu }(t)\rangle ^{p}$. Therefore, in the case under
consideration the leading contributions to $J_{\mu }^{p}\left( t\right) $
and $\tau _{\mu \nu }^{p}\left( t\right) $ are given by%
\begin{equation}
J_{\mu }^{p}\left( t\right) =\func{Re}\langle j_{\mu }(t)\rangle ^{p},\;\tau
_{\mu \nu }^{p}\left( t\right) =\func{Re}\langle T_{\mu \nu }(t)\rangle ^{p}.
\label{emt74}
\end{equation}

For $\varepsilon _{\mathbf{p}}^{(\zeta )}<\left| \mu ^{(\zeta )}\right| $,
with a sufficiently large $\left| \mu ^{(\zeta )}\right| \gg \sqrt{\left|
qE\right| }K$, contributions from $\mathrm{\func{Re}}\langle j_{\mu }\left(
t\right) \rangle _{\theta }^{p}$ and $\mathrm{\func{Re}}\langle T_{\mu \nu
}\left( t\right) \rangle _{\theta }^{p}$ are comparable with contributions
from $\func{Re}\langle j_{\mu }(t)\rangle ^{p}$ and $\func{Re}\langle T_{\mu
\nu }(t)\rangle ^{p}$, respectively. For example, at $\left| \mu ^{(\zeta
)}\right| \gtrsim \left| qE\right| \left( t+T/2\right) $ we have 
\begin{equation*}
\mathrm{\func{Re}}\langle j_{\mu }\left( t\right) \rangle _{\theta }^{p}=-2%
\func{Re}\langle j_{\mu }(t)\rangle ^{p},\;\;\mathrm{\func{Re}}\langle
T_{\mu \nu }\left( t\right) \rangle _{\theta }^{p}=-2\func{Re}\langle T_{\mu
\nu }(t)\rangle ^{p}\,.
\end{equation*}%
Then%
\begin{equation}
J_{\mu }^{p}\left( t\right) =-\func{Re}\langle j_{\mu }(t)\rangle
^{p},\;\tau _{\mu \nu }^{p}\left( t\right) =-\func{Re}\langle T_{\mu \nu
}(t)\rangle ^{p}\,.  \label{emt75}
\end{equation}%
The negative sign in these expressions implies that the total number-density
of particles decreases due to electron-positron annihilation in electric
field, and, as a consequence, $\langle j_{\mu }\left( t\right) \rangle $ and 
$\langle T_{\mu \nu }(t)\rangle _{ren}\,$in (\ref{emt71}) are less than $%
\mathrm{\func{Re}}\langle j_{\mu }\left( t\right) \rangle _{\theta }^{c}$
and $\langle T_{\mu \nu }(t)\rangle _{ren}^{c}+\mathrm{\func{Re}}\langle
T_{\mu \nu }\left( t\right) \rangle _{\theta }^{c}$, respectively. This\
fact is in agreement with the fact that, due to particle creation, there
holds the relation 
\begin{equation*}
N_{m}^{(\zeta )}\left( out\right) =\left( 1-\aleph _{m}\right) N_{m}^{(\zeta
)}(in)+\aleph _{m}\left[ 1-N_{m}^{(-\zeta )}(in)\right] \,.
\end{equation*}%
between the initial $N_{m}^{(\zeta )}(in)$ and final $N_{m}^{(\zeta )}\left(
out\right) $ differential mean numbers; see \cite{GavGT06}. Thus, the
differential mean numbers of created particles are given by the difference $%
\Delta N_{m}^{(\zeta )}=N_{m}^{(\zeta )}\left( out\right) -N_{m}^{(\zeta
)}(in)$. As a result, $\Delta N_{m}^{(\zeta )}$ is negative in case $%
N_{m}^{(+)}(in)+N_{m}^{(-)}(in)>1$.

At high temperatures, $\beta \left[ \left| qE\right| \left( t+T/2\right)
-\mu ^{(\pm )}\right] \ll 1$, we find 
\begin{eqnarray}
&&J_{\mu }^{p}\left( t\right) =\frac{\beta }{2}\left[ \frac{1}{2}\left\{
\left| qE\right| \left( t+T/2\right) -\mu ^{(+)}-\mu ^{(-)}\right\} +\sqrt{%
\left| qE\right| }O\left( K\right) \right] \func{Re}\langle j_{\mu
}(t)\rangle ^{p}\,,  \notag \\
&&\tau _{00}^{p}\left( t\right) =\tau _{33}^{p}\left( t\right) =\frac{\beta 
}{2}\left[ \frac{1}{3}\left| qE\right| \left( t+T/2\right) -\frac{1}{2}%
\left( \mu ^{(+)}+\mu ^{(-)}\right) +\sqrt{\left| qE\right| }O\left(
K\right) \right] \func{Re}\langle T_{00}(t)\rangle ^{p}\,,  \notag \\
&&\tau _{11}^{p}\left( t\right) =\tau _{22}^{p}\left( t\right) =\frac{\beta 
}{2}\left[ \left| qE\right| \left( t+T/2\right) -\frac{1}{2}\left( \mu
^{(+)}+\mu ^{(-)}\right) \right]  \notag \\
&&\times \left[ 1+\sqrt{\left| qE\right| }O\left( \frac{\ln K}{\ln \left[ 
\sqrt{\left| qE\right| }\left( t+T/2\right) \right] }\right) \right] \func{Re%
}\langle T_{11}(t)\rangle ^{p}\,.  \label{emt76}
\end{eqnarray}

We can see that all the components of $J_{\mu }^{p}\left( t\right) $ and $%
\tau _{\mu \nu }^{p}\left( t\right) $ at high temperatures are far less than
the corresponding components of the vacuum contributions $\langle j_{\mu
}(t)\rangle ^{p}$ and $\langle T_{\mu \nu }(t)\rangle ^{p}$, which is
natural for Fermi particles.

The resulting expressions for the mean energy density of created particles
make it possible to conclude that the restriction (\ref{val}) for the
external constant electric field, established for vacuum, is also valid at
the initial low-temperature state. In case the temperature of the initial
state is sufficiently high in comparison with the mean kinetic energy of
created particles, $\beta \left| qE\right| T\ll 1$, this restriction changes
considerably. In this case, the mean energy density of created particles
after the electric field turns off, 
\begin{equation}
\tau _{00}^{p}\left( t_{2}\right) =\frac{1}{6}\beta \left| qE\right| T\func{%
Re}\langle T_{00}(t_{2})\rangle ^{p}\,,  \label{nnt}
\end{equation}%
in accordance with\ (\ref{emt76}), is much smaller than the corresponding
vacuum expression $\func{Re}\langle T_{00}(t_{2})\rangle ^{p}$\ in (\ref{nn}%
). Repeating the arguments of Subsec. 3.3, we find that one can neglect the
back-reaction of created pairs in electric field only in case $\tau
_{00}^{p}\left( t_{2}\right) $ is far less than $E^{2}/8\pi $, which implies%
\begin{equation}
\beta \left| qE\right| ^{2}T^{3}\ll \frac{3\pi ^{2}}{q^{2}}\,\exp \left( \pi 
\frac{M^{2}}{\left| qE\right| }\right) .  \label{valt}
\end{equation}%
This is much less restrictive than (\ref{val}).

\subsection{Conventional contribution}

We are now going to calculate the temperature-dependent contributions to the
mean current density and EMT, which are labelled by the upperscript $c.$ We
refer to these contributions as conventional contributions. Such
contributions are not related to either particle creation or vacuum
polarization. Thus, it is natural to say that they are determined only by
the quantum statistical behavior of the initial thermal gas of charged
particles in the constant external field under consideration.

\subsubsection{Proper-time representation}

Let us represent $\langle j_{\mu }\left( t\right) \rangle _{\theta }^{c}$\
and $\langle T_{\mu \nu }\left( t\right) \rangle _{\theta }^{c}$\ in (\ref%
{emt71}) as integrals over the proper time. To this end, we relate $%
S_{\theta ,\zeta }$ in (\ref{emt54}) (see the Appendix) to the proper-time
representation (\ref{emt31}) of $S^{c}$; for details, see Subsec. 2 of the
Appendix. For simplicity, we assume that the chemical potentials are not
very large, $\left| \mu ^{(\zeta )}\right| <M$. As always, we consider large 
$T$, namely, 
\begin{equation}
\sqrt{\left| qE\right| }T\gg 1+M^{2}/\left| qE\right| \,.  \label{tcond}
\end{equation}

Using (\ref{emt65.3}), (\ref{emt65.2}) and (\ref{emt59}), as well as the
explicit form (\ref{emt60.3}) for $g(x,x^{\prime },s,\tau )$, and
calculating all the derivatives and traces, we find the following
expressions for $\func{Re}\langle j_{\mu }\left( t\right) \rangle _{\theta
}^{c}$\ and $\func{Re}\langle T_{\mu \nu }\left( t\right) \rangle _{\theta
}^{c}$: 
\begin{eqnarray}
&&\func{Re}<j_{\mu }\left( t\right) >_{\theta }^{c}=\sum_{\zeta =\pm
}\sum_{l=1}^{\infty }\left( -1\right) ^{l+1}e^{\beta l\mu ^{(\zeta )}}\func{%
Re}J_{\mu }^{(l)}\,,  \notag \\
&&J_{\mu }^{(l)}=\int_{0}^{\infty }ds\int_{0}^{\infty }du\,h\left(
s,u\right) Y\left( s,u\right) j_{\mu }\left( s,u\right) \,,\;j_{\mu }\left(
s,u\right) =i\delta _{\mu }^{3}q^{2}E\left( t+T/2\right) \frac{\tau }{%
a-i\tau }\,,  \notag \\
&&h\left( s,u\right) =-\frac{i}{8\pi ^{2}\sqrt{\pi }}\frac{q^{2}EB\cot \left[
qB\left( s-i\tau \right) \right] }{u^{1/2}\sinh \left( qEs\right) }\left( 
\frac{a}{a-i\tau }\right) ^{1/2}e^{-u/4}e^{-M^{2}\left( is+\tau \right) }\,,
\notag \\
&&Y\left( s,u\right) =\exp \left[ -\left( qE\right) ^{2}\left( t+T/2\right)
^{2}\frac{\tau a}{a-i\tau }\right] \,,\;\;\tau =\left( \beta l\right)
^{2}/u\,,  \label{emt61.1}
\end{eqnarray}%
and 
\begin{eqnarray}
&&\func{Re}\langle T_{\mu \nu }\left( t\right) \rangle _{\theta
}^{c}=\sum_{\zeta =\pm }\sum_{l=1}^{\infty }\left( -1\right) ^{l+1}e^{\beta
l\mu ^{(\zeta )}}\func{Re}T_{\mu \nu }^{(l)}\,,  \notag \\
&&T_{\mu \nu }^{(l)}=\int_{0}^{\infty }ds\int_{0}^{\infty }du\;h\left(
s,u\right) Y\left( s,u\right) t_{\mu \nu }\left( s,u\right) \,,
\label{emt61.2}
\end{eqnarray}%
where it is only the diagonal values $t_{\mu \nu }\left( s,u\right) $\ that
are nonzero:%
\begin{eqnarray}
t_{00}\left( s,u\right) &=&\sum_{k=1,2,3}t_{kk}\left( s,u\right) +M^{2}\cosh
\left( qEs\right) \,,\;\;t_{11}\left( s,u\right) =t_{22}\left( s,u\right) =%
\frac{-iqB}{\sin \left[ 2qB\left( s-i\tau \right) \right] }\,,  \notag \\
t_{33}\left( s,u\right) &=&\frac{-i}{2\left( a-i\tau \right) \cosh \left(
qEs\right) }-\frac{\left( qE\right) ^{2}\left( t+T/2\right) ^{2}}{\cosh
\left( qEs\right) }\left( \frac{\tau }{a-i\tau }\right) ^{2}\,.
\label{emt61.3}
\end{eqnarray}

Note that the integrals over $s$ and $u$ in the representations (\ref%
{emt61.1}) and (\ref{emt61.2}) are finite. Such representations hold true
for any $E$ and temperature for $t\in \lbrack t_{1},t_{2}]$ within the
restriction (\ref{tcond}) on the interval $T$.

Representations (\ref{emt61.1}) and (\ref{emt61.2}) could be transformed to
other forms, being more convenient for regions near $s=0$ in the integrands;
for details, see Subsec. 3 of the Appendix. The result is that $\func{Re}%
J_{\mu }^{(l)}$ and $\func{Re}T_{\mu \nu }^{(l)}$ in (\ref{emt61.1}) and (%
\ref{emt61.2}) can be presented as%
\begin{equation}
\func{Re}J_{\mu }^{(l)}=\tilde{J}_{\mu }^{a}+\func{Re}J_{\mu }^{b}\,,\;\;%
\func{Re}T_{\mu \nu }^{(l)}=\tilde{T}_{\mu \nu }^{a}+\func{Re}T_{\mu \nu
}^{b}\,,  \label{emt61.10}
\end{equation}%
where%
\begin{eqnarray}
&&\tilde{J}_{\mu }^{a}=\int_{0}^{\pi /\left\vert 2qE\right\vert
-0}ds^{\prime }\int_{0}^{\infty }du^{\prime }\;\tilde{h}\left( s^{\prime
},u^{\prime }\right) \tilde{Y}\left( s^{\prime },u^{\prime }\right) \tilde{j}%
_{\mu }\left( s^{\prime },u^{\prime }\right) \,,  \notag \\
&&\tilde{T}_{\mu \nu }^{a}=\int_{0}^{\pi /\left\vert 2qE\right\vert
-0}ds^{\prime }\int_{0}^{\infty }du^{\prime }\;\tilde{h}\left( s^{\prime
},u^{\prime }\right) \tilde{Y}\left( s^{\prime },u^{\prime }\right) \tilde{t}%
_{\mu \iota }\left( s^{\prime },u^{\prime }\right) \,,  \label{emt61.7}
\end{eqnarray}%
and 
\begin{eqnarray}
&&\tilde{h}\left( s^{\prime },u^{\prime }\right) =-ih\left( i\left(
c_{0}-s^{\prime }\right) ,u^{\prime }+u_{0}\right)  \notag \\
&&\,=\frac{1}{8\pi ^{2}\sqrt{\pi }}\frac{q^{2}EB\coth \left[ qB\left(
s^{\prime }+\delta \tau \right) \right] }{u^{\prime 1/2}\sin \left(
qEs^{\prime }\right) }\exp \left[ -\frac{u^{\prime }+u_{0}}{4}-M^{2}\left(
s^{\prime }+\delta \tau \right) \right] \,,  \notag \\
&&\tilde{Y}\left( s^{\prime },u^{\prime }\right) =Y\left( i\left(
c_{0}-s^{\prime }\right) ,u^{\prime }+u_{0}\right) =\exp \left[ -\frac{%
\left( qE\right) ^{2}\left( t+T/2\right) ^{2}\tau u^{\prime }}{\left(
u^{\prime }+u_{0}\right) \left[ 1+\left( qE\tau \right) ^{2}\right] }\right]
\,,  \notag \\
&&\delta \tau =\tau -\left( qE\right) ^{-1}\arctan \left( qE\tau \right)
,\;\;\tau =\frac{\left( \beta l\right) ^{2}}{u^{\prime }+u_{0}}\,,
\label{emt61.8}
\end{eqnarray}%
\begin{eqnarray}
\tilde{j}_{\mu }\left( s^{\prime },u^{\prime }\right) &=&j_{\mu }\left(
i\left( c_{0}-s^{\prime }\right) ,u^{\prime }+u_{0}\right) \,,\;\;\tilde{t}%
_{\mu \nu }\left( s^{\prime },u^{\prime }\right) =t_{\mu \nu }\left( i\left(
c_{0}-s^{\prime }\right) ,u^{\prime }+u_{0}\right) \,,  \notag \\
\tilde{j}_{\mu }\left( s^{\prime },u^{\prime }\right) &=&-\delta _{\mu }^{3}%
\frac{q\left( qE\right) ^{2}\left( t+T/2\right) \tau \left[ 1+qE\tau \tan
\left( qEs^{\prime }\right) \right] }{\tan \left( qEs^{\prime }\right) \left[
1+\left( qE\tau \right) ^{2}\right] }\,,  \notag \\
\tilde{t}_{11}\left( s^{\prime },u^{\prime }\right) &=&\tilde{t}_{22}\left(
s^{\prime },u^{\prime }\right) =\frac{qB}{\sinh \left[ 2qB\left( s^{\prime
}+\delta \tau \right) \right] }\,,  \notag \\
\tilde{t}_{33}\left( s^{\prime },u^{\prime }\right) &=&\frac{qE}{2\sin
\left( qEs^{\prime }\right) \left[ 1+\left( qE\tau \right) ^{2}\right] ^{1/2}%
}+\frac{\left( qE\right) ^{4}\left( t+T/2\right) ^{2}\tau ^{2}\left[
1+qE\tau \tan \left( qEs^{\prime }\right) \right] }{\sin \left( qEs^{\prime
}\right) \tan \left( qEs^{\prime }\right) \left[ 1+\left( qE\tau \right) ^{2}%
\right] ^{3/2}}\,,  \notag \\
\tilde{t}_{00}\left( s^{\prime },u^{\prime }\right) &=&\sum_{k=1,2,3}\tilde{t%
}_{kk}\left( s^{\prime },u^{\prime }\right) +M^{2}\cos \left( qEs^{\prime
}\right) \frac{1+qE\tau \tan \left( qEs^{\prime }\right) }{\left[ 1+\left(
qE\tau \right) ^{2}\right] ^{1/2}}\,.  \label{emt61.9}
\end{eqnarray}%
Note that $\tau \leq \left( qE\right) ^{-1}\tan \left( qEs^{\prime }\right) $
in\ $\tilde{J}_{\mu }^{a}$ and $\tilde{T}_{\mu \nu }^{a}$. The expressions
for $J_{\mu }^{b}$ and $T_{\mu \nu }^{b}$ are given by (\ref{emt61.4}) and (%
\ref{emt61.5}), respectively. The form (\ref{emt61.10}) is convenient for
the study of both low-temperature approximations to a weak electric field
and high-temperature approximations to an arbitrary electric field.

In a weak electric field, $\left| qE\right| /M^{2}\ll 1$, the quantities $%
\func{Re}J_{\mu }^{b}$ and $\func{Re}T_{\mu \nu }^{b}$ are exponentially
small, so that in order to find the expansions of $\func{Re}J_{\mu }^{(l)}$
and $\func{Re}T_{\mu \nu }^{(l)}$\ in the powers of\emph{\ }$\left|
qE\right| /M^{2}$,we use the relations 
\begin{equation}
\func{Re}J_{\mu }^{(l)}=\tilde{J}_{\mu }^{a},\;\;\func{Re}T_{\mu \nu }^{(l)}=%
\tilde{T}_{\mu \nu }^{a},  \label{emt61.11}
\end{equation}%
where the exponentially small contributions have been neglected.

\subsubsection{Zero electric field}

When the electric field is zero, the mean current density is zero as well.
Considering the limit $E\rightarrow 0$, one can see that the integral (\ref%
{emt61.7}) over $s^{\prime }$ for $\tilde{T}_{\mu \nu }^{a}$ is taken from
zero to infinity. In addition, the dependence on the variable $u^{\prime }$
in the kernel of the matrix elements $\tilde{T}_{\mu \nu }^{a}$ factors out
as the multiplier $e^{-u^{\prime }/4}/\sqrt{u^{\prime }}$. Calculating the
integral over $u^{\prime }$ by using (\ref{emt55.2}), in case $E=0$, we find 
\begin{eqnarray}
&&\tilde{T}_{\mu \nu }^{a}=T_{\mu \nu }^{0}=\int_{0}^{\infty }ds\rho
_{0}\left( s\right) t_{\mu \iota }^{0}\left( s\right) \,,  \notag \\
&&\rho _{0}\left( s\right) =\frac{qB\coth \left( qBs\right) }{4\pi ^{2}s}%
\exp \left[ -M^{2}s-\frac{\left( \beta l\right) ^{2}}{4s}\right]
\,,\;\;t_{33}^{0}\left( s\right) =\frac{1}{2s}\,,  \notag \\
&&t_{11}^{0}\left( s\right) =t_{22}^{0}\left( s\right) =\frac{qB}{\sinh
\left( 2qBs\right) }\,,\;\;t_{00}^{0}\left( s\right)
=\sum_{k=1,2,3}t_{kk}^{0}\left( s\right) +M^{2}\,.  \label{emt77}
\end{eqnarray}%
The quantities $T_{\mu \nu }^{0}$ are real-valued. Integrating by parts, one
can see that $T_{00}^{0}$ has the form (\ref{emt77}), where 
\begin{equation*}
t_{00}^{0}\left( s\right) =-\frac{1}{2s}+\frac{\left( \beta l\right) ^{2}}{%
4s^{2}}\,.
\end{equation*}

In the case under consideration, the vacuum is stable, while the
renormalized EMT (\ref{emt71}) is real-valued, time-independent, and is
given only by the conventional terms%
\begin{eqnarray}
\left. \langle T_{\mu \nu }(t)\rangle _{ren}\right| _{E=0} &=&\left. \langle
T_{\mu \nu }(t)\rangle _{ren}^{c}\right| _{E=0}+\left. \langle T_{\mu \nu
}\left( t\right) \rangle _{\theta }^{c}\right| _{E=0}\,,  \notag \\
\left. \langle T_{\mu \nu }\left( t\right) \rangle _{\theta }^{c}\right|
_{E=0} &=&\sum_{\zeta =\pm }\sum_{l=1}^{\infty }\left( -1\right)
^{l+1}e^{\beta l\mu ^{(\zeta )}}T_{\mu \nu }^{0}\,,  \label{emt80}
\end{eqnarray}%
where $\left. \langle T_{\mu \nu }(t)\rangle _{ren}^{c}\right| _{E=0}$ is
determined by (\ref{HEL}) at $E=0$.

The system under consideration is a relativistic fermionic gas in an
external magnetic field, $B$, immersed in a sufficiently large quantization
volume $V$. This system is described by the partition function%
\begin{equation*}
Z=\mathrm{Tr}\exp \left\{ \beta \left( \sum_{\zeta =\pm }\mu ^{(\zeta
)}N^{(\zeta )}-H\right) \right\} ,
\end{equation*}%
that follows from (\ref{emt1}), $H=H\left( t_{1}\right) $, and by the
thermodynamic potential $\Omega =-\theta \ln Z$. At the state of
equilibrium, the temperature- and field-dependent effective Lagrangian $%
\mathcal{L}_{ren}^{\theta }$ is related to $\Omega $ as follows:%
\begin{equation}
\mathcal{L}_{ren}^{\theta }=-\Omega /V.  \label{emt81}
\end{equation}%
The mean values of energy density $\langle H\rangle /V$ and particle density 
$n^{(\zeta )}$ in the state (\ref{emt1}) can be expressed in terms of $%
\mathcal{L}_{ren}^{\theta }$ as 
\begin{eqnarray}
\frac{\langle H\rangle }{V} &=&-\mathcal{L}_{ren}^{\theta }-\beta \frac{%
\partial \mathcal{L}_{ren}^{\theta }}{\partial \beta }+\sum_{\zeta =\pm }\mu
^{(\zeta )}n^{(\zeta )}\,,  \label{emt82} \\
n^{(\zeta )} &=&\frac{\partial \mathcal{L}_{ren}^{\theta }}{\partial \mu
^{(\zeta )}}\,.  \label{emt83}
\end{eqnarray}%
Note that the energy density\ has been already calculated as a component of (%
\ref{emt80}), namely, 
\begin{equation*}
\langle H_{0}\rangle /V=\left. \langle T_{00}(t)\rangle _{ren}\right|
_{E=0}\,.
\end{equation*}%
Thus, we find that the effective Lagrangian has the form%
\begin{eqnarray}
\mathcal{L}_{ren}^{\theta } &=&\left. \mathcal{L}_{ren}\right| _{E=0}+\Delta 
\mathcal{L}_{\theta }\,,  \notag \\
\Delta \mathcal{L}_{\theta } &=&\frac{1}{8\pi ^{2}}\sum_{\zeta =\pm
}\sum_{l=1}^{\infty }\left( -1\right) ^{l+1}e^{\beta l\mu ^{(\zeta
)}}\int_{0}^{\infty }\frac{ds}{s^{2}}\exp \left[ -M^{2}s-\frac{\left( \beta
l\right) ^{2}}{4s}\right] qB\coth (qBs)\,,  \label{emt84}
\end{eqnarray}%
where $\left. \mathcal{L}_{ren}\right| _{E=0}$ is the renormalized
Heisenberg--Euler effective Lagrangian (\ref{HEL1}) at $E=0$.

It turns out that the space components of $\left. \langle T_{ik}(t)\rangle
_{ren}\right| _{E=0}$ (\ref{emt80}) can be derived from the effective
Lagrangian $\mathcal{L}_{ren}^{\theta }$ as follows: 
\begin{eqnarray}
\left. \langle T_{11}(t)\rangle _{ren}\right| _{E=0} &=&\left. \langle
T_{22}(t)\rangle _{ren}\right| _{E=0}=\mathcal{L}_{ren}^{\theta }-B\frac{%
\partial \mathcal{L}_{ren}^{\theta }}{\partial B}\,,  \notag \\
\left. \langle T_{33}(t)\rangle _{ren}\right| _{E=0} &=&\mathcal{L}%
_{ren}^{\theta }\,.  \label{emt84a}
\end{eqnarray}

It should be noted that the finite temperature and density effects are
contained only in $\Delta \mathcal{L}_{\theta }$. In particular, the
thermodynamic potential $\Omega _{free}$ of perfect free Fermi gas at $B=0$
is related to $\left. \Delta \mathcal{L}_{\theta }\right| _{B=0}$ in the
form $\Omega _{free}=-\left. \Delta \mathcal{L}_{\theta }\right| _{B=0}V$ .
Expressing the proper-time integral for $\left. \Delta \mathcal{L}_{\theta
}\right| _{B=0}$ in terms of the modified Bessel function of the third kind
(Macdonald's function) $K_{\nu }\left( z\right) $ \cite{BatE74}, and using
formula 3.471.9 in \cite{GR94},%
\begin{equation}
\int_{0}^{\infty }x^{\nu -1}e^{-a/x-bx}dx=2\left( \frac{a}{b}\right) ^{\nu
/2}K_{\nu }\left( 2\sqrt{ab}\right) ,  \label{emt94}
\end{equation}%
we find 
\begin{equation}
\left. \Delta \mathcal{L}_{\theta }\right| _{B=0}=\left( \frac{M}{\beta \pi }%
\right) ^{2}\sum_{\zeta =\pm }\sum_{l=1}^{\infty }\frac{\left( -1\right)
^{l+1}}{l^{2}}e^{\beta l\mu ^{(\zeta )}}K_{2}\left( M\beta l\right) \,.
\label{emt84.1}
\end{equation}%
Thus, we obtain for the quantity $\Omega _{free}$ the expression well-known
from textbooks.

Considering the usual case $\mu ^{(+)}=-\mu ^{(-)}=\mu $, we can see that (%
\ref{emt84}) is identical with the results obtained earlier in the case $E=0$%
: see Eq. (52) in the paper \cite{CanD96}, where the imaginary-time
formalism was used, and Eq. (5.20) in \cite{ElmPS93}, where the real-time
formalism was used (we disregard the factor $1/2$ in Eq. (5.20) of \cite%
{ElmPS93} as an misprint). In the case $\mu =0$, Eq. (\ref{emt84}) is in
agreement with the result obtained in \cite{Dit79}.

It should be noted that our analysis deals primarily with the effects of
electric field at finite initial temperatures, and therefore we do not
examine in detail the particular case of magnetic field (at $E=0$). Note,
however, that the leading magnetic field-dependent terms follow immediately
from (\ref{emt84}). We present these terms for comparison with a nonzero
electric field: in the weak-field limit, when either $\left| qB\right|
/M^{2}\ll 1$ or $\left| qB\right| \beta ^{2}\ll 1$, we have 
\begin{equation}
\mathcal{L}_{1}=\Delta \mathcal{L}_{\theta }-\left. \Delta \mathcal{L}%
_{\theta }\right| _{B=0}=\frac{\left( qB\right) ^{2}}{12\pi ^{2}}\sum_{\zeta
=\pm }\sum_{l=1}^{\infty }\left( -1\right) ^{l+1}e^{\beta l\mu ^{(\zeta
)}}K_{0}\left( M\beta l\right) \,.  \label{emt84.2}
\end{equation}%
In the strong-field limit, $\left| qB\right| \gg M^{2},\beta ^{-2}$, we have 
\begin{equation*}
\mathcal{L}_{1}=\frac{\left| qB\right| M}{2\pi ^{2}\beta }\sum_{\zeta =\pm
}\sum_{l=1}^{\infty }\frac{\left( -1\right) ^{l+1}}{l}e^{\beta l\mu ^{(\zeta
)}}K_{1}\left( M\beta l\right) \,.
\end{equation*}%
The latter quantity is small in comparison with the corresponding vacuum
polarization term, which is given by the following expression:%
\begin{equation*}
\left. \mathcal{L}_{ren}\right| _{E=0}-\left. \mathcal{L}_{ren}\right|
_{E=0,B=0}=\frac{\left( qB\right) ^{2}}{24\pi ^{2}}\ln \frac{\left|
qB\right| }{M^{2}}\,.
\end{equation*}

In the weak-field limit, we can see that the leading field-dependent thermal
contributions to EMT have the form 
\begin{eqnarray}
\tau _{11}^{B} &=&\tau _{22}^{B}=\mathcal{L}_{1}-B\frac{\partial \mathcal{L}%
_{1}}{\partial B}=-\mathcal{L}_{1}\,,\;\;\tau _{33}^{B}=\mathcal{L}_{1}\,, 
\notag \\
\tau _{00}^{B} &=&-\mathcal{L}_{1}-\beta \frac{\partial \mathcal{L}_{1}}{%
\partial \beta }+\sum_{\zeta =\pm }\mu ^{(\zeta )}\frac{\partial \mathcal{L}%
_{1}}{\partial \mu ^{(\zeta )}}\,.  \label{emt84.3}
\end{eqnarray}

Note that Fermi systems with an external magnetic field have been under
intensive study, and various asymptotic expansions of the effective
Lagrangian at finite temperatures have been made: for a review, see \cite%
{CanD96,ElmPS93}; a review on finite-temperature free relativistic systems
can be found, e.g., in \cite{LanW87,Bel96,Actor86}.

\subsubsection{Weak electric field and low temperatures}

In a weak electric field, $\left| qE\right| /M^{2}\ll 1$, the quantities $%
\func{Re}J_{\mu }^{(l)}$ and $\func{Re}T_{\mu \nu }^{(l)}$ in (\ref{emt61.1}%
) and (\ref{emt61.2}) are given by $\tilde{J}_{\mu }^{a}$ and $\tilde{T}%
_{\mu \nu }^{a}$ (\ref{emt61.7})--(\ref{emt61.9}); see (\ref{emt61.11}). The
dependence of expressions (\ref{emt61.7})--(\ref{emt61.9}) on the parameters 
$E$ and $\beta $ is described by $qE\tau $, which is a function of two
variables: $s^{\prime }$ and $u^{\prime }$. Approximations of this function
and of the corresponding integral over $u^{\prime }$ at low temperatures and
at high temperatures are quite different. Therefore, unlike the case $E=0$,
these limiting cases must be studied separately. In this subsection, we
shall examine the case of low temperatures, $M\beta \gg 1$.

Contributions to integrals (\ref{emt61.7}), given by the region $%
M^{2}s^{\prime }\gg 1$ of large $s^{\prime }$, are exponentially small.
Therefore, we limit the range of integration over $s^{\prime }$ from above
by the value $\chi /M^{2}$, so that $1\ll \chi \ll M^{2}/\left| qE\right| $.
At low temperatures, the leading contributions to the integral over $%
s^{\prime }$ are formed at the region $M^{2}s^{\prime }\sim M\beta l/2\gg 1$%
, provided that $M\beta l/2\ll \chi $. It is interesting to examine an
approximation which admits the limiting process $E\rightarrow 0$. For every $%
l\leq l_{\max }$ in a sufficiently weak electric field, there exists such $%
\chi $ that $M\beta l_{\max }/2\ll \chi $. Under this condition, we obtain
the approximation for finite values of $l$. Note that at low temperatures
the terms $e^{\mu ^{(\zeta )}\beta l}\func{Re}J_{\mu }^{(l)}$ and $e^{\mu
^{(\zeta )}\beta l}\func{Re}T_{\mu \nu }^{(l)}$ of order $e^{-\left( M-\mu
^{(\zeta )}\right) \beta l}l^{-\alpha }$ in case $\alpha \geq 1/2$ decrease
when $l$ increases. Therefore, any accuracy of calculation for the sums in (%
\ref{emt61.1}) and (\ref{emt61.2}) can be provided by a finite number of
summands. Accordingly, at $\left| \mu ^{(\zeta )}\right| \ll M$, in order to
estimate the leading contributions it is sufficient to take into account
only the first summand, at $l=1$. However, even for large values of chemical
potentials, $\left| \mu ^{(\zeta )}\right| \sim M$, a sufficient accuracy is
provided by summation up to a corresponding finite number $l_{est}$.
Accordingly, we suppose that $l_{est}\leq l_{\max }$. Thus, the approximated
expressions for $\func{Re}\langle j_{\mu }\left( t\right) \rangle _{\theta
}^{c}$ and $\func{Re}\langle T_{\mu \nu }\left( t\right) \rangle _{\theta
}^{c}$ at low temperatures have the form 
\begin{eqnarray}
&&\func{Re}\langle j_{\mu }\left( t\right) \rangle _{\theta
}^{c}=\sum_{\zeta =\pm }\sum_{l=1}^{l_{est}}\left( -1\right) ^{l+1}e^{\beta
l\mu ^{(\zeta )}}J_{\mu }^{\chi }\,,  \notag \\
&&J_{\mu }^{\chi }=\int_{0}^{\chi /M^{2}}ds\int_{0}^{\infty }du\;\tilde{h}%
\left( s,u\right) \tilde{Y}\left( s,u\right) \tilde{j}_{\mu }\left(
s,u\right) \,;  \notag \\
&&\func{Re}\langle T_{\mu \nu }\left( t\right) \rangle _{\theta
}^{c}=\sum_{\zeta =\pm }\sum_{l=1}^{l_{est}}\left( -1\right) ^{l+1}e^{\beta
l\mu ^{(\zeta )}}T_{\mu \nu }^{\chi }\,,  \notag \\
&&T_{\mu \nu }^{\chi }=\int_{0}^{\chi /M^{2}}ds\int_{0}^{\infty }du\;\tilde{h%
}\left( s,u\right) \tilde{Y}\left( s,u\right) \tilde{t}_{\mu \iota }\left(
s,u\right) \,,  \label{emt85}
\end{eqnarray}%
where $\tilde{h}\left( s,u\right) $, $\tilde{Y}\left( s,u\right) $, $\tilde{j%
}_{\mu }\left( s,u\right) $, and $\tilde{t}_{\mu \iota }\left( s,u\right) $
are defined in (\ref{emt61.8}) and (\ref{emt61.9}).

Let us find the leading contributions depending on the electric field for
expressions (\ref{emt85}). These quantities depend on two dimensionless
parameters that characterize the influence of electric field: $\left|
qE\right| /M^{2}$ is a parameter determined by the strength of constant
electric field, whereas $qE\left( t+T/2\right) /M$ is a parameter depending
on time $t$ and determined by the increment in the kinetic momentum of a
particle in electric field by the time instant $t$ (i.e., the increment that
accumulates during the time interval $t-t_{1}=t+T/2$). The first parameter $%
\left| qE\right| /M^{2}$ is small in a weak electric field. We shall expand
all the functions in the integrals (\ref{emt85}) as series in the powers of
this parameter and examine only the leading terms of the expansion. In
addition, the value $\left| qE\right| \left( t+T/2\right) /M$ is not
necessarily small for a sufficiently large time interval, $t+T/2$. Thus, the
dependence on this value in the following expression is taken into account
precisely.\ For simplicity, we assume that the magnetic field is weak, so we
can neglect the biquadratic terms of the order $q^{2}\left| EB\right| /M^{4}$%
. Therefore, with accuracy up to the leading term depending on the electric
field, we obtain%
\begin{eqnarray}
J_{\mu }^{\chi } &=&\int_{0}^{\infty }ds\int_{0}^{\infty
}du\;h^{0}Y^{W}j_{\mu }^{W}\,,  \notag \\
T_{\mu \nu }^{\chi } &=&\int_{0}^{\infty }ds\int_{0}^{\infty }du\;\left(
h^{0}+\delta h\right) Y^{W}\left( t_{\mu \iota }^{0}+\delta t_{\mu \iota
}\right) \,.  \label{emt86}
\end{eqnarray}%
Here, we have neglected the exponentially small contribution and extended
the range of integration over $s$ to infinity, $+\infty $; besides, 
\begin{eqnarray}
&&h^{0}=\frac{e^{-u/4}}{2\sqrt{\pi u}}\rho _{0},\;\;\delta h=\frac{e^{-u/4}}{%
2\sqrt{\pi u}}\frac{\left( qE\right) ^{2}}{4\pi ^{2}}\exp \left[ -M^{2}s-%
\frac{\left( \beta l\right) ^{2}}{4s}\right] D\,,  \notag \\
&&D=\left[ \frac{\left( \beta l\right) ^{2}}{12s}-\frac{\tau _{W}^{3}}{3s^{2}%
}\left( \frac{1}{s}+M^{2}\right) +\frac{1}{6}\right] ,\;\;Y^{W}=\exp \left[ -%
\frac{\left( qE\right) ^{2}\left( t+T/2\right) ^{2}\tau _{W}u}{u+\left(
\beta l\right) ^{2}/s}\right] \,,  \notag \\
&&\tau _{W}=\frac{\left( \beta l\right) ^{2}}{u+\left( \beta l\right) ^{2}/s}%
\,,\;\;\;j_{\mu }^{W}=-\delta _{\mu }^{3}q^{2}E\left( t+T/2\right) \tau
_{W}/s\,,\;\;\delta t_{11}=\delta t_{22}=-\frac{\left( qE\right) ^{2}\tau
_{W}^{3}}{6s^{2}}\,,  \notag \\
&&\delta t_{33}=\frac{\left( qE\right) ^{2}}{4}\left( \frac{s}{3}-\frac{\tau
_{W}^{2}}{s}\right) +\left( qE\right) ^{2}\left( t+T/2\right) ^{2}\left( 
\frac{\tau _{W}}{s}\right) ^{2}\,,  \notag \\
&&\delta t_{00}=\sum_{k=1,2,3}\delta t_{kk}+\delta t_{M},\;\;\delta t_{M}=-%
\frac{1}{2}M^{2}\left( qE\right) ^{2}\left( s-\tau _{W}\right) ^{2}\,,
\label{emt87}
\end{eqnarray}%
where $t_{\mu \iota }^{0}$ and $\rho _{0}$ are given by (\ref{emt77}). Note
that in expressions (\ref{emt87}) we have not imposed any conditions on
temperature. Let us do it now. Namely, since at low temperatures the leading
contributions to the integral over $s$ are formed at the region of $s\sim
\beta l/\left( 2M\right) $, the function $\tau _{W}$ can be expanded as a
series in the powers of $u$, 
\begin{equation}
\tau _{W}=s\left\{ 1-\frac{su}{\left( \beta l\right) ^{2}}+\left[ \frac{su}{%
\left( \beta l\right) ^{2}}\right] ^{2}-\cdots \right\} \,.  \label{emt88}
\end{equation}%
We consider only the leading contribution to the expansion in inverse powers
of $\beta M$. To calculate this contribution, it is sufficient to take into
account the first three terms of expansion (\ref{emt88}); moreover, in the
coefficients at time-dependent functions it is sufficient to take into
account only the first term of the expansion. Therefore, at low temperatures
the function $Y^{W}$ in the leading approximation has the form%
\begin{equation}
Y^{W}=Y^{WL}=\exp \left[ -\left( qE\right) ^{2}\left( t+T/2\right) ^{2}\frac{%
s^{2}u}{\left( \beta l\right) ^{2}}\right] \,.  \label{emt89}
\end{equation}

Using expansion (\ref{emt88})\ and expression (\ref{emt89}),\ we can
calculate the integrals over $u$\ in (\ref{emt86}). To this end, it is
convenient to use the generating function $\Phi _{0}\left( \alpha \right) $,
defined by the integral%
\begin{equation}
\Phi _{0}\left( \alpha \right) =\frac{1}{2\sqrt{\pi }}\int_{0}^{\infty }%
\frac{du}{u^{1/2}}e^{-u\left( 1/4+\alpha \right) }\,.  \label{emt90}
\end{equation}%
Calculating this integral according to formula (\ref{emt55.2}), we obtain $%
\Phi _{0}\left( \alpha \right) =\left( 1+4\alpha \right) ^{-1/2}$. Then,
taking derivatives with respect to the parameter $\alpha $, we obtain a
formula we need for calculation: 
\begin{equation}
\Phi _{n}\left( \alpha \right) =\frac{1}{2\sqrt{\pi }}\int_{0}^{\infty }%
\frac{du}{u^{1/2}}e^{-u\left( 1/4+\alpha \right) }u^{n}=\left( -1\right) ^{n}%
\frac{\partial ^{n}\Phi _{0}\left( \alpha \right) }{\partial \alpha ^{n}}\,.
\label{emt91}
\end{equation}%
In the case under consideration, we have $\alpha =\left( qE\right)
^{2}\left( t+T/2\right) ^{2}\left( \frac{s}{\beta l}\right) ^{2}$.

Until the $T$-constant field has switched on, the system in question is at
the state of thermal equilibrium. When the $T$-constant field switches on at
the instant $t_{1}$ the particles at the initial state begin to accelerate.
The corresponding current density and EMT change with time until the moment $%
t_{2}$, when the $T$-constant field turns off. Nevertheless, such a
non-equilibrium state during a certain interval of time until the increment
of kinetic momentum is sufficiently small, $\left| qE\right| \left(
t+T/2\right) /M\ll 1$, is not very different from the initial state of
thermal equilibrium, and it makes sense to compare these states. It is
exactly this condition that guarantees the existence of a limiting process $%
E\rightarrow 0$. Therefore, it is important to examine such a case in more
detail. On the other hand, it is of special interest to examine the case of
asymptotically large values of time, when $\left| qE\right| \left(
t+T/2\right) /M\gg 1$. In this case, the presence of electric field
manifests itself to the maximum. In what follows, we examine two limiting
cases.

\paragraph{Small increment of kinetic momentum}

When the increment of kinetic momentum is small, $\left| qE\right| \left(
t+T/2\right) /M\ll 1$, the kernels in (\ref{emt86}) can be expanded in its
powers. To this end, we take into account the fact that $\alpha =0$ implies
that the function $\Phi _{n}\left( \alpha \right) $ is expressed in terms of
binomial coefficients, $\Phi _{0}\left( 0\right) =1$ and $\Phi _{n}\left(
0\right) =2^{n}\cdot 1\cdot 3\cdot 5\cdot \ldots \cdot \left( 2n-1\right) $,
for $n\geq 1$. Using (\ref{emt86}) after integrating over $u$, and
considering only the leading time-dependent contributions, we obtain 
\begin{equation}
J_{\mu }^{\chi }=j_{\mu }^{0}\int_{0}^{\infty }ds\rho _{0}\,,\;\;T_{\mu \nu
}^{\chi }=T_{\mu \nu }^{0}+\Delta T_{\mu \nu }\,,  \label{emt92}
\end{equation}%
where $T_{\mu \nu }^{0}$ and $\rho _{0}$ are independent of electric field
and given by (\ref{emt77}); besides 
\begin{equation}
j_{\mu }^{0}=-\delta _{\mu }^{3}q^{2}E\left( t+T/2\right) .  \label{emt92a}
\end{equation}%
In the above representation of $T_{\mu \nu }^{\chi }$, the part that depends
on the electric field is separated as $\Delta T_{\mu \nu }$ and has the form 
\begin{eqnarray}
\Delta T_{\mu \nu } &=&\frac{1}{4\pi ^{2}}\int_{0}^{\infty }ds\exp \left[
-M^{2}s-\frac{\left( \beta l\right) ^{2}}{4s}\right] \Delta t_{\mu \nu }\,, 
\notag \\
\Delta t_{11} &=&\Delta t_{22}=\frac{\left( qE\right) ^{2}}{2s}\left( D_{0}-%
\frac{1}{3}\right) -\frac{\left( qE\right) ^{2}\left( t+T/2\right) ^{2}}{%
s\left( \beta l\right) ^{2}}\,,  \notag \\
\Delta t_{33} &=&\Delta t_{11}+\frac{\left( qE\right) ^{2}}{2\left( \beta
l\right) ^{2}}+\frac{\left( qE\right) ^{2}\left( t+T/2\right) ^{2}}{s^{2}}\,,
\notag \\
\Delta t_{00} &=&\left( qE\right) ^{2}M^{2}D_{0}+\frac{\left( qE\right)
^{2}\left( t+T/2\right) ^{2}}{\left( \beta l\right) ^{2}}\left[ \frac{\left(
\beta l\right) ^{2}}{s^{2}}-2M^{2}\right] \,,  \notag \\
D_{0} &=&-\frac{1}{6}+\frac{\left( \beta l\right) ^{2}}{12s}-\frac{M^{2}s}{3}%
+2\frac{M^{2}s^{2}}{\left( \beta l\right) ^{2}}-24\frac{M^{2}s^{3}}{\left(
\beta l\right) ^{4}}\,.  \label{emt93}
\end{eqnarray}%
Integration in (\ref{emt93}) leads to the modified Bessel functions of the
third kind $K_{\nu }\left( z\right) $, according to formula (\ref{emt94}).
For simplicity, we assume that the chemical potentials are such that $%
e^{-\left( M-\mu ^{(\zeta )}\right) \beta }\ll 1$. We then use the
asymptotic expansion 7.13.1.(7) \cite{BatE74}: 
\begin{equation}
K_{\nu }\left( z\right) =\left( \frac{\pi }{2z}\right) ^{1/2}e^{-z}\left[
1+\left( \nu ^{2}-\frac{1}{4}\right) \frac{1}{2z}+\left( \nu ^{2}-\frac{1}{4}%
\right) \left( \nu ^{2}-\frac{9}{4}\right) \frac{1}{8z^{2}}+O\left( \left|
z\right| ^{-3}\right) \right] .  \label{emt95}
\end{equation}%
As a result of the asymptotic expansion at low temperatures, we obtain the
leading contributions of $\Delta T_{\mu \nu }$ in the form 
\begin{eqnarray}
\Delta T_{11} &=&\Delta T_{22}=-C_{l}\frac{2}{M\beta l}\left[ 1+\frac{%
M\left( t+T/2\right) ^{2}}{\beta l}\right] \left[ 1+O\left( \frac{1}{M\beta l%
}\right) \right] ,  \notag \\
\Delta T_{33} &=&C_{l}\left[ -\frac{3}{2M\beta l}+4\frac{M\left(
t+T/2\right) ^{2}}{\beta l}\right] \left[ 1+O\left( \frac{1}{M\beta l}%
\right) \right] ,  \notag \\
\Delta T_{00} &=&C_{l}\left[ -\frac{235}{256}+2\frac{M\left( t+T/2\right)
^{2}}{\beta l}\right] \left[ 1+O\left( \frac{1}{M\beta l}\right) \right] , 
\notag \\
C_{l} &=&\frac{\left( qE\right) ^{2}}{4\pi ^{2}}\left( \frac{\pi }{2M\beta l}%
\right) ^{1/2}e^{-M\beta l}.  \label{emt96}
\end{eqnarray}%
This makes it possible to present the expression for EMT $\func{Re}\langle
T_{\mu \nu }\left( t\right) \rangle _{\theta }^{c}$ in (\ref{emt85}) as
follows: 
\begin{equation}
\func{Re}\langle T_{\mu \nu }\left( t\right) \rangle _{\theta }^{c}=\left.
\langle T_{\mu \nu }\left( t\right) \rangle _{\theta }^{c}\right|
_{E=0}+\sum_{\zeta =\pm }\sum_{l=1}^{l_{est}}\left( -1\right) ^{l+1}e^{\beta
l\mu ^{(\zeta )}}\Delta T_{\mu \nu },  \label{emt97}
\end{equation}%
where we have explicitly separated the part $\left. \langle T_{\mu \nu
}\left( t\right) \rangle _{\theta }^{c}\right| _{E=0}$, independent of the
electric field, which is defined in (\ref{emt80}), while the terms $\Delta
T_{\mu \nu }$ depending on the electric field are given by (\ref{emt96}).

With the help of the expression for $J_{\mu }^{\chi }$ in (\ref{emt92}), the
current density $\func{Re}\langle j_{\mu }\left( t\right) \rangle _{\theta
}^{c}$ in (\ref{emt85}), with the corresponding accuracy, can be expressed
in the form 
\begin{equation}
\func{Re}\langle j_{\mu }\left( t\right) \rangle _{\theta }^{c}=-\delta
_{\mu }^{3}\frac{q^{2}E\left( t+T/2\right) }{M}\sum_{\zeta =\pm }n^{(\zeta
)}\,,  \label{emt98}
\end{equation}%
where the initial particle density $n^{(\zeta )}$ is given by (\ref{emt83}).
Considering only the leading asymptotic contributions, we obtain $n^{(\zeta
)}$ as follows: 
\begin{equation}
n^{(\zeta )}=2\left( \frac{M}{2\pi \beta }\right)
^{3/2}\sum_{l=1}^{l_{est}}\left( -1\right) ^{l+1}e^{\beta l\left( \mu
^{(\zeta )}-M\right) }\frac{1}{l^{3/2}}\frac{qB\beta l}{2M}\coth \frac{%
qB\beta l}{2M}\,.  \label{emt99}
\end{equation}%
If the chemical potential is small, $\left| \mu ^{(\zeta )}\right| \ll M$,
the leading contribution in the sums (\ref{emt97}), (\ref{emt99}) yields one
term, at $l=1$. In this case, the leading magnetic field-dependent thermal
contributions (\ref{emt84.3}) in $\func{Re}\langle T_{\mu \nu }\left(
t\right) \rangle _{\theta }^{c}$ are $\tau _{11}^{B}=\tau _{22}^{B}=\mathcal{%
-}\tau _{33}^{B}<0$, $\tau _{00}^{B}>0$, and all the time-independent
contributions for electric field-dependent terms in (\ref{emt97}) are
negative.

\paragraph{Large increment of kinetic momentum}

When the increment of kinetic momentum is large, $\left| qE\right| \left(
t+T/2\right) /M\gg 1$, the parameter $\alpha $ in (\ref{emt90}) is large as
well. Then, we calculate the integrals over $u$ in (\ref{emt86}), by using
the asymptotics of $\Phi _{0}\left( \alpha \right) $, $\Phi _{0}\left(
\alpha \right) \simeq \alpha ^{-1/2}/2$, and obtain 
\begin{eqnarray}
J_{\mu }^{\chi } &=&\frac{j_{\mu }^{0}\beta l}{2\left| qE\right| \left(
t+T/2\right) }\int_{0}^{\infty }ds\rho _{0}s^{-1}\,,  \notag \\
T_{\mu \nu }^{\chi } &=&\frac{\beta l}{2\left| qE\right| \left( t+T/2\right) 
}\int_{0}^{\infty }ds\rho _{0}s^{-1}\left( t_{\mu \iota }^{0}+\left. \delta
t_{\mu \iota }\right| _{\tau _{W}=s}\right) \,,  \label{emt100}
\end{eqnarray}%
where we have used only the first term of expansion (\ref{emt88}). At low
temperatures, the main contribution to the integral over $s$ is formed at
the region $s\sim \beta l/\left( 2M\right) $, and thus the leading
contributions in (\ref{emt100}) can be found immediately: 
\begin{eqnarray}
J_{\mu }^{\chi } &=&\frac{j_{\mu }^{0}M}{\left| qE\right| \left(
t+T/2\right) }\int_{0}^{\infty }ds\rho _{0}\,,  \notag \\
T_{\mu \nu }^{\chi } &=&\frac{M}{\left| qE\right| \left( t+T/2\right) }%
\int_{0}^{\infty }ds\rho _{0}\left( t_{\mu \iota }^{0}+\delta ^{0}t_{\mu
\iota }\right) \,,  \label{emt101}
\end{eqnarray}%
where 
\begin{equation*}
\delta ^{0}t_{11}=\delta ^{0}t_{22}=0\,,\;\;\delta ^{0}t_{00}=\delta
^{0}t_{33}=\left( qE\right) ^{2}\left( t+T/2\right) ^{2}\,.
\end{equation*}%
Consequently, using expressions (\ref{emt101}) we obtain from representation
(\ref{emt85}) that 
\begin{eqnarray}
\func{Re}\langle j_{\mu }\left( t\right) \rangle _{\theta }^{c} &=&-\delta
_{\mu }^{3}\left| q\right| \mathrm{sgn}\left( E\right) \sum_{\zeta =\pm
}n^{(\zeta )}\,,  \notag \\
\func{Re}\langle T_{11}\left( t\right) \rangle _{\theta }^{c} &=&\func{Re}%
\langle T_{22}\left( t\right) \rangle _{\theta }^{c}=\frac{M}{\left|
qE\right| \left( t+T/2\right) }\left. \func{Re}\langle T_{11}\left( t\right)
\rangle _{\theta }^{c}\right| _{E=0}\,,  \notag \\
\func{Re}\langle T_{00}\left( t\right) \rangle _{\theta }^{c} &=&\func{Re}%
\langle T_{33}\left( t\right) \rangle _{\theta }^{c}=\left| qE\right| \left(
t+T/2\right) \sum_{\zeta =\pm }n^{(\zeta )}\,,  \label{103}
\end{eqnarray}%
where\ $n^{(\zeta )}$ is the initial particle density, given by (\ref{emt99}%
), and $\left. \func{Re}\langle T_{11}\left( t\right) \rangle _{\theta
}^{c}\right| _{E=0}$ is the initial value of $\func{Re}\langle T_{11}\left(
t\right) \rangle _{\theta }^{c}$, defined by (\ref{emt80}) at $t<t_{1}$.

\subsubsection{High temperatures}

Let us find the leading field-dependent contributions for $\func{Re}\langle
j_{\mu }\left( t\right) \rangle _{\theta }^{c}$\ and $\func{Re}\langle
T_{\mu \nu }\left( t\right) \rangle _{\theta }^{c}$ at high temperatures. In
the presence of external magnetic and electric fields, the conditions of the
applicability of high-temperature asymptotic include the following
conditions for the field strengths:%
\begin{equation}
M\beta \ll 1\,,\;\;\sqrt{\left| qB\right| }\beta \ll 1\,,\;\;\sqrt{\left|
qE\right| }\beta \ll 1\,.  \label{emt104}
\end{equation}%
In the case of fields that are not very strong, $\left| qB\right| /M^{2}\leq
1$, $\left| qE\right| /M^{2}\leq 1$, these subsidiary conditions are not
relevant; however, they become necessary in the case of strong fields, $%
\left| qB\right| /M^{2}>1$, $\left| qE\right| /M^{2}>1$. There are no other
necessary restrictions for the field strengths. Under these restrictions, it
is convenient to use the representation (\ref{emt61.10}). At high
temperatures, the leading contributions into the expressions for $\func{Re}%
J_{\mu }^{(l)}$ and $\func{Re}T_{\mu \nu }^{(l)}$ are given by the terms $%
\tilde{J}_{\mu }^{a}$ and $\tilde{T}_{\mu \nu }^{a}$, respectively. These
terms emerge from the integrals (\ref{emt61.7}). In turn, at high
temperatures, the leading contributions to these integrals emerge from small
values of $s^{\prime }$, namely, $s^{\prime }\sim \left( \beta l\right) ^{2}$%
. Therefore, in calculating the leading contributions to all the
field-dependent functions one can assume that $\left| qB\right| s^{\prime
}\ll 1$ and $\left| qE\right| s^{\prime }\ll 1$. This allows one to use, in
the integrals (\ref{emt61.7}), expansions in the powers of $\left| qB\right|
s^{\prime }$ and $\left| qE\right| s^{\prime }$, and then to extend the
range of integration over $s^{\prime }$ to $+\infty $, while neglecting the
exponentially small contribution. As a result, one arrives at the conclusion
that $\tilde{J}_{\mu }^{a}$ and $\tilde{T}_{\mu \nu }^{a}$, are approximated
by the expressions $J_{\mu }^{\chi }$ and $T_{\mu \nu }^{\chi }$ in (\ref%
{emt86}), respectively. Note that at high temperatures the leading
contribution into the sums (\ref{emt61.1}) and (\ref{emt61.2}) is formed by
a finite number of summands, when $l$ belongs to the range $1\leq l\leq $ $%
l_{\max }$. A choice for the number $l_{\max }$ is determined by a given
accuracy of calculation, and by the fact that $l_{\max }\ll $ $\left( \sqrt{%
\left| qE\right| }\beta \right) ^{-1}$\ and $l_{\max }\ll $ $\left( \sqrt{%
\left| qB\right| }\beta \right) ^{-1}$. The terms (\ref{emt86}) are
well-decreasing functions of the number $l$, and therefore in the
high-temperature limit one can assume that $l_{\max }\rightarrow \infty $.
Thus, the sum is taken over positive integers, as in the case of thermal
equilibrium, which happens in the absence of electric field. As a result, we
obtain the following asymptotic expressions for $\func{Re}\langle j_{\mu
}\left( t\right) \rangle _{\theta }^{c}$ and $\func{Re}\langle T_{\mu \nu
}\left( t\right) \rangle _{\theta }^{c}$: 
\begin{eqnarray}
\func{Re}\langle j_{\mu }\left( t\right) \rangle _{\theta }^{c}
&=&\sum_{\zeta =\pm }\sum_{l=1}^{\infty }\left( -1\right) ^{l+1}e^{\beta
l\mu ^{(\zeta )}}J_{\mu }^{\chi }\,,  \notag \\
\func{Re}\langle T_{\mu \nu }\left( t\right) \rangle _{\theta }^{c}
&=&\sum_{\zeta =\pm }\sum_{l=1}^{\infty }\left( -1\right) ^{l+1}e^{\beta
l\mu ^{(\zeta )}}T_{\mu \nu }^{\chi }\,,  \label{emt105}
\end{eqnarray}%
where $J_{\mu }^{\chi }$ and $T_{\mu \nu }^{\chi }$ are given by expressions
(\ref{emt86}) and (\ref{emt87}). As usual, in these asymptotics, we suppose $%
\left| \mu ^{(\zeta )}\right| \beta \ll 1$.

As a whole, we are interested in the leading electric-field-dependent
contributions into expressions (\ref{emt105}), as well as in the limiting
process $E\rightarrow 0$, and in the possibility to compare the leading
electric-field-dependent term with the leading magnetic-field-dependent
term. Therefore, we assume for simplicity that the magnetic field is not
stronger than the electric field, $\left\vert B/E\right\vert \leq 1$; then,
within the given accuracy of calculation, we can neglect the contributions
that contain the magnetic field at powers higher than the second, as well as
the contributions depending on the product $B^{2}E^{2}$.

At high temperatures, the dimensionless parameter that depends on time ($t$)
and characterizes the influence of electric field has the form $qE\left(
t+T/2\right) \beta $. As in the low-temperature case, this parameter is
determined by the relative increment of the kinetic momentum of a particle
in electric field reached by the time instant $t$; however, the scale on
which one estimates this increment is now the temperature $\theta $ rather
than the mass $M$.

When the $T$-constant field switches on, the particles found at the initial
state begin to accelerate. Nevertheless, such a non-equilibrium state is not
very different from the initial state of thermal equilibrium when $\left|
qE\right| \left( t+T/2\right) \beta \ll 1$, and it makes sense to compare
these states. It is exactly this condition that guarantees the existence of
the limiting process $E\rightarrow 0$. It is clear that at any finite value $%
\left| qE\right| \left( t+T/2\right) $ the given condition takes place when $%
\beta \rightarrow 0$, i.e., it is exactly in this case that we deal with an
asymptotic expression at high temperatures. Therefore, this case is the most
interesting one. Let us note, in addition, that this condition implies
comparatively weaker restrictions for the value of kinetic momenta, $\left|
qE\right| \left( t+T/2\right) \ll 1/\beta $ than the values at low
temperatures when one requires that $\left| qE\right| \left( t+T/2\right)
\ll M$. For sufficiently high temperatures, ultrarelativistic kinetic
momenta can be regarded as sufficiently small. The opposite case of
asymptotically large values of time at large but finite $\beta $, when $%
\left| qE\right| \left( t+T/2\right) \beta \gg 1$, is a complicated case,
when, according to conditions (\ref{emt104}), the temperature is
sufficiently high in comparison with $M$, $B$, and $E$; however, at the same
time, the temperature is low in comparison with the kinetic momentum $%
qE\left( t+T/2\right) $. This case is also interesting, since the presence
of electric field manifests itself to the maximum. In what follows, we
examine these two limiting cases.

\paragraph{Small increment of kinetic momentum}

When the increment of kinetic momentum is small, $\left| qE\right| \left(
t+T/2\right) \beta \ll 1$, the function $Y^{W}$ in (\ref{emt87}) can be
expanded as 
\begin{equation*}
Y^{W}=1-\delta Y\,,\;\;\delta Y=\left( qE\right) ^{2}\left( t+T/2\right)
^{2}\left( \tau _{W}-s^{-1}\tau _{W}^{2}\right) \,.
\end{equation*}%
Let us examine the following approximation:%
\begin{eqnarray}
J_{\mu }^{\chi } &=&j_{\mu }^{0}\int_{0}^{\infty }\frac{ds}{s}%
\int_{0}^{\infty }du\;\left. h^{0}\right| _{B=0}\tau _{W}\,,  \notag \\
T_{\mu \nu }^{\chi } &=&T_{\mu \nu }^{0}+\Delta T_{\mu \nu }\,,
\label{emt106}
\end{eqnarray}%
where $T_{\mu \nu }^{0}$ and $h^{0}$ are expressions being independent of
electric field; $T_{\mu \nu }^{0}$ is given by (\ref{emt77}); $h^{0}$ and $%
\tau _{W}$ are given by (\ref{emt87}); $j_{\mu }^{0}$ is given by (\ref%
{emt92a}); besides%
\begin{eqnarray}
\Delta T_{\mu \nu } &=&\frac{1}{4\pi ^{2}}\int_{0}^{\infty }ds\exp \left[
-M^{2}s-\frac{\left( \beta l\right) ^{2}}{4s}\right] \left( \Delta t_{\mu
\nu }^{1}+\Delta t_{\mu \nu }^{2}\right) \,,  \notag \\
\Delta t_{\mu \nu }^{1} &=&\frac{1}{2\sqrt{\pi }}\int_{0}^{\infty }\frac{du}{%
u^{1/2}}e^{-u/4}\frac{1}{s^{2}}\left. \delta t_{\mu \nu }\right| _{B=0}\,, 
\notag \\
\Delta t_{\mu \nu }^{2} &=&\frac{1}{2\sqrt{\pi }}\int_{0}^{\infty }\frac{du}{%
u^{1/2}}e^{-u/4}\left[ \left( qE\right) ^{2}D-\delta Y\frac{1}{s^{2}}\right]
\left. t_{\mu \nu }^{0}\right| _{B=0}\,,  \label{emt107}
\end{eqnarray}%
where $t_{\mu \nu }^{0}$ is given by (\ref{emt77}), $\delta t_{\mu \nu }$
and\ $D$ are given by (\ref{emt87}), respectively.

To calculate the integrals in (\ref{emt106}) and (\ref{emt107}), we need the
following functions of a real argument $p$: 
\begin{equation*}
I_{n}\left( p\right) =\frac{1}{2\sqrt{\pi }}\int_{0}^{\infty }\frac{du}{%
u^{1/2}}e^{-u/4}\frac{1}{\left( u+p\right) ^{n}},\;\;n\geq 1.
\end{equation*}
By using formula (3.363.2) in \cite{GR94}, the function $I_{1}\left(
p\right) $ can be expressed in terms of the incomplete gamma-function: 
\begin{equation}
I_{1}\left( p\right) =\frac{1}{2p^{1/2}}e^{p/4}\Gamma \left( \frac{1}{2},%
\frac{p}{4}\right) \,.  \label{emt108}
\end{equation}
Differentiating this relation with respect to the variable $p$, one can find
the corresponding representation for the remaining functions $I_{n}\left(
p\right) $ at $n\geq 2$. In particular, we have 
\begin{equation}
I_{2}\left( p\right) =\frac{1}{2}\left( \frac{1}{p}-\frac{1}{2}\right)
I_{1}\left( p\right) +\frac{1}{4p}\,.  \label{emt109}
\end{equation}
In the case under consideration, $p=\left( \beta l\right) ^{2}/s$. Let us
express $J_{\mu }^{\chi }$ and $\Delta t_{kk}$ for $k=1,2,3$ in terms of the
functions $I_{n}\left( p\right) $: 
\begin{eqnarray}
&&J_{\mu }^{\chi }=\frac{j_{\mu }^{0}\left( \beta l\right) ^{2}}{4\pi ^{2}}%
\int_{0}^{\infty }\frac{ds}{s^{3}}\exp \left[ -M^{2}s-\frac{\left( \beta
l\right) ^{2}}{4s}\right] I_{1}\left( \frac{\left( \beta l\right) ^{2}}{s}%
\right) \,,  \notag \\
&&\Delta t_{11}^{1}=\Delta t_{22}^{1}=-\frac{\left( qE\right) ^{2}\left(
\beta l\right) ^{6}}{6s^{4}}I_{3}\left( \frac{\left( \beta l\right) ^{2}}{s}%
\right) \,,  \notag \\
&&\Delta t_{33}^{1}=\frac{\left( qE\right) ^{2}}{4}\left[ \frac{1}{3s}-\frac{%
\left( \beta l\right) ^{4}}{s^{3}}I_{2}\left( \frac{\left( \beta l\right)
^{2}}{s}\right) \right] +\left( qE\right) ^{2}\left( t+T/2\right) ^{2}\frac{%
\left( \beta l\right) ^{4}}{s^{4}}I_{2}\left( \frac{\left( \beta l\right)
^{2}}{s}\right) \,,  \notag \\
&&\Delta t_{kk}^{2}=\frac{\left( qE\right) ^{2}}{6}\left[ \frac{1}{2s}+\frac{%
\left( \beta l\right) ^{2}}{4s^{2}}-\frac{\left( \beta l\right) ^{6}}{s^{4}}%
I_{3}\left( \frac{\left( \beta l\right) ^{2}}{s}\right) \right]  \notag \\
&&-\left( qE\right) ^{2}\left( t+T/2\right) ^{2}\frac{\left( \beta l\right)
^{2}}{2s^{3}}\left[ I_{1}\left( \frac{\left( \beta l\right) ^{2}}{s}\right) -%
\frac{\left( \beta l\right) ^{2}}{s}I_{2}\left( \frac{\left( \beta l\right)
^{2}}{s}\right) \right] \,,  \label{emt110}
\end{eqnarray}
where the relation $\Phi _{0}\left( 0\right) =1$ for the integral (\ref%
{emt90}) has been used, and we have neglected the terms smaller that the
remaining ones when $s\sim \left( \beta l\right) ^{2}$.

Using the representation 9.1(2) in \cite{BatE74}, 
\begin{equation*}
\Gamma \left( \frac{1}{2},\frac{p}{4}\right) =\frac{p^{1/2}}{2}%
\int_{1}^{\infty }\frac{du}{u^{1/2}}e^{-up/4}\,,
\end{equation*}%
and formula (\ref{emt94}), one can calculate the following integrals:%
\begin{equation*}
\chi _{n}=\int_{0}^{\infty }s^{-N}\exp \left[ -M^{2}s-\frac{\left( \beta
l\right) ^{2}}{4s}\right] I_{n}\left( \frac{\left( \beta l\right) ^{2}}{s}%
\right) ds\,,\;\;N\geq 2\,.
\end{equation*}%
\ Let us first examine the case $n=1$. Calculating the integral over $s$, we
obtain%
\begin{equation}
\chi _{1}=\frac{2^{N-1}M^{2N-3}}{\beta l}\int_{M\beta l}^{\infty
}z^{1-N}K_{N-1}\left( z\right) dz\,.  \label{emt111}
\end{equation}%
This representation allows one to extract the leading contribution for $%
M\beta \rightarrow 0$. The function $K_{N-1}\left( z\right) $ decreases
sufficiently fast at large $z$, and it is easy to see that the leading
contribution is formed at the region of small $z$, $z\leq \delta _{z}$,
where $\delta _{z}$ is a small value, such that $M\beta l\ll \delta _{z}\ll
1 $. It is sufficient to take into account such $l$ that $M\beta l\ll 1$,
since the contributions $\chi _{1}$ into the sum (\ref{emt105}) due to\ the
higher values of $l$ can be neglected.\footnote{%
As an alternative, one can sum over $l$ the contribution $\chi _{1}$ by
using the representation (\ref{emt111}) and the summation formula with the
function $K_{n}\left( xl\right) $ \cite{LanW87}. This approach is more
effective when one needs to obtain high-temperature expansions beyond the
leading terms.} Neglecting the small contribution given by the region of $%
z>\delta _{z}$, we now represent $\chi _{1}$ in the form 
\begin{equation*}
\chi _{1}=\frac{2^{N-1}M^{2N-3}}{\beta l}\int_{M\beta l}^{\delta
_{z}}z^{1-N}K_{N-1}\left( z\right) dz\,.
\end{equation*}%
This integral can be calculated due to expansion 8.446 \cite{GR94} for $%
K_{N-1}\left( z\right) $ at $z\ll 1$, 
\begin{equation}
K_{N-1}\left( z\right) =\frac{1}{2}\left( \frac{2}{z}\right) ^{N-1}\left(
N-2\right) !\,.  \label{emt111a}
\end{equation}%
Then, we finally obtain%
\begin{equation}
\chi _{1}=\frac{2^{2N-1}\left( N-2\right) !}{\left( 2N-3\right) \left( \beta
l\right) ^{2N-2}}\,.  \label{emt113a}
\end{equation}%
in the leading approximation. Expressing $I_{2}\left( p\right) $ through $%
I_{1}\left( p\right) $ by using relation (\ref{emt109}) and following the
same procedure as in the case of calculating $\chi _{1}$, we obtain the
leading term for $\chi _{2}$ at $N\geq 3$, namely,%
\begin{equation}
\chi _{2}=\frac{2^{2N-3}}{\left( \beta l\right) ^{2N-2}}\left[ \frac{\left(
N-3\right) !}{8}+\frac{1}{2\left( 2N-5\right) }-\frac{1}{2N-3}\right] \,.
\label{emt113}
\end{equation}%
In a similar way, we can estimate the leading contribution for $\chi _{3}$.
For our purposes, it is sufficient to estimate the orders of $\chi _{3}$: $%
\chi _{3}\sim \left( \beta l\right) ^{-4}\ln \left( M\beta l\right) $ at $%
N=3 $, and $\chi _{3}\sim \left( \beta l\right) ^{-6}$ at $N=4$. Using these
results for a calculation of the expressions defined by (\ref{emt107}) and (%
\ref{emt110}), and leaving only the leading contributions, we find 
\begin{eqnarray}
&&J_{\mu }^{\chi }=\frac{8}{3\pi ^{2}}\frac{j_{\mu }^{0}}{\left( \beta
l\right) ^{2}},  \notag \\
&&\Delta T_{11}=\Delta T_{22}=\frac{1}{4\pi ^{2}}\left[ \frac{\left(
qE\right) ^{2}}{6}K_{0}\left( M\beta l\right) -\frac{58}{15}\frac{\left(
qE\right) ^{2}\left( t+T/2\right) ^{2}}{\left( \beta l\right) ^{2}}\right] ,
\notag \\
&&\Delta T_{33}=\frac{1}{4\pi ^{2}}\left[ \frac{\left( qE\right) ^{2}}{3}%
K_{0}\left( M\beta l\right) -\frac{14}{15}\frac{\left( qE\right) ^{2}\left(
t+T/2\right) ^{2}}{\left( \beta l\right) ^{2}}\right] .  \label{emt114}
\end{eqnarray}%
Proceeding in the same way, we find the leading contribution to $\Delta
T_{00}$ as follows: 
\begin{equation}
\Delta T_{00}=\sum_{k=1,2,3}\Delta T_{kk}.  \label{emt114a}
\end{equation}

Taking the sum of the terms (\ref{emt114}), in accordance with the
representation (\ref{emt105}), we use the alternating zeta-function%
\begin{equation*}
\eta \left( s\right) =\sum_{l=1}^{\infty }\left( -1\right) ^{l+1}\frac{1}{%
l^{s}}
\end{equation*}%
and the summation formula (see, e.g., the Appendix of \cite{LanW87}) 
\begin{equation}
\sum_{l=1}^{\infty }\left( -1\right) ^{l+1}K_{0}\left( M\beta l\right) =-%
\frac{1}{2}\ln \left( M\beta \right) +O\left( 1\right) ,  \label{emt115}
\end{equation}%
in which we have fixed only the leading term, in accordance with our
approximation. Since $\eta \left( 2\right) =\pi ^{2}/12$, we obtain the
final result as follows:%
\begin{eqnarray}
&&\func{Re}\langle j_{\mu }\left( t\right) \rangle _{\theta }^{c}=-\frac{4}{9%
}\delta _{\mu }^{3}q^{2}E\left( t+T/2\right) /\beta ^{2}\,,  \notag \\
&&\func{Re}\langle T_{\mu \nu }\left( t\right) \rangle _{\theta }^{c}=\left. 
\func{Re}\langle T_{\mu \nu }\left( t\right) \rangle _{\theta }^{c}\right|
_{E=0}+\tau _{\mu \nu }^{\theta }\left( t\right) \,,  \notag \\
&&\tau _{11}^{\theta }\left( t\right) =\tau _{22}^{\theta }\left( t\right) =%
\frac{\left( qE\right) ^{2}}{12\pi ^{2}}\left[ -\frac{1}{2}\ln \left( M\beta
\right) +O\left( 1\right) -\frac{29\pi ^{2}}{15}\frac{\left( t+T/2\right)
^{2}}{\beta ^{2}}\right] \,,  \notag \\
&&\tau _{33}^{\theta }\left( t\right) =\frac{\left( qE\right) ^{2}}{12\pi
^{2}}\left[ -\ln \left( M\beta \right) +O\left( 1\right) -\frac{7\pi ^{2}}{15%
}\frac{\left( t+T/2\right) ^{2}}{\beta ^{2}}\right] \,,  \notag \\
&&\tau _{00}^{\theta }\left( t\right) =\frac{\left( qE\right) ^{2}}{12\pi
^{2}}\left[ -2\ln \left( M\beta \right) +O\left( 1\right) -\frac{13\pi ^{2}}{%
3}\frac{\left( t+T/2\right) ^{2}}{\beta ^{2}}\right] \,.  \label{emt116}
\end{eqnarray}

Even though we examine the case of a small kinetic-momentum increment, the
time-dependent terms in $\tau _{\mu \nu }^{\theta }\left( t\right) $ (\ref%
{emt116}) may be larger that the logarithmic contributions, because they are
limited only by the condition that $\left( t+T/2\right) ^{2}/\beta ^{2}\ll
1/\left( qE\beta ^{2}\right) ^{2}$. The derivative of these terms with
respect to time $t$ determines the rate of change for the components $\func{%
Re}\langle T_{\mu \nu }\left( t\right) \rangle _{\theta }^{c}$.

Let us find the relation of the resulting expression for the current density 
$\func{Re}\langle j_{\mu }\left( t\right) \rangle _{\theta }^{c}$ with
respect to the initial particle density $n^{(\zeta )}$. The leading terms of
the effective Lagrangian $\left. \Delta \mathcal{L}_{\theta }\right| _{B=0}$
(\ref{emt84.1}) can be expressed as 
\begin{equation*}
\left. \Delta \mathcal{L}_{\theta }\right| _{B=0}=\frac{2}{\pi ^{2}\beta ^{4}%
}\sum_{\zeta =\pm }\left[ \eta \left( 4\right) +\eta \left( 3\right) \beta
\mu ^{\left( \zeta \right) }\right] ,
\end{equation*}%
where formula (\ref{emt111a}) has been used. Then, using (\ref{emt83}) we
obtain 
\begin{equation*}
n^{(+)}=n^{(-)}=\frac{2\eta \left( 3\right) }{\pi ^{2}\beta ^{3}}\,,
\end{equation*}%
where $\eta \left( 3\right) =0,9015426774$, and we have%
\begin{equation}
\func{Re}\langle j_{\mu }\left( t\right) \rangle _{\theta }^{c}=-\frac{\pi
^{2}}{9\eta \left( 3\right) }\delta _{\mu }^{3}q^{2}E\left( t+T/2\right)
\beta \sum_{\zeta =\pm }n^{(\zeta )}\,.  \label{emt117}
\end{equation}

Let us compare the time-independent terms in (\ref{emt116}) with the
magnetic field-dependent terms in $\left. \langle T_{\mu \nu }\left(
t\right) \rangle _{\theta }^{c}\right| _{E=0}$ (\ref{emt84.3}). Using
formula (\ref{emt115}), we find the leading field-dependent term of the
effective Lagrangian $\Delta \mathcal{L}_{\theta }$ (\ref{emt84.2}) at high
temperatures:%
\begin{equation*}
\mathcal{L}_{1}=-\frac{\left( qB\right) ^{2}}{12\pi ^{2}}\left[ \ln \left(
M\beta \right) +O\left( 1\right) \right] \,,
\end{equation*}%
which coincides with previously known results (see, e.g., \cite{CanD96}).
Then, using formula (\ref{emt84.3}), we find that the magnetic
field-dependent terms of EMT are given by%
\begin{equation*}
\tau _{00}^{B}=\tau _{11}^{B}=\tau _{22}^{B}=-\tau _{33}^{B}=-\mathcal{L}%
_{1}\,.
\end{equation*}%
We can see that the equation of state for the magnetic field in a medium at
high temperature is identical with the equation of state for this field in
vacuum. The increment of the energy density of magnetic field caused by
vacuum polarization, $\left. \langle T_{00}(t)\rangle _{ren}^{c}\right|
_{E=0}$, and the increment $\tau _{00}^{B}$ are both negative, which holds
true in the case of a field of any intensity. On the contrary, the increment
of the energy density of electric field caused by vacuum polarization, $%
\langle T_{00}(t)\rangle _{ren}^{c}$, is positive in a weak field and is
negative in a strong one. The time-independent terms of $\tau _{\mu \nu
}^{\theta }\left( t\right) $ (\ref{emt116}) are positive. Therefore, even at
the initial period of the existence of electric field, when the
time-dependent terms of $\tau _{\mu \nu }^{\theta }\left( t\right) $ can be
neglected, its equation of state is essentially different from the vacuum
equation and has different forms in the cases of weak and strong fields.

\paragraph{Large increment of kinetic momentum}

When the increment of kinetic momentum is large, $\left| qE\right| \left(
t+T/2\right) \beta \gg 1$, the leading contribution to the integral over $u$
in (\ref{emt105}) is formed at the region $u\gg 1$. In order to find the
asymptotic behavior of expressions (\ref{emt105}), it is sufficient to know
a rough approximation of the function $Y^{W}$, namely, 
\begin{equation*}
Y^{W}=Y_{0}=\exp \left[ -\left( qE\right) ^{2}\left( t+T/2\right) ^{2}\left(
\beta l\right) ^{2}/u\right] .
\end{equation*}%
Then $J_{\mu }^{\chi }$ and $T_{\mu \nu }^{\chi }$ can be approximated as
follows:%
\begin{eqnarray}
J_{\mu }^{\chi } &=&\int_{0}^{\infty }ds\int_{0}^{\infty
}du\;h^{0}Y_{0}j_{\mu }^{W}\,,  \notag \\
T_{\mu \nu }^{\chi } &=&\int_{0}^{\infty }ds\int_{0}^{\infty }du\;\left(
h^{0}+\delta h\right) Y_{0}\left( t_{\mu \iota }^{0}+\delta t_{\mu \iota
}\right) \,,  \label{emt119}
\end{eqnarray}%
with $t_{\mu \iota }^{0}$ given by (\ref{emt77}), and $h^{0}$, $\delta h$, $%
\delta t_{\mu \iota }$ given by (\ref{emt87}), where the approximation $\tau
_{W}=\left( \beta l\right) ^{2}/u$ has been used. It is convenient to carry
out the integration over $u$ in (\ref{emt119}) with the help of the
generating function $Y_{0}\left( \sigma \right) =\exp \left[ -\left(
qE\right) ^{2}\left( t+T/2\right) ^{2}\left( \beta l\right) ^{2}\sigma /u%
\right] $, $\sigma >0$, which is identical with $Y_{0}$ at $\sigma =1$.
Using formula (\ref{emt55.2}), we obtain 
\begin{equation*}
\frac{1}{2\sqrt{\pi }}\int_{0}^{\infty }\frac{du}{u^{1/2}}%
e^{-u/4}Y_{0}\left( \sigma \right) =\exp \left[ -\left| qE\right| \left(
t+T/2\right) \beta l\sigma ^{1/2}\right] \,.
\end{equation*}%
This formula, along with the first and second derivative with respect to the
parameter $\sigma $, allows one to carry out the integration over $u$ in (%
\ref{emt119}). As a result, we find, with accuracy up to the pre-exponential
factor, that $J_{\mu }^{\chi }$ and $T_{\mu \nu }^{\chi }$ are of the order $%
\exp \left[ -\left| qE\right| \left( t+T/2\right) \beta l\right] $.
Consequently, the leading contribution is given by the terms $J_{\mu }^{\chi
}$ and $T_{\mu \nu }^{\chi }$ for $l=1$. Therefore, considering
asymptotically large kinetic momenta, one can see that nonzero $\func{Re}%
\langle j_{\mu }\left( t\right) \rangle _{\theta }^{c}$ and $\func{Re}%
\langle T_{\mu \nu }\left( t\right) \rangle _{\theta }^{c}$ decrease
exponentially: 
\begin{eqnarray}
\func{Re}\langle j_{3}\left( t\right) \rangle _{\theta }^{c} &\sim &\exp %
\left[ -\left| qE\right| \left( t+T/2\right) \beta \right] ,  \notag \\
\func{Re}\langle T_{\mu \mu }\left( t\right) \rangle _{\theta }^{c} &\sim
&\exp \left[ -\left| qE\right| \left( t+T/2\right) \beta \right] .
\label{emt120}
\end{eqnarray}%
A calculation of the pre-exponential factor requires a more accurate
approximation and some very tedious calculations.

\subsubsection{Strong electric field}

In the limit of a strong electric field, $\left\vert qE\right\vert \gg
M^{2},\beta ^{-2},\left\vert qB\right\vert $, the leading contributions
depending on the electric field for the integrals $\func{Re}J_{\mu }^{(l)}$
and $\func{Re}T_{\mu \nu }^{(l)}$ in (\ref{emt61.1}) and (\ref{emt61.2})
emerge from large values of $\left\vert qE\right\vert s$ and $\left\vert
qE\right\vert \tau $, $\left\vert qE\right\vert \tau \gg \left\vert
qE\right\vert s\gg 1$. For simplicity, we assume that $B=0$. In addition, we
know from Subsec. 3.2 that the leading time-dependent vacuum contributions, $%
\func{Re}\langle j_{\mu }(t)\rangle ^{p}$ and $\func{Re}\langle T_{\mu \nu
}(t)\rangle ^{p}$ do not depend on the details of switching on and off when
the time interval $t-t_{1}=t+T/2$ is sufficiently large, so that condition (%
\ref{unistab}) holds true. Then, it is instructive to examine thermal
contributions for the same large interval. In this case, we can use the
following approximation: 
\begin{eqnarray}
&&\func{Re}\langle j_{\mu }\left( t\right) \rangle _{\theta
}^{c}=\sum_{\zeta =\pm }\sum_{l=1}^{\infty }\left( -1\right) ^{l+1}e^{\beta
l\mu ^{(\zeta )}}\func{Re}J_{\mu }^{(l)}\,,  \notag \\
&&J_{\mu }^{(l)}=j_{\mu }^{0}\int_{0}^{\infty }ds\int_{0}^{\infty
}du\;h^{s}Y^{s}\,;  \label{emt121}
\end{eqnarray}%
and 
\begin{eqnarray}
&&\func{Re}\langle T_{\mu \nu }\left( t\right) \rangle _{\theta
}^{c}=\sum_{\zeta =\pm }\sum_{l=1}^{\infty }\left( -1\right) ^{l+1}e^{\beta
l\mu ^{(\zeta )}}\func{Re}T_{\mu \nu }^{(l)}\,,  \notag \\
&&T_{\mu \nu }^{(l)}=\int_{0}^{\infty }ds\int_{0}^{\infty
}du\;h^{s}Y^{s}t_{\mu \nu }^{s}\,,  \notag \\
&&t_{00}^{s}=\sum_{k=1,2,3}t_{kk}^{s}+M^{2}e^{\left\vert qE\right\vert
s}/2\,,\;\;t_{11}^{s}=t_{22}^{s}=\frac{u}{2\left( \beta l\right) ^{2}}\,, 
\notag \\
&&t_{33}^{s}=2e^{-\left\vert qE\right\vert s}\left[ \frac{u}{2\left( \beta
l\right) ^{2}}+\left( qE\right) ^{2}\left( t+T/2\right) ^{2}\right] \,;
\label{emt124}
\end{eqnarray}%
where 
\begin{eqnarray}
&&h^{s}=\frac{1}{2}D_{l}\left\vert qE\right\vert u\exp \left[ \frac{i\pi }{4}%
-\left( \left\vert qE\right\vert +iM^{2}\right) s-\frac{u}{4}-\frac{\left(
M\beta l\right) ^{2}}{u}\right] \,,  \notag \\
&&D_{l}=\left[ 2\pi ^{5/2}\sqrt{\left\vert qE\right\vert }\left( \beta
l\right) ^{3}\right] ^{-1}\,,  \notag \\
&&Y^{s}=\exp \left[ -i\left\vert qE\right\vert \left( t+T/2\right)
^{2}-\left( t+T/2\right) ^{2}\left( \beta l\right) ^{-2}u\right] \,,
\label{emt125}
\end{eqnarray}%
and $j_{\mu }^{0}$ is given by expression (\ref{emt92a}). In this
representation, the kernels are products: a function of the variable $s$
times a function of the variable $u$; therefore, these double integrals can
be represented as a product: an integral over $s$ times an integral over $u$%
. Having calculated the integral over $s$, and using formula (\ref{emt94})
for a calculation of the integral over $u$, we find that 
\begin{eqnarray}
\func{Re}J_{\mu }^{(l)} &=&2j_{\mu }^{0}\cos \left[ \frac{\pi }{4}%
-\left\vert qE\right\vert \left( t+T/2\right) ^{2}\right] D_{l}f_{1}\,, 
\notag \\
\func{Re}T_{00}^{(l)} &=&\func{Re}T_{33}^{(l)}+\sin \left[ \frac{\pi }{4}%
-\left\vert qE\right\vert \left( t+T/2\right) ^{2}\right] D_{l}\left\vert
qE\right\vert f_{1}\,,  \notag \\
\func{Re}T_{11}^{(l)} &=&\func{Re}T_{22}^{(l)}=\cos \left[ \frac{\pi }{4}%
-\left\vert qE\right\vert \left( t+T/2\right) ^{2}\right] \frac{D_{l}f_{2}}{%
\left( \beta l\right) ^{2}}\,,  \notag \\
\func{Re}T_{33}^{(l)} &=&\cos \left[ \frac{\pi }{4}-\left\vert qE\right\vert
\left( t+T/2\right) ^{2}\right] D_{l}\left[ \frac{f_{2}}{\left( \beta
l\right) ^{2}}+2\left( qE\right) ^{2}\left( t+T/2\right) ^{2}f_{1}\right] \,,
\label{emt126}
\end{eqnarray}%
where 
\begin{equation*}
f_{n}=\left( \frac{M\beta l}{\sqrt{1/4+\left( t+T/2\right) ^{2}\left( \beta
l\right) ^{-2}}}\right) ^{n+1}K_{n+1}\left( \sqrt{\left( M\beta l\right)
^{2}+4M^{2}\left( t+T/2\right) ^{2}}\right) \,.
\end{equation*}

We note that the dimensionless parameter $M\left( t+T/2\right) $ is equal to
unity for the time period taken by light to cross the Compton radius of a
particle of mass $M$. Thus, for a macroscopic period of time, $M\left(
t+T/2\right) \gg 1$, one can use the asymptotic expansion (\ref{emt95}),
namely, 
\begin{eqnarray*}
&&K_{n+1}\left( \sqrt{\left( M\beta l\right) ^{2}+4M^{2}\left( t+T/2\right)
^{2}}\right) \\
&&\,=\sqrt{\frac{\pi }{2\sqrt{\left( M\beta l\right) ^{2}+4M^{2}\left(
t+T/2\right) ^{2}}}}\exp \left[ -\sqrt{\left( M\beta l\right)
^{2}+4M^{2}\left( t+T/2\right) ^{2}}\right] .
\end{eqnarray*}%
At low initial temperatures, $M\beta \gg 1$, this asymptotic expansion is
valid for any period of time. We can see that in the limit of a strong
electric field the behavior of the current density $\func{Re}\langle j_{\mu
}\left( t\right) \rangle _{\theta }^{c}$ and the EMT components $\func{Re}%
\langle T_{\mu \nu }\left( t\right) \rangle _{\theta }^{c}$ is described by
relaxation oscillations. The amplitudes of these oscillations are small in
comparison with the increasing vacuum contributions $\func{Re}\langle j_{\mu
}(t)\rangle ^{p}$ and $\func{Re}\langle T_{\mu \nu }(t)\rangle ^{p}$ in (\ref%
{emt44}), respectively, starting from the instant $t$ when conditions (\ref%
{unistab}) are fulfilled.

\section{Discussion and summary}

We have obtained non-perturbative one-loop representations for the mean
current density $\,\langle j_{\mu }\left( t\right) \rangle $ and
renormalized EMT $\langle T_{\mu \nu }(t)\rangle _{ren}$ of a Dirac field in
a constant electric-like background. Two cases of initial states are
considered, the vacuum state and a thermal equilibrium state. In the general
case, each of the obtained expressions consists of three characteristic
terms: 
\begin{eqnarray*}
\,\langle j_{\mu }\left( t\right) \rangle &=&\mathrm{\func{Re}}\langle
j_{\mu }\left( t\right) \rangle _{\theta }^{c}+J_{\mu }^{p}\left( t\right) ,
\\
\,\langle T_{\mu \nu }(t)\rangle _{ren} &=&\func{Re}\langle T_{\mu \nu
}(t)\rangle _{ren}^{c}+\mathrm{\func{Re}}\langle T_{\mu \nu }\left( t\right)
\rangle _{\theta }^{c}+\tau _{\mu \nu }^{p}\left( t\right) \,;
\end{eqnarray*}%
see (\ref{emt71}).

The components $\func{Re}\langle T_{\mu \nu }(t)\rangle _{ren}^{c}$ given by
(\ref{HEL}) describe the contribution due to vacuum polarization. The
corresponding components of current density are zero, $\langle j_{\mu
}\left( t\right) \rangle ^{c}=0$. These components $\func{Re}\,\langle
T_{\mu \nu }\left( t\right) \rangle ^{c}$ can be related with the real part
of Heisenberg--Euler Lagrangian. They are local, i.e., they depend on $t$,
but do not depend on the history of the process.

The components $\mathrm{\func{Re}}\langle j_{\mu }\left( t\right) \rangle
_{\theta }^{c}$ and $\mathrm{\func{Re}}\langle T_{\mu \nu }\left( t\right)
\rangle _{\theta }^{c}$ given by (\ref{emt61.1}), (\ref{emt61.2}), (\ref%
{emt61.10}) describe the contribution due to the work of the external field
on the particles at the initial state. The components $J_{\mu }^{p}\left(
t\right) $ and $\tau _{\mu \nu }^{p}\left( t\right) $ given by (\ref{emt.44}%
), (\ref{emt44}), (\ref{emt68}) describe the contribution due to the
creation of real particles from vacuum. These contributions depend on the
time interval $t-t_{1}$ that passes since the instant the electric field
turns on, and therefore they are global quantities.

All these contributions are investigated in detail in different regimes,
limits of weak and strong fields and low and high temperatures. In all these
limiting cases, we have obtained the leading contributions, which are given
by elementary functions of the basis dimensionless parameters.

The action of the electric field manifests itself differently for different
components of the EMT $\langle T_{\mu \nu }(t)\rangle _{ren}$. For instance,
in a strong electric field, or in a field which is not strong but acts for a
sufficiently long time interval, there is the following correspondence: a
coincidence of the leading terms for energy density,\ $\mathrm{\func{Re}}%
\langle T_{00}\left( t\right) \rangle _{\theta }^{c}+\tau _{00}^{p}\left(
t\right) $, and the pressure{\large \ }along the direction of the electric
field, $\mathrm{\func{Re}}\langle T_{33}\left( t\right) \rangle _{\theta
}^{c}+\tau _{33}^{p}\left( t\right) $. Note that for vacuum polarization the
terms in the mentioned correspondence have the opposite signs: $\func{Re}%
\langle T_{00}(t)\rangle _{ren}^{c}=-\func{Re}\langle T_{33}(t)\rangle
_{ren}^{c}$. This happens because in the former relation we deal with the
equation of state for the relativistic fermions, while in the latter
relation we deal with the equation of state for the electromagnetic field.
The dependence of the electric field and its duration for the transversal
component $\mathrm{\func{Re}}\langle T_{11}\left( t\right) \rangle _{\theta
}^{c}+\tau _{11}^{p}\left( t\right) $ is quite different from the dependence
of the longitudinal component $\mathrm{\func{Re}}\langle T_{33}\left(
t\right) \rangle _{\theta }^{c}+\tau _{33}^{p}\left( t\right) $. This means
that one cannot define a universal functional for the electric field, whose
variations should produce all the components of the EMT. This is valid not
only in case there exist particles at the initial state, but also when the
initial state is vacuum. Therefore, we can see that there cannot be any
generalization of the Heisenberg--Euler Lagrangian for the problem of
calculating the mean values. In our opinion, this fact takes place for any
pair-creating background, and is related, in such a case, to the existence
of nonlocal contributions to the EMT. This explains the failure of numerous
attempts to examine one-loop effects on the basis of different variants \cite%
{GanKP95,Gie99} of such a generalization.

Making a comparison of the constant temperature-dependent components of the
mean energy density and pressure in a weak electric field immediately after
it switches on, we can see that in this case any generalization of the
Heisenberg--Euler Lagrangian to the case of finite temperature is
impossible. In other words, even a seemingly small disturbance of thermal
equilibrium, produced by a weak electric field, exceeds the limits of
applicability of the approaches based on thermal equilibrium.

We have established the restriction $\left| qE\right| T^{2}\ll \frac{\pi ^{2}%
}{2q^{2}}$ (which takes place both for the initial vacuum state and for a
low-temperature initial thermal state) for the strength and duration of
electric field under which QED with a strong constant electric field remains
consistent. Under this restriction, one can neglect the back-reaction of
particles created by the electric field. On the other hand, there exists
another restriction, $1\ll \left| qE\right| T^{2}$, see (\ref{unistab}),
that allows one to disregard the details of switching the electric field on
and off. Gathering both restrictions, we obtain the following range of the
dimensionless parameter $\left| qE\right| T^{2}$ under which QED with a
strong constant electric field is consistent: 
\begin{equation*}
1\ll \left| qE\right| T^{2}\ll \pi ^{2}/2q^{2}\,.
\end{equation*}%
This inequality is consistent due to $\pi ^{2}/2q^{2}\gg 1$, and one can be
certain that QED with a $T$-constant field does exist.

Similar restrictions can be obtained when the initial thermal equilibrium is
taken at sufficiently high temperatures, $\beta \left| qE\right| T\ll 1$. In
this case, we have two inequalities, $\beta \left| qE\right| ^{2}T^{3}\ll 
\frac{3\pi ^{2}}{q^{2}}\,$\ and $1\ll \left| qE\right| T^{2}$, which imply%
\begin{equation*}
1\ll \left| qE\right| T^{2}\ll \frac{3\pi ^{2}}{q^{2}\beta \left| qE\right| T%
}\,.
\end{equation*}%
We can see that the upper restriction for $\left| qE\right| T^{2}$ is weaker
than in the low-temperature case.

We believe that a similar consideration for an electric-like non-Abelian
external field will lead to the same restrictions when the created Fermi
particles (partons) can be treated as weakly coupled. In the case of high
temperatures, one can neglect the back-reaction of the created particles on
the chromoelectric field, in comparison with the contribution from the
bosons, which are influenced by temperature in a different way (the case of
bosons will be examined in another work).

\appendix

\section*{Appendix}

\subsection{Separation of particle creation contributions}

Using relations (\ref{emt18}), we find 
\begin{eqnarray}
\pm \tilde{S}_{\theta }^{\mp }\left( x,x^{\prime }\right) &=&S_{\theta ,\pm
}\left( x,x^{\prime }\right) +S_{\theta ,\pm }^{p}\left( x,x^{\prime
}\right) \,,  \notag \\
S_{\theta ,+}\left( x,x^{\prime }\right) &=&-i\sum_{n,m}N_{m}^{(+)}\left(
in\right) \,^{+}\psi _{n}\left( x\right) G\left( \left. _{+}\right|
^{+}\right) _{nm}^{-1}\ _{+}\bar{\psi}_{m}\left( x^{\prime }\right) \,, 
\notag \\
S_{\theta ,-}\left( x,x^{\prime }\right) &=&i\sum_{n,m}N_{n}^{(-)}\left(
in\right) \,_{-}\psi _{n}\left( x\right) \left[ G\left( \left. _{-}\right|
^{-}\right) ^{-1}\right] _{nm}^{\dagger }\ ^{-}\bar{\psi}_{m}\left(
x^{\prime }\right) \,,  \notag \\
S_{\theta ,+}^{p}\left( x,x^{\prime }\right) &=&-i\sum_{nm}N_{m}^{(+)}\left(
in\right) \,_{-}{\psi }_{n}(x)\,\left[ G(_{+}|^{-})G(_{-}|^{-})^{-1}\right]
_{nm}^{\dagger }\,{_{+}\bar{\psi}}_{m}(x^{\prime })\,,  \notag \\
S_{\theta ,-}^{p}\left( x,x^{\prime }\right) &=&-i\sum_{nm}N_{n}^{(-)}\left(
in\right) \ _{-}{\psi }_{n}(x)\,\left[ G(_{+}|^{-})G(_{-}|^{-})^{-1}\right]
_{nm}^{\dagger }\,{_{+}\bar{\psi}}_{m}(x^{\prime })\,.  \label{emt54}
\end{eqnarray}%
Taking into account (\ref{a.2b}), we can see that $S_{\theta ,\pm }^{p}$\
describe the contributions due to particle-creation.

Then, we represent the real-valued terms $\langle j_{\mu }\left( t\right)
\rangle ^{\theta }$\ and $\langle T_{\mu \nu }\left( t\right) \rangle
^{\theta }$\ in (\ref{emt24.2}) as follows:%
\begin{eqnarray}
&\langle j_{\mu }\left( t\right) \rangle ^{\theta }=&\mathrm{\func{Re}}%
\langle j_{\mu }\left( t\right) \rangle _{\theta }^{c}+\mathrm{\func{Re}}%
\langle j_{\mu }\left( t\right) \rangle _{\theta }^{p}\,,  \notag \\
&\langle T_{\mu \nu }\left( t\right) \rangle ^{\theta }=&\mathrm{\func{Re}}%
\langle T_{\mu \nu }\left( t\right) \rangle _{\theta }^{c}+\mathrm{\func{Re}}%
\langle T_{\mu \nu }\left( t\right) \rangle _{\theta }^{p}\,,  \label{emt64}
\end{eqnarray}%
where 
\begin{eqnarray}
&\langle j_{\mu }\left( t\right) \rangle _{\theta }^{c}=&iq\sum_{\zeta =\pm
}\left. \mathrm{tr}\left[ \gamma _{\mu }S_{\theta ,\zeta }(x,x^{\prime })%
\right] \right| _{x=x^{\prime }}\,,\;\langle T_{\mu \nu }\left( t\right)
\rangle _{\theta }^{c}=i\sum_{\zeta =\pm }\left. \mathrm{tr}\left[ A_{\mu
\nu }S_{\theta ,\zeta }(x,x^{\prime })\right] \right| _{x=x^{\prime }}\,, 
\notag \\
&\langle j_{\mu }\left( t\right) \rangle _{\theta }^{p}=&iq\sum_{\zeta =\pm
}\left. \mathrm{tr}\left[ \gamma _{\mu }S_{\theta ,\zeta }^{p}(x,x^{\prime })%
\right] \right| _{x=x^{\prime }}\,,\;\langle T_{\mu \nu }\left( t\right)
\rangle _{\theta }^{p}=i\sum_{\zeta =\pm }\left. \mathrm{tr}\left[ A_{\mu
\nu }S_{\theta ,\zeta }^{p}(x,x^{\prime })\right] \right| _{x=x^{\prime }}\,.
\label{emt65}
\end{eqnarray}

Because of the symmetry $\gamma ^{0}\tilde{S}_{\theta }^{\mp }\left(
x,x^{\prime }\right) ^{\dagger }\gamma ^{0}=\tilde{S}_{\theta }^{\mp }\left(
x^{\prime },x\right) $, which can be observed in representation (\ref{emt13}%
), we can calculate the real-valued parts of the right-hand sides of (\ref%
{emt65}) at $x=x^{\prime }$\ as follows:%
\begin{eqnarray}
\left. \mathrm{tr}\left[ \gamma _{\mu }S_{\theta ,+}(x,x^{\prime })\right]
\right| _{x=x^{\prime }} &=&\left. \lim_{x_{0}\rightarrow x_{0}^{\prime }+0}%
\mathcal{\,}\mathrm{tr}\left[ \gamma _{\mu }S_{\theta ,+}(x,x^{\prime })%
\right] \right| _{\mathbf{x=x}^{\prime }}\mathrm{\,},  \notag \\
\left. \mathrm{tr}\left[ \gamma _{\mu }S_{\theta ,-}(x,x^{\prime })\right]
\right| _{x=x^{\prime }} &=&\left. \lim_{x_{0}\rightarrow x_{0}^{\prime }-0}%
\mathcal{\,}\mathrm{tr}\left[ \gamma _{\mu }S_{\theta ,-}(x,x^{\prime })%
\right] \right| _{\mathbf{x=x}^{\prime }}\,,  \label{emt65.1}
\end{eqnarray}%
\emph{\bigskip }and so on.

\subsection{Proper-time representation of $S_{\protect\theta ,\protect\zeta %
} $}

As has been mentioned, we know the form of the function $\tilde{S}_{\theta
}^{\mp }$ in (\ref{emt13}) until the moment $t_{1}$ the electric field turns
on. Therefore, we continue our analysis in the case of $t>t_{1},$ $t^{\prime
}>t_{1}$, which we have began in the previous section. We are now interested
in the components of this functions, which are denoted in (\ref{emt54}) as $%
S_{\theta ,\zeta }$.

Because of the cutting-off factor, the integrals $S_{\theta ,\zeta }$ are
finite. However, they have derivatives proportional to the time interval, $%
t-t_{1}$, which may be large. Therefore, in the study of conventional
contributions we have to pay attention to the contributions that grow as $%
t-t_{1}$ increases. Note that, as distinct from the contributions $J_{\mu
}^{p}\left( t\right) $ and $\tau _{\mu \nu }^{p}\left( t\right) $, which
have been estimated only for sufficiently large intervals $t-t_{1}$,
conventional contributions are calculated for any $t$, $0\leq t-t_{1}\leq T$.

We remind that the functions $_{\zeta }\psi _{n}(x)$\ and $^{\zeta }\psi
_{n}\left( x\right) $ are solutions of the Dirac equation in a constant
field, described in Subsec. 3.1. In the relation (\ref{e40a}) of that
subsection, we have presented the expressions for the coefficients $G\left(
\left. _{-}\right| ^{+}\right) $, obtained for the $T$-constant field, which
determine all the coefficients $G\left( \left. _{\zeta }\right| ^{\zeta
}\right) $ via the unitarity condition. These expressions for $G\left(
\left. _{-}\right| ^{+}\right) $ at longitudinal momenta $p_{3}$ being large
and comparable with $\left| qE\right| T$ are different from the asymptotic
form at $T\rightarrow \infty $. Accordingly, for such large values $p_{3}$
the coefficients $G\left( \left. _{\zeta }\right| ^{\zeta }\right) $ are
different from the corresponding asymptotic forms. However, it can be shown
that under condition (\ref{tcond}) these differences in expressions for $%
S_{\theta ,\zeta }$ can be neglected, and one can use, for every $p_{3}$,
the asymptotic form of the coefficients $G\left( \left. _{\zeta }\right|
^{\zeta }\right) $, which holds true for a constant field. On the other
hand, we need to attract attention to the above-mentioned fact, which will
be used to obtain the necessary representation, whereas a manifest form of
these coefficients in a constant field will be unnecessary.

The energy spectrum of $in$-particles at $t<t_{1}$\ has the form 
\begin{equation*}
\varepsilon _{n}^{(\zeta )}=\varepsilon _{n}=\sqrt{M^{2}+\omega +\left( \pi
_{3}\right) ^{2}},\;,\;\pi _{3}=\left( \frac{qET}{2}+p_{3}\right) ,
\end{equation*}%
where 
\begin{eqnarray*}
&&n=(p_{1},n_{B},p_{3},r),\;\omega =|qB|(2n_{B}+1-r)\,,\;n_{B}=0,1,\ldots
,\;B\neq 0\,; \\
&&n=(\mathbf{p},r),\;\omega =p_{1}^{2}+p_{2}^{2}\,,\;\;B=0\,.
\end{eqnarray*}%
We expand $N_{n}^{\left( \zeta \right) }\left( in\right) $\ (\ref{in}) as a
power series of $\exp \left\{ -\beta \left( \varepsilon _{n}^{(\zeta )}-\mu
^{(\zeta )}\right) \right\} $\ and make use of the formula (see \cite{GR94},
Eqs. 3.471.9 and 8.469.3)\emph{\ }%
\begin{equation}
\exp \left\{ -\beta l\varepsilon _{n}\right\} =\frac{1}{2\sqrt{\pi }}%
\int_{0}^{\infty }\frac{du}{u^{1/2}}\exp \left( -\frac{u}{4}-\frac{\beta
^{2}l^{2}\varepsilon _{n}^{2}}{u}\right) \,.  \label{emt55.2}
\end{equation}%
Since the operator $\left( \mathcal{H}\left( t_{1}\right) \right)
^{2}=M^{2}-\left( \boldsymbol{\gamma }\mathbf{P}_{\perp }\right) ^{2}+\left(
i\partial _{3}+qET/2\right) ^{2}$\ is an integral of motion in the $T$%
-constant field, the relation 
\begin{equation*}
\left( \mathcal{H}\left( t_{1}\right) \right) ^{2}\,_{\pm }\psi _{n}\left(
x\right) =\varepsilon _{n}^{2}\,_{\pm }\psi _{n}\left( x\right)
\end{equation*}%
holds true at any time instant $t$. Then, $S_{\theta ,\zeta }$\ (\ref{emt54}%
) can be represented as follows: 
\begin{eqnarray}
S_{\theta ,+}\left( x,x^{\prime }\right) &=&i\sum_{n,m}\Upsilon ^{\left(
+\right) }{}^{+}\psi _{n}\left( x\right) G\left( \left. _{+}\right\vert
^{+}\right) _{nm}^{-1}\ _{+}\bar{\psi}_{m}\left( x^{\prime }\right) \,, 
\notag \\
S_{\theta ,-}\left( x,x^{\prime }\right) &=&-i\sum_{n,m}\Upsilon ^{\left(
-\right) }\,_{-}\psi _{n}\left( x\right) \left[ G\left( \left.
_{-}\right\vert ^{-}\right) ^{-1}\right] _{nm}^{\dagger }\ ^{-}\bar{\psi}%
_{m}\left( x^{\prime }\right) \,,  \label{emt56.1}
\end{eqnarray}%
where 
\begin{eqnarray*}
\Upsilon ^{\left( \zeta \right) } &=&\sum_{l=1}^{\infty }\left( -1\right)
^{l}e^{\beta l\mu ^{(\zeta )}}\frac{1}{2\sqrt{\pi }}\int_{0}^{\infty }\frac{%
du}{u^{1/2}}\exp \left( -\frac{u}{4}-\frac{\beta ^{2}l^{2}\mathcal{E}_{\zeta
}^{2}}{u}\right) \,, \\
\mathcal{E}_{+}^{2} &=&M^{2}-\left( \boldsymbol{\gamma }\mathbf{P}_{\perp
}^{\prime \ast }\right) ^{2}+\left( -i\partial _{3}^{\prime }+qET/2\right)
^{2}, \\
\mathcal{E}_{-}^{2} &=&M^{2}-\left( \boldsymbol{\gamma }\mathbf{P}_{\perp
}\right) ^{2}+\left( i\partial _{3}+qET/2\right) ^{2}\,.
\end{eqnarray*}

Representations (\ref{emt56.1}) and (\ref{emt19}) are formally related as
follows:\emph{\ }%
\begin{equation}
S_{\theta ,\pm }\left( x,x^{\prime }\right) =\pm \Upsilon ^{\left( \pm
\right) }S^{\mp }\left( x,x^{\prime }\right) \,.  \label{emt56.2}
\end{equation}

To calculate $\langle j_{\mu }\left( t\right) \rangle _{\theta }^{c}$\ and $%
\langle T_{\mu \nu }\left( t\right) \rangle _{\theta }^{c}$\ in (\ref{emt65}%
) according to the prescription (\ref{emt65.1}), we need the expressions for 
$S_{\theta ,+}\left( x,x^{\prime }\right) $\ at $x_{0}>x_{0}^{\prime }$\ and 
$S_{\theta ,-}\left( x,x^{\prime }\right) $\ at $x_{0}<x_{0}^{\prime }$\
only. Under this condition, the proper-time representation for $S^{\mp
}(x,x^{\prime })$\ has the form\emph{\ }%
\begin{eqnarray}
S^{\mp }(x,x^{\prime }) &=&(\gamma P+M)\Delta ^{\mp }(x,x^{\prime })\,, 
\notag \\
\pm \Delta ^{\mp }(x,x^{\prime }) &=&\Delta ^{c}(x,x^{\prime
})\,,\;x_{0}-x_{0}^{\prime }\gtrless 0\,,  \label{emt56.3}
\end{eqnarray}%
where $\Delta ^{c}(x,x^{\prime })$\ is defined by the proper-time
representations (\ref{emt31}) and (\ref{emt33}); see \cite{N79,Nik70,GGG98}.
For the function $S_{\theta ,\zeta }\left( x,x^{\prime }\right) $, we find%
\emph{\ }%
\begin{eqnarray}
S_{\theta ,\zeta }\left( x,x^{\prime }\right) &=&(\gamma P+M)\Delta _{\theta
,\zeta }^{c}\left( x,x^{\prime }\right) \,,  \notag \\
\Delta _{\theta ,\zeta }^{c}(x,x^{\prime }) &=&\int_{0}^{\infty }f_{\theta
,\zeta }(x,x^{\prime },s)ds\,,\;\;f_{\theta ,\zeta }(x,x^{\prime
},s)=\Upsilon ^{\left( \zeta \right) }f(x,x^{\prime },s)\,,  \label{emt65.3}
\end{eqnarray}%
where $f(x,x^{\prime },s)$\ is given by (\ref{emt33}).

Thus, we also have proper-time representations for the quantities $\langle
j_{\mu }\left( t\right) \rangle _{\theta }^{c}$\ and $\langle T_{ik}\left(
t\right) \rangle _{\theta }^{c}$. The latter quantities are finite at finite 
$t-t_{1}.$\ This implies that representation (\ref{emt65.3}) can be used in
calculating the quantities $\langle j_{\mu }\left( t\right) \rangle _{\theta
}^{c}$\ and $\langle T_{ik}\left( t\right) \rangle _{\theta }^{c}.$\ Then,
the quantity $\langle T_{00}\left( t\right) \rangle _{\theta }^{c}$\ must be
calculated via $\langle T_{ik}\left( t\right) \rangle _{\theta }^{c}$\ as
follows:%
\begin{equation}
\langle T_{00}\left( t\right) \rangle _{\theta }^{c}=\sum_{k=1,2,3}\langle
T_{kk}\left( t\right) \rangle _{\theta }^{c}+iM\sum_{\zeta =\pm }\left. 
\mathrm{tr\,}S_{\theta ,\zeta }(x,x^{\prime })\right| _{x=x^{\prime }}\;,
\label{emt56.4}
\end{equation}%
The latter follows from (\ref{emt65}), with allowance for the fact that $%
S_{\theta ,\zeta }$\ are solutions of the Dirac equation. Then, we represent
expressions in (\ref{emt65}) and (\ref{emt56.4}) in the form\emph{\ }%
\begin{eqnarray}
&\langle j_{\mu }\left( t\right) \rangle _{\theta }^{c}=&iq\sum_{\zeta =\pm
}\left. \mathrm{tr}\,\left[ \gamma _{\mu }\gamma P\Delta _{\theta ,\zeta
}^{c}(x,x^{\prime })\right] \right| _{x=x^{\prime }}\,,  \notag \\
&\langle T_{\mu \nu }\left( t\right) \rangle _{\theta }^{c}=&i\sum_{\zeta
=\pm }\left. \mathrm{tr}\,\left[ A_{\mu \nu }\gamma P\Delta _{\theta ,\zeta
}^{c}(x,x^{\prime })\right] \right| _{x=x^{\prime }}\,,\;\mathrm{\;}\mu ,\nu
\neq 0\,,  \notag \\
&\langle T_{00}\left( t\right) \rangle _{\theta }^{c}=&\sum_{k=1,2,3}\langle
T_{kk}\left( t\right) \rangle _{\theta }^{c}+iM^{2}\sum_{\zeta =\pm }\left. 
\mathrm{tr}\,\left[ \Delta _{\theta ,\zeta }^{c}(x,x^{\prime })\right]
\right| _{x=x^{\prime }}\,.  \label{emt65.2}
\end{eqnarray}

Following Schwinger, see \cite{S51}, the kernel $f(x,x^{\prime },s)$\ can be
treated as a matrix element of an operator in the $x$-representation:%
\begin{equation}
f(x,x^{\prime },s)=i\langle x\left| \exp \left\{ -is\left[ M^{2}-\left(
\gamma \hat{P}\right) ^{2}\right] \right\} \right| x^{\prime }\rangle \,.
\label{emt58}
\end{equation}%
Here, $|x^{\prime }\rangle $\ is a complete set of eigenvectors of some
commuting operators $\hat{X}^{\mu }$, such that $\hat{X}^{\mu }|x\rangle
=x^{\mu }|x\rangle $, $\langle x|x^{\prime }\rangle =\delta
^{(4)}(x-x^{\prime })$. There exist canonically conjugate operators $\hat{P}%
_{\mu }$\ and the commutation relations $\left[ \hat{X}^{\mu },\hat{P}_{\nu }%
\right] =i\delta _{\nu }^{\mu }$, $\left[ \hat{P}_{\nu },\hat{P}_{\nu }%
\right] =-iqF_{\mu \nu }$, where $F_{\mu \nu }$\ is{\Large \ }the field
strength of the external field under consideration. Using the representation
(\ref{emt58}),\ we rewrite $f_{\theta ,\zeta }(x,x^{\prime },s)$\ (\ref%
{emt65.3}) as follows:\emph{\ }%
\begin{equation}
f_{\theta ,\zeta }(x,x^{\prime },s)=\sum_{l=1}^{\infty }\left( -1\right)
^{l}e^{\beta l\mu ^{(\zeta )}}\frac{1}{2\sqrt{\pi }}\int_{0}^{\infty }\frac{%
du}{u^{1/2}}e^{-u/4}g(x,x^{\prime },s,\tau ),  \label{emt59}
\end{equation}%
where 
\begin{eqnarray}
g(x,x^{\prime },s,\tau ) &=&\exp \left[ -\tau \left( i\partial
_{3}+qET/2\right) ^{2}\right] f(x,x^{\prime },s,\tau ),\;\;\tau =\left(
\beta l\right) ^{2}/u\,,  \notag \\
f(x,x^{\prime },s,\tau ) &=&i\exp \left\{ -\tau \left[ M^{2}-\left( \gamma 
\hat{P}_{\perp }\right) ^{2}\right] \right\} \langle x\left| \exp \left\{ -is%
\left[ M^{2}-\left( \gamma \hat{P}\right) ^{2}\right] \right\} \right|
x^{\prime }\rangle ,  \label{gfun}
\end{eqnarray}%
and 
\begin{equation*}
\left( \gamma \hat{P}\right) ^{2}=\left( \gamma \hat{P}_{\parallel }\right)
^{2}+\left( \gamma \hat{P}_{\bot }\right) ^{2},\;\;\hat{P}_{\perp }=\left( 0,%
\hat{P}_{1},\hat{P}_{2},0\right) ,\;\hat{P}_{\parallel }=\left( \hat{P}%
_{0},0,0,\hat{P}_{3}\right) .
\end{equation*}

Since the operators $\left( \gamma \hat{P}_{\parallel }\right) ^{2}$\ and $%
\left( \gamma \hat{P}_{\bot }\right) ^{2}$\ commute, we present $%
f(x,x^{\prime },s,\tau )$\ as follows%
\begin{equation*}
f(x,x^{\prime },s,\tau )=i\langle x\left| \exp \left\{ is\left( \gamma \hat{P%
}_{\parallel }\right) ^{2}-i\left( s-i\tau \right) \left[ M^{2}-\left(
\gamma \hat{P}_{\perp }\right) ^{2}\right] \right\} \right| x^{\prime
}\rangle \,.
\end{equation*}%
Applying the Schwinger operator techniques \cite{S51} to this expression, we
find 
\begin{eqnarray}
&&f(x,x^{\prime },s,\tau )=f_{\Vert }(x_{\Vert },x_{\Vert }^{\prime
},s)f_{\bot }(\mathbf{x}_{\bot },\mathbf{x}_{\bot }^{\prime },s-i\tau )\,, 
\notag \\
&&f_{\Vert }(x_{\Vert },x_{\Vert }^{\prime },s)=\exp \left[ iq\Lambda
_{\parallel }+\frac{i}{4}\left( x_{3}-x_{3}^{\prime }\right) ^{2}qE\coth
\left( qEs\right) \right] f_{0}(x_{0},x_{0}^{\prime },s)\,,  \notag \\
&&f_{0}(x_{0},x_{0}^{\prime },s)=\frac{iqE}{4\pi \sinh \left( qEs\right) }%
\exp \left[ qEs\gamma ^{0}\gamma ^{3}-\frac{i}{4}\left( x_{0}-x_{0}^{\prime
}\right) ^{2}qE\coth \left( qEs\right) \right] \,,  \notag \\
&&f_{\bot }(\mathbf{x}_{\bot },\mathbf{x}_{\bot }^{\prime },s-i\tau )=\frac{%
-iqB}{4\pi \sin \left[ qB\left( s-i\tau \right) \right] }\exp \left[
iq\Lambda _{\perp }-i\left( s-i\tau \right) \left( M^{2}-qB\Sigma
^{3}\right) +\frac{iqB\left( \mathbf{x}_{\perp }-\mathbf{x}_{\perp }^{\prime
}\right) ^{2}}{4\tan \left[ qB\left( s-i\tau \right) \right] }\right] \,, 
\notag \\
&&iq\Lambda _{\perp }=-i\frac{qB}{2}\left( x_{1}-x_{1}^{\prime }\right)
\left( x_{2}+x_{2}^{\prime }\right) \,.  \label{emt60.1}
\end{eqnarray}

The action of the exponential operator in (\ref{gfun}) is determined by the
expression 
\begin{equation*}
\varphi (s,\tau )=\exp \left[ -\tau \left( i\partial _{3}+qET/2\right) ^{2}%
\right] \exp \left[ iq\Lambda _{\parallel }+\frac{i}{4}\left(
x_{3}-x_{3}^{\prime }\right) ^{2}qE\coth \left( qEs\right) \right] .
\end{equation*}%
Using the integral representation 
\begin{equation*}
\int_{-\infty }^{+\infty }e^{-ip^{2}a+ibp}dp=\left( e^{-i\pi /2}\pi
a^{-1}\right) ^{1/2}e^{ib^{2}/4a}\,,
\end{equation*}%
we find 
\begin{eqnarray}
\varphi (s,\tau ) &&\,=\left( \frac{a}{a-i\tau }\right) ^{1/2}\exp \left\{
iq\Lambda _{\parallel }-\left( x_{0}+x_{0}^{\prime }+T\right) ^{2}\left( 
\frac{qE}{2}\right) ^{2}\frac{\tau a}{a-i\tau }\right.  \notag \\
&&+\left. i\frac{b^{2}+i2b\tau qE\left( x_{0}+x_{0}^{\prime }+T\right) }{%
4\left( a-i\tau \right) }\right\} \,,  \label{emt60.2}
\end{eqnarray}%
where 
\begin{equation*}
a=a\left( s\right) =\left( qE\right) ^{-1}\tanh \left( qEs\right)
\,,\;b=x_{3}-x_{3}^{\prime }\,.
\end{equation*}%
Therefore, 
\begin{equation}
g(x,x^{\prime },s,\tau )=\varphi (s,\tau )f_{0}(x_{0},x_{0}^{\prime
},s)f_{\bot }(x_{\bot },x_{\bot }^{\prime },s-i\tau )\,.  \label{emt60.3}
\end{equation}

\subsection{A convenient representation for weak-field approximation}

First of all, we note that at finite temperatures the kernels in (\ref%
{emt61.1}) and (\ref{emt61.2}) have singularities due to the zeros of $%
\left( a-i\tau \right) $. If we introduce a new variable $s^{\prime },$%
\begin{equation*}
s^{\prime }=is+c_{0}\,,\;c_{0}=\left( qE\right) ^{-1}\arctan \left( qE\tau
\right) \,,
\end{equation*}%
then 
\begin{equation}
\left( a-i\tau \right) ^{-1}=\frac{qE\left[ 1+\tan \left( qEs^{\prime
}\right) qE\tau \right] }{\tan \left( qEs^{\prime }\right) \left[ 1+\left(
qE\tau \right) ^{2}\right] }\,.  \label{sing}
\end{equation}%
The singular points of $\left( a-i\tau \right) ^{-1}$ are $s^{\prime }=\pm
\pi k,$ $k=0,1,2,\ldots $. Let us deform the integration contour over $s$ in
integrals (\ref{emt61.1}) and (\ref{emt61.2}) so that it leaves the origin
and proceeds downwards along the imaginary axis until the point $-ic_{1}$, $%
c_{1}=\left( \pi /2-c_{0}\right) \left| qE\right| ^{-1}$, without touching
it, and then proceeds in parallel to the axis $\func{Re}s$ towards the
positive infinity. Then, expressions (\ref{emt61.1}) and (\ref{emt61.2}) can
be reorganized as follows: 
\begin{eqnarray}
J_{\mu }^{(l)} &=&J_{\mu }^{a}+J_{\mu }^{b}\,,  \notag \\
J_{\mu }^{a} &=&-i\int_{0}^{c_{1}-0}d\tilde{s}\int_{0}^{\infty }du\;\left. %
\left[ h\left( s,u\right) Y\left( s,u\right) j_{\mu }\left( s,u\right) %
\right] \right| _{s=-i\tilde{s}}\,,  \notag \\
J_{\mu }^{b} &=&\int_{0}^{\infty }d\tilde{s}\int_{0}^{\infty }du\;\left. %
\left[ h\left( s,u\right) Y\left( s,u\right) j_{\mu }\left( s,u\right) %
\right] \right| _{s=\tilde{s}-i\left( c_{1}-0\right) }\,,  \label{emt61.4}
\end{eqnarray}%
and 
\begin{eqnarray}
T_{\mu \nu }^{(l)} &=&T_{\mu \nu }^{a}+T_{\mu \nu }^{b}\,,  \notag \\
T_{\mu \nu }^{a} &=&-i\int_{0}^{c_{1}-0}d\tilde{s}\int_{0}^{\infty
}du\;\left. \left[ h\left( s,u\right) Y\left( s,u\right) t_{\mu \nu }\left(
s,u\right) \right] \right| _{s=-i\tilde{s}}\,,  \notag \\
T_{\mu \nu }^{b} &=&\int_{0}^{\infty }d\tilde{s}\int_{0}^{\infty }du\;\left. %
\left[ h\left( s,u\right) Y\left( s,u\right) t_{\mu \nu }\left( s,u\right) %
\right] \right| _{s=\tilde{s}-i\left( c_{1}-0\right) }\,,  \label{emt61.5}
\end{eqnarray}

One can verify that the quantities 
\begin{eqnarray*}
\delta J_{\mu }^{a} &=&-i\int_{0}^{\infty }du\;\int_{e^{i\pi }\left(
c_{0}-0\right) }^{0}d\tilde{s}\left. \left[ h\left( s,u\right) Y\left(
s,u\right) j_{\mu }\left( s,u\right) \right] \right| _{s=-i\tilde{s}}\,, \\
\delta T_{\mu \nu }^{a} &=&-i\int_{0}^{\infty }du\;\int_{e^{i\pi }\left(
c_{0}-0\right) }^{0}d\tilde{s}\left. \left[ h\left( s,u\right) Y\left(
s,u\right) t_{\mu \nu }\left( s,u\right) \right] \right| _{s=-i\tilde{s}}\,,
\end{eqnarray*}
are imaginary; therefore the real-valued parts of $J_{\mu }^{a}$ and $T_{\mu
\nu }^{a}$ can be represented as 
\begin{eqnarray}
&&\func{Re}J_{\mu }^{a}=\func{Re}J_{\mu }^{\prime a}\,,\;\;\func{Re}T_{\mu
\nu }^{a}=\func{Re}T_{\mu \nu }^{\prime a}\,,  \notag \\
&&J_{\mu }^{\prime a}=J_{\mu }^{a}+\delta J_{\mu }^{a}\,,\;\;T_{\mu \nu
}^{\prime a}=T_{\mu \nu }^{a}+\delta T_{\mu \nu }^{a}\,.  \label{emt61.6}
\end{eqnarray}

We now change the variable $\tilde{s}$ by $s^{\prime }=\tilde{s}+c_{0}$ in
the integrals for $J_{\mu }^{\prime a}$ and $T_{\mu \nu }^{\prime a}$. Then,
one can see that the contributions from the integration region\emph{\ }$%
0\leq u<u_{0}$ over $u$ in these integrals are imaginary. Then, changing the
variable $u$ by $u^{\prime },$%
\begin{equation*}
u^{\prime }=u-u_{0}\,,\;\;u_{0}=qE\left( \beta l\right) ^{2}\cot \left(
qEs^{\prime }\right) \,,
\end{equation*}%
and introducing the notation $\func{Re}J_{\mu }^{\prime a}=\tilde{J}_{\mu
}^{a}\,,\;\;\func{Re}T_{\mu \nu }^{\prime a}=\tilde{T}_{\mu \nu }^{a}\,$, we
represent\ $\tilde{J}_{\mu }^{a}$ and $\tilde{T}_{\mu \nu }^{a}$ as (\ref%
{emt61.7}).

\subparagraph{\protect\large Acknowledgement}

S.P.G. thanks FAPESP for support and Universidade de São Paulo for
hospitality. D.M.G. acknowledges FAPESP and CNPq for permanent support.

\end{document}